\documentclass[fleqn,usenatbib]{mnras}
\usepackage{newtxtext,newtxmath}
\usepackage[T1]{fontenc}
\usepackage{graphicx}	
\usepackage{amsmath}	
\usepackage{CJKutf8}
\usepackage{array}
\usepackage[figuresright]{rotating}
\usepackage{tabularx}

\title{Spectroscopic confirmation of high-amplitude eruptive YSOs and dipping giants from the VVV survey}

\author[Z. Guo et al.]{Zhen Guo$^{1, 2, 3, 4}$\thanks{E-mail: zhen.guo@uv.cl},
P. W. Lucas$^{3}$, R. Kurtev$^{1,5}$, J. Borissova$^{1,5}$ , C. Contreras Pe{\~n}a$^{6, 7}$,
\newauthor
 S. N. Yurchenko$^{8}$, L. C. Smith$^{9}$, D. Minniti$^{10, 11, 12}$, R. K. Saito$^{12}$, A. Bayo$^{13}$,  M. Catelan$^{14, 5}$,
\newauthor
  J. Alonso-Garc\'{i}a$^{15,5}$, A. Caratti o Garatti$^{16,17}$, C. Morris$^{3}$,  D. Froebrich$^{18}$, J. Tennyson$^{7}$, 
 \newauthor
 K. Mauc\'{o}$^{13, 2}$, A. Aguayo$^{1,2}$, N. Miller$^{3}$, and H. D. S. Muthu$^{3}$
\\
$^{1}$Instituto de F{\'i}sica y Astronom{\'i}a, Universidad de Valpara{\'i}so, ave. Gran Breta{\~n}a, 1111, Casilla 5030, Valpara{\'i}so, Chile\\
$^{2}$N\'ucleo Milenio de Formaci\'on Planetaria (NPF), ave. Gran Breta{\~n}a, 1111, Casilla 5030, Valpara{\'i}so, Chile\\
$^{3}$Centre for Astrophysics Research, University of Hertfordshire, Hatfield AL10 9AB, UK\\
$^{4}$Departamento de F{\'i}sica, Universidad Tecnic{\'a} Federico Santa Mar{\'i}a, Avenida Espa{\~n}a 1680, Valpara{\'i}so, Chile\\
$^{5}$Millennium Institute of Astrophysics, Nuncio Monse{\~n}or Sotero Sanz 100, Of. 104, Providencia, Santiago, Chile\\
$^{7}$Research Institute of Basic Sciences, Seoul National University, Seoul 08826, Republic of Korea\\
$^{8}$Department of Physics and Astronomy, University College London, Gower Street, WC1E 6BT London, UK\\
$^{9}$Institute of Astronomy, University of Cambridge, Madingley Road, Cambridge, CB3 0HA, UK\\
$^{10}$Departamento de Ciencias Fisicas, Universidad Andres Bello, Republica 220, 8320000 Santiago, Chile\\
$^{11}$Vatican Observatory, V00120 Vatican City State, Italy\\
$^{12}$Departamento de F{\'i}sica, Universidade Federal de Santa Catarina, Trindade 88040-900, Florianopol{\'i}s, SC, Brazil\\
$^{13}$ European Organisation for Astronomical Research in the Southern Hemisphere (ESO), Karl-Schwarzschild-Str. 2, 85748 Garching bei München, Germany\\
$^{14}$Instituto de Astrof{\'i}sica, Facultad de F{\'i}sica, Pontificia Universidad Cat{\'o}lica de Chile, Av. Vicu{\~n}a Mackenna 4860, 7820436 Macul, Santiago, Chile\\
$^{15}$Centro de Astronom{\'i}a (CITEVA), Universidad de Antofagasta, Av. Angamos 601, 02800 Antofagasta, Chile\\
$^{16}$INAF - Osservatorio Astronomico di Capodimonte, salita Moiariello 16, 80131, Napoli, Italy\\
$^{17}$Dublin Institute for Advanced Studies, School of Cosmic Physics, Astronomy and Astrophysics Section, 31 Fitzwilliam Place, Dublin 2, Ireland\\
$^{18}$Centre for Astrophysics and Planetary Science, University of Kent, Canterbury CT2 7NH, UK\\
}

\date{Accepted XXX. Received YYY; in original form ZZZ}
\pubyear{2023}

\begin{document}
\label{firstpage}
\maketitle

\begin{abstract}

During the pre-main-sequence (pre-MS) evolution stage of a star, significant amounts of stellar mass are accreted during episodic accretion events, such as multi-decade FUor-type outbursts. Here, we present a near-infrared spectroscopic follow-up study of 33 high-amplitude (most with $\Delta K_s >$~4~mag) variable sources discovered by the Vista Variables in the Via Lactea (VVV) survey. Based on the spectral features, 25 sources are classified as eruptive young stellar objects (YSOs), including 15 newly identified FUors, six with long-lasting but EXor-like bursts of magnetospheric accretion and four displaying outflow-dominated spectra. By examining the photometric behaviours of eruptive YSOs, we found most FUor-type outbursts have higher amplitudes ($\Delta K_s$ and $\Delta W2$), faster eruptive timescales and bluer infrared colours than the other outburst types. In addition, we identified seven post-main sequence variables apparently associated with deep dipping events and an eruptive star with deep AlO absorption bands resembling those seen in the V838 Mon stellar merger.



\end{abstract}


\section{Introduction}
\label{sec:intro}

The mass accretion process plays a critical role on many astrophysical scales, from young stellar objects (YSOs) to supermassive black holes, with variability as a main character \citep[see][]{Scaringi2015}. The protostellar and pre-MS stellar evolution is dominated by the mass accretion \citep[reviewed by][]{Hartmann2016} with unstable nature on timescales from hours to decades \citep[see][]{Hillenbrand2015}. Stellar evolutionary models \citep{Hartmann1996} have predicted multiple accretion bursts throughout the protostellar and pre-MS stage, so-called episodic accretion, suggesting that most stellar mass is accumulated during several episodes of accretion outbursts. The episodic accretion model provides a solution to the problem of large luminosity spreads among low-mass protostars \citep{Dunham2012}, as well as the luminosity problems in these objects \citep{Kenyon1990}, though the latter issue has been partially solved by a longer duration of the Class I stage of evolution \citep{Fischer2017}. Moreover, episodic accretion events can efficiently heat the inner accretion disc, which results in the expansion of the snowlines of various molecules that leave impacts on the grain growth and planet formation \citep[e.g.][]{Lee2019NatAs, Jorgensen2020, Kospal2023}. 

Theoretical studies have explored various physical mechanisms triggering episodic accretion on pre-MS sources, such as combinations of gravitational instability (GI) and magneto-hydrodynamic turbulence at the inner accretion disc \citep{Armitage2001, Kratter2016, Bourdarot2023}, self-regulated thermal instability \citep{Bell1994}, thermal instability introduced by a massive young planet inside the accretion disc \citep{Lodato2004, Clarke2005}, the imbalance between gravitational and magneto-rotational instability  \citep[MRI,][]{Zhu2009b, Elbakyan2021}, fragmentation of massive protostellar discs \citep{Vorobyov2005, Vorobyov2010}, infalling of piled-up disc materials outside the star-disc co-rotation radius \citep{DAngelo2010, DAngelo2012}, and flybys from stellar perturber \citep{Cuello2019, Borchert2022}. Nevertheless, there is no universal solution that responds to trigger all types of episodic accretion on YSOs, and more observational evidence is desired to differentiate these mechanisms. 

%
%

\begin{table*} 
\caption{Basic information and photometric behaviours of our targets }
\renewcommand\arraystretch{1.3}
\begin{tabular}{l c c c c c c c c c c c c c}
\hline
\hline
Name & RA (J2000) & Dec (J2000) & $H-K_s$ & $H-K_s$ & LC type & $\Delta K_s  $  &  t$_{\rm var}$ & $\alpha_{\rm class}$ & $d_{\rm SFR}$ & $W1 - W2$ \\
& deg & deg & faint stage & bright stage & &~mag & day & & kpc & faint stage\\ 
L222\_1 &	175.78945 &	-62.35367& 1.49 & 1.18 & Eruptive &  3.9 & >2451 & 0.82  & 9.6 & 2.0 \\
L222\_4	& 185.22517	& -62.63942 & 1.09 & 2.44	& Eruptive & 6.2 & >2651  & 0.59 & 7.6 & 0.9\\ 
L222\_6  &	193.73921 &	-61.04417& - & 1.53 & Eruptive & 4.0 & >2251  &  2.07 & 3.2 & 2.2\\ 
L222\_10   &	201.45950 &	-62.79639& 1.42 & 1.34 & Eruptive & 4.0 & >2953 & -0.58   & - & - \\ 
L222\_13 & 203.04039	& -62.73005& 0.83 &	0.80 & Eruptive & 4.1 & >1921 & - & 3.0 & - \\ 
L222\_15 & 204.54729	& -62.48267& 2.92 & 0.93	& Eruptive & 4.8 & >2951   & -& 2.0 & -\\
L222\_18  &	214.07488 &	-61.37306& 1.97 & 1.47 & Eruptive & 4.0 & >3051  &  -0.03 & - & 0.2\\
L222\_25&	230.39117 &   -57.88889& -0.09 & 1.10  & Eruptive & 3.3 & >2951 & - & 3.1 & - \\  
L222\_28 &	232.57471 &	-55.58203& 2.79 & 1.48 & Dipper & 4.5  & 2548 & 0.09 & - & 1.4\\ 
L222\_32 & 239.45987	& -53.95964& - & 1.97	& Eruptive & 4.2 & >2151 & 1.29 & 2.8 & 2.6\\ 
L222\_33& 239.85950	& -51.95328&  2.15 & 1.58 & Eruptive & 4.2 & >3601 &  0.39$^\star$ & 3.4 & 0.3\\
L222\_37 &	241.77933 &	-49.40261& - & 1.66 & Eruptive & >4.3 & >4151 & 1.36 &  - & 3.1\\  
VVV1636-4744 &	249.15808 &	-47.74558&  2.27  & 2.03 & Eruptive (MTV) & 3.7 & $\sim$300 & 0.96$^\star$ & 4.5 & -\\
VVV1640-4846 &	250.04900 &	-48.78150& 2.33 & 1.52 & Eruptive & 3.2 & >1627  & 0.96$^\star$ & 2.1 & 0.6 \\
L222\_59 & 253.43487	& -43.47205&  1.81 & 1.55 & Eruptive & 5.7 & >2251 & 0.89 & - & 3.1 \\ 
VVVv800 & 258.19183	& -38.42350 & - &  1.88 & Eruptive & 3.2 & 1851  & 0.84 & 1.4 & 2.3\\ 
L222\_73 & 258.65950 &	-38.49140 & - & 3.02 & Eruptive &  >3.9 & >4151  & 1.39 & 6.2 & 0.7 \\  
L222\_78 & 259.58187	& -32.38142 &  0.33 & 0.46 & Eruptive & 4.6 & >2642   & - & 1.1 & 0.4\\ 
L222\_93 & 261.73025	& -34.14661&  1.37 & 1.18 & Eruptive & 4.5 & >3151 & - & 6.5 & -\\ 
L222\_95 & 262.28616	& -33.52969& 1.27 & 0.43 & Eruptive & 5.3 & >2751  & 1.41 &  6.6 & 0.6\\
L222\_120$\dagger$ &	265.37971 &	-31.43672& 3.50 & 1.79  & Eruptive & 5.3 & > 2463  & 0.82 & 4.5 & 2.5\\
L222\_130 & 266.11000 & -29.45197 & 3.91 & 3.62  &  Dipper? & >4.2 & >2451 & - & - & -\\ 
L222\_144 & 266.53671 & -28.68197& - & 3.56 & Dipper &  4.2 & >1450 & 2.08$^\star$ & - & -\\ 
L222\_145 &	266.58596 &	-28.56781 & 3.83  & 2.91 & Dipper & 4.7 & >2551 & 0.85$^\star$ & - & - \\
L222\_148 & 266.64095 & -29.37922& 3.30 &  0.38	& Eruptive & >4.5 & >2751  & 1.67  & 9.1 & 2.5\\
L222\_149 & 266.82162 & -28.32786 & - & 3.98	& Dipper & 5.4 & 1157  & -0.93 & - & 2.5 \\ 
L222\_154 &	266.95925 &	-28.24306 & -  & 4.19 & Dipper & 4.8 & >2165  & 2.16$^\star$ & -& -\\ 
L222\_165 & 267.41287 & -28.44864&  0.85 & 0.95	& Eruptive &  4.3 & >2151  & - &  1.6 & - \\
L222\_167 & 267.60937 & -28.87519& - & 2.78 & Eruptive &  5.7 & >2501   & 2.33 &  7.8  & 3.9  \\
L222\_168 &	267.66867 &	-28.06083 & 2.79  & 1.68  & Dipper & 4.4 & $\sim$400   & -0.93$^\star$ & 8.3 & 1.4 \\ 
L222\_172 &	267.86138 &	-26.12528& - & 3.13  & Dipper? & >5.2 & >4151  &  1.87$^\star$ & - & -\\ 
L222\_192 & 269.43525 & -24.34244&  1.07 & 0.85 & Eruptive &  4.8 & >1881  &   0.40$^\star$ & 6.1 & 0.3\\
L222\_210 & 271.91033 &  -21.76933&  2.37 & 2.21 & Eruptive &  3.6 & >2951  &  0.62$^\star$ & 5.0 & 1.2\\
  \hline
\hline
\end{tabular}
\flushleft{$\star$: targets without $W4$ (22 $\mu$m) and MIPS (24 $\mu$m) detections. $\dagger$: also known as DR4\_v67 from \citet{Guo2021}.}
\flushleft{Multi-timescale variable (MTV): Defined in \citet{Guo2020}, multi-timescale variables have a general variation trend, but their short-term variability (i.e. timescales < 1 yr) is comparable to the long-term variability.}
\flushleft{The $H - K_s$ colours are measured from the VVV photometry.}

\label{tab: info}
\end{table*}

Large-amplitude eruptive events, often attributed to episodic accretion, are observed on YSOs via time-domain surveys from optical to infrared. Dozens of such events have been captured among Class I and II YSOs with extensive diversity in the variation amplitude and timescale \citep[reviewed by][]{Audard2014, Fischer2022}. Traditionally, these events are sorted into two main categories, FUor-type and EXor-type, separated by distinct observational and physical characteristics. Named after FU~Orionis, FUor-type objects experience high-amplitude outbursts with rapid-rising ($t_{\rm rise} <$ 1000~d) and long-lasting (duration some decades) morphologies \citep{Hartmann1996, Kenyon2000}. Until recently, there have been only 30 FUors that have been confirmed by both photometric eruption and spectroscopic observations \citep[see examples in][and the list from Contreras Pe\~na in prep]{Connelley2018}, including recent discoveries via {\it Gaia} time series \citep[e.g.][]{Hillenbrand2018, Szegedi-Elek2020, Hillenbrand2021}. The occurrence rate of such events is about once per 10$^5$ years on T Tauri stars \citep[][]{Scholz2013, Hillenbrand2015, Contreras2019}. During a FUor-type event, the stellar mass accretion rate reaches an extremely high value ($10^{-5}$ to $10^{-4}$ $M_{\odot}{\rm yr}^{-1}$), and the viscous heated inner accretion disc becomes self-luminous and over-shines the stellar photosphere \citep{Zhu2009, Hartmann2011, Liu2022}. Molecular absorption features, such as water and CO bands, are seen in the near-infrared spectra of FUors, arising in the cool surface of the disc.  By contrast, EXor-type outbursts have typical timescales of several hundred days and are sometimes repeatable \citep{Herbig1989, Herbig2008, Kuhn2023}, with significant diversity on a case-by-case basis \citep[e.g.][]{Aspin2009, Lorenzetti2012, Hodapp2020, Park2022}. Spectroscopic follow-ups show that most EXors still maintained the magnetospheric accretion mode during the outburst, with emission features such as H{\sc i} recombination lines and CO bandhead emission \citep[e.g.][]{Lorenzetti2009}. 




The decade-long VISTA Variables in the Via Lactea (VVV) survey and its extension VVVX survey obtained near-infrared photometry of the inner Galactic plane and the Galactic bulge from 2010 to 2020 \citep{Minniti2010, Saito2012, Minniti2016}. Most of the VVV images are observed through the $K_s$ filter with tens of visits per year during 2010 - 2015 and a lower cadence after 2015. In the past years, thousands of high-amplitude near-infrared variables have been identified from the VVV $K_s$ time series, including eruptive YSOs, Miras, symbiotic systems, and ``blinking'' post-main-sequence giants \citep[e.g.][]{Contreras2017, Guo2021, Smith2021, Nikzat2022}. Specifically, \citet{Contreras2017} discovered that large-amplitude eruptive events ($\Delta K_s > 1$~mag, named therein with the prefix ``VVVv'') are an order of magnitude more commonly detected among embedded protostars than more evolved disc-bearing T Tauri systems. In \citet{Contreras2017b}, a new sub-category of young eruptive objects was identified as MNors, which presented mixed photometric and spectroscopic features of FUors and EXors. In \citet{Guo2021} we summarised photometric and spectroscopic behaviours of 61 highly variable YSOs discovered from the VVV survey. We found that despite all FUor-type outbursts having long duration ($t \ge 5$~yr), emission line objects (under the magnetically controlled accretion mode) still predominate among the long-lasting events, with $K_s$ amplitudes between 2 to 4 mag. Questions have been raised on the physical origination and the frequency of these long-lasting magnetospheric controlled accretion bursts, and their impacts on the stellar evolution history. 

In this series of works, Lucas et al. (submitted, hereafter {LSG23}) identified 222 high-amplitude ($\Delta K_s \geq 4$~mag) variable sources from the latest VVV catalogues. In this paper, we present the near-infrared spectroscopic follow-up observation of 33 sources, most of which are eruptive YSO candidates from the aforementioned catalogue. The primary goal of this work is to provide a distinctive spectroscopically confirmed sample of high-amplitude and long-lasting eruptive YSOs, to examine their accretion nature during the eruptive stage, and to link the episodic accretion behaviours to their astrophysical origins. Furthermore, we aim to provide comprehensive statistical views of these long-lasting eruptive events, by combining near- to mid-infrared light curves and spectroscopic characteristics. Highlighted sources will be further studied in follow-up works.

This paper is organised as follows: basic information on target selection, the design of spectroscopic observation and data reduction, and archival photometric data are presented in \S\ref{sec:info}.  Spectroscopic classifications of eruptive events and other sources are described in \S\ref{sec:res}. We present discussions on different types of eruptive behaviours in \S\ref{sec:dis}, including statistical views of eruptive timescales and variation amplitudes. An unusual source is described in \S\ref{sec: L222_148}. Finally, this paper is concluded in \S\ref{sec:con}.

\section{Observation and data reduction}
\label{sec:info}
\subsection{Target selection and basic information}
\label{sec:target}

In the previous work of this series, 222 high-amplitude ($\Delta K_s \geq 4 $ mag) variable or transient sources were selected from a pre-release version of the VVV Infrared Astrometric Catalogue \citep[VIRAC2;][and in prep]{Smith2018}.
{In this work, we selected 31 new targets for observation from {LSG23}, of which 30 were regarded as eruptive YSO candidates, based on their red near-infrared colours and a VVV light curve morphology that indicated a possible outburst, or possibly an extinction-driven dip in some cases that were ambiguous. Among them, 29 sources have high variation amplitude ($\Delta K_s > 4$ mag, with suffix L222\_), and the other two sources have slightly lower amplitudes but also with eruptive light curve morphologies (VVV1640-4846 and VVV1636-4744).

As discussed in {LSG23}, there was a cluster of sources near the Galactic centre with ambiguous light curves that were thought to be either eruptive YSOs, dipping YSOs or dipping giant stars, the latter two options involving variable extinction. The 31st source, L222\_149, clearly had a dip in the VVV light curve so it was observed as an exemplar for comparison. An additional criterion was that the targets were estimated to be bright enough ($K_s <$ 14.5 mag), based on photometry taken in previous years, that a useful spectrum could be obtained in a short observation.

We also observed two additional sources, VVVv800 \citep{Contreras2017b} and  L222\_120 \citep[=DR4\_v67][]{Guo2021} that were previously confirmed as probable eruptive YSOs but had been observed in a relatively faint state and showed only strong outflow signatures such as  H$_2$ and [Fe {\sc ii}], rather than the spectral features associated specifically with FUor-type or EXor-type outbursts. Unfortunately, both targets were again found to be in a faint state at the time the spectra were taken so the new spectra added little to earlier observations.}

Some key information about these targets is presented in Table~\ref{tab: info}, including coordinates, photometric amplitudes, and $d_{\rm SFR}$ as the star-forming region (SFR) distance. As defined in \citet{Contreras2017b}, $d_{\rm SFR}$ is the distance to the SFRs that are within 5$'$ from each target. The list of SFRs around each target is presented in the online supplementary data. Although, the spatially nearby SFR might be a projection effect, $d_{\rm SFR}$ provides us with a rough estimation of the distance to the target. Based on the VVV light curves, we classified our targets into three groups: eruptive, dipper and multi-timescale variables \citep[MTVs; see][]{Guo2020}. Specifically, light curves in the eruptive category have $\Delta K_s > 2$ mag and an overall rising trend. Plus, there is no large-amplitude decaying trend before the ``outburst´´. The total duration of the variation (either eruption or dip) is described by $t_{\rm var}$. The majority of our eruptive sources are either in a post-outbursting decaying stage or still on their brightness plateau, hence we only have the lower limit of $t_{\rm var}$. The $t_{\rm var}$ of dippers is measured between the beginning and the ending of the dipping events. The simultaneous $H - K_s$ colours measured near photometric maxima and minima (when available) are also presented in Table~\ref{tab: info}.

\begin{table}
\caption{Spectroscopic observations of YSOs}
\renewcommand\arraystretch{1.4}
\begin{tabular}{p{1.2cm} p{1cm} p{0.4cm} c p{0.4cm} c c}
\hline
\hline
Name & t$_{\rm obs}$ &  t$_{\rm exp}$ & type & $A_{K_s}$ & RV & $d_k$\\
 & d/m/y &\centering s & &~mag & km/s & kpc \\
 \hline
\multicolumn{2}{l}{XSHOOTER/VLT}\\
\hline
L222\_1 &  16/04/22   & 600   & FUor & 1.1 & 40 & 10.2$^{+1.0}_{-0.8}$ \\
L222\_4  & 28/04/21  &  960  & FUor? & - & -20 & 6.6$^{+0.9}_{-0.9}$ \\
L222\_10   &  16/04/22 & 600  & FUor & 1.5 & -47 &  3.7$^{+0.5}_{-1.0}$\\
L222\_13   & 28/04/21 	& 1200 & FUor  & 0.7 & -30 & 8.4$^{+0.4}_{-1.1}$ \\
L222\_15 & 03/05/21	& 960  & FUor &  0.7 & -15 & 1.2$^{+0.4}_{-0.7}$ \\
L222\_18  &  16/04/22  & 960  & FUor  & 1.7 & -30 & 2.0$^{+0.5}_{-0.6}$ \\
L222\_25& 15/04/22   & 960  & FUor & 0.9 & -10 & 0.8$^{+0.3}_{-0.6}$ \\
L222\_33& 29/04/21	& 960  & FUor & 1.9 & -70 & 4.2$^{+0.5}_{-0.5}$ \\
VVV1640-4846 & 15/05/22  & 960  & FUor  & 2.0 &  -40 & 2.8$^{+0.3}_{-0.2}$\\
L222\_73 & 15/05/22   & 960  & FUor? & - & -30 & 3.8$^{+0.6}_{-1.1}$ \\
L222\_78 & 16/04/21	& 600  & FUor & 0.1 & -10 & 4.2$^{+1.2}_{-1.5}$ \\
L222\_93 & 29/04/21&  960  & FUor& 1.0 & -95 & 7.8$^{+0.3}_{-0.4}$ \\
L222\_95 & 29/04/21 	& 960  & FUor & 1.8 & -50 & 6.9$^{+0.4}_{-1.0}$ \\
L222\_165 &  04/05/21	& 960  & FUor & 0.4 & -30 & - \\
L222\_192 &  01/07/21 	& 960  & FUor & 0.9 & 5 & 3.0$^{+1.1}_{-2.0}$ \\
L222\_6  & 16/04/22   & 960  & Em. Line & 2.4& - & - \\
L222\_28 &   03/05/22  & 960  & Em. Line & - & -10 & 0.9$^{+0.3}_{-0.8}$\\
VVV1636-4744&  11/05/22  & 960  & Em. Line & - & - \\
L222\_148 &  04/05/21 	& 960  & Em. Line & 1.4 & -10  & 8.3$^{+0.2}_{-0.2}$\\
L222\_167 &  29/04/21	& 960  & Em. Line  & 0.6 & -5  & 7.8$^{+0.8}_{-3.1}$\\
L222\_167 & 09/06/21 	& 960  & Em. Line & 0.4  & -5 & 7.8$^{+0.8}_{-3.1}$\\
L222\_210 &  10/07/21 	& 960  & Em. Line  & 0.1 & - & -\\
L222\_32 &  29/04/21 	& 960   & Outflow & 1.5 & - & - \\
	& 08/06/21 	& 960  & Outflow & 1.8 & -  & - \\
L222\_37 & 07/05/22   & 960  & Outflow & - & - \\
VVVv800 & 29/04/21 	& 960  & Outflow & 1.2 & - & - \\
L222\_120 & 11/05/22   & 960  & Outflow & 3.4 & - & - \\

\hline
\multicolumn{2}{l}{FIRE/Magellan} \\
\hline
L222\_4	& 20/05/19  & 475 & FUor & 3.1 & - & - \\
	& 21/05/19  & 1056 & FUor & 3.1 & - & -\\
 L222\_15 &22/05/19	& 634 & FUor & 0.6  & - &  - \\
\hline
\hline
\end{tabular}
\flushleft{$d_k$: kinematic distance calculated using the online tool provided in \citet{Wenger2018}. }
\label{tab:obs}
\end{table} 

\subsection{Spectroscopic observation and data reduction}

Two spectroscopic campaigns were obtained by the XSHOOTER spectrograph on the European Southern Observatory Very Large Telescope (ESO VLT, program ID: 105.20CJ, 109.233U, PI: Lucas) in the years 2021 and 2022. During the target acquisition, blind offsetting was applied as most targets are not visible in the $z$-band acquisition image. All targets were observed simultaneously with 0.9$''$ slit width in optical and 0.6$''$ slit width in near-infrared arms. The exposure time of each target was estimated by the expected infrared brightness. Some targets were observed multiple times due to poor weather conditions. The typical spectral resolution is $R = 8000$ in $K_s$-bandpass. The XSHOOTER spectra were extracted using the pipeline built on the Recipe Flexible Execution Workbench software \citep[{\it Reflex};][]{Freudling2013}. The telluric features were further corrected by the {\it molecfit} software \citep{Smette2015, Kausch2015} that provides a better calibration than simply applying the telluric standard provided per night. In \citet{Guo2021}, a constant 83 km s$^{-1}$ shift {was} found between the linear wavelength solution from the pipeline and the measured wavelengths of the skylines. Here, we {removed} this shift by setting the {\sc wavelength\_frame} parameter to {\sc air} in {\it molecfit} and applying a 2nd-order polynomial function to generate the wavelength solution\footnote{Originally proposed in the master thesis of A. Aguayo.}. During the final examination of the spectra, some non-astrophysical peaks, likely caused by cosmic rays, {were removed by replacing them with the median value of neighbouring pixels.} The accurate flux calibration could not be performed, since most targets were acquired by blind offsetting and some were observed during cloudy weather conditions. The information on spectroscopic observations is presented in Table~\ref{tab:obs}. The extracted spectra are presented in Figure~\ref{fig:xshooter_spec_1} to \ref{fig:xshooter_spec_5}.

 Near-infrared spectra of two sources (L222\_4  and L222\_15) were observed using the Folded-port Infra-Red Echellette (FIRE) spectrograph \citep{Simcoe2013} on the Magellan Baade Telescope in 2019 (PI: Kurtev). A uniform 0.6\arcsec slit was used, providing a spectral resolution of $R = 6000$ in $K$-bandpass. The data {was} reduced using the {\sc firehose v2.0} pipeline designed by \citet{Gagne2015}. Some breaks were seen at the joint of orders, especially between the last two orders in $K$-bandpass, caused by the mistracking of spectra towards the edge of the detector. We fixed these breaks by assuming a constant slope at the edge of the two orders ($<$0.005 $\mu$m wide). We first applied polynomial functions to fit the curved edges between the two orders and then replaced the curved continuum with a linear slope throughout the edge area \citep[see][for more details]{Guo2020, Guo2021}. Finally, we used the OH skylines from the scientific exposures as wavelength calibration.


\subsection{Photometric data}
\subsubsection{Near-infrared photometric data}

The pre-release VIRAC2$\beta$ version of the VIRAC2 catalogue provides near-infrared photometry that contains dozens of $K_s$ band measurements and a few multi-colour epochs (Smith et al., in prep). Additional processing of a few sources was conducted to correct for saturation or spatially resolved nebulosity. The first multi-colour epoch was taken in 2010 when most eruptive sources were at the quiescent stage, the second coloured epoch was taken in 2015 and sources in the inner bulge benefit from additional subsequent epochs. To increase the photometric accuracy, the $K_s$ light curves were resampled into 1-day bins by the median brightness after removing outlier detections (3$\sigma$ in each bin).

In April 2021 and May 2022, two epochs of $J, H, K_s$ photometric follow-ups were carried out on the ESO 3.58~m New Technology Telescope (NTT), using the imaging mode of the Son OF ISAAC (SOFI) instrument. As extensions of the VVV survey, these two epochs targeted highlighted variables that were discovered from the VVV light curves \citep[][and LSG23]{Contreras2017, Teixeira2018}. In total, 136 targets were observed in the 2021 campaign, and 97 were observed in 2022. The photometric data were extracted by the aperture photometry methods, similar to the routine designed in \citet{Guo2018a}. Photometric calibration stars were selected based on the stability of their $K_s$ brightness. 

\subsubsection{Archival-infrared photometric data}

We obtained photometric data from {\it WISE} and {\it Spizter} surveys via the NASA/IPAC Infrared Science Archive (IRSA\footnote{\url{https://irsa.ipac.caltech.edu}}). The {\it WISE} $W1$ to $W4$ photometry is adopted from the {\it All-Sky} data release when most sources were at the quiescent stage \citep{Wright2010}. The $W1$- and $W2$-band light curves were obtained from the multi-epoch photometry table of the {\it ALLWISE} and the {\it NEOWISE} surveys \citep[][2014-2022]{Mainzer2014}. The $W1$ and $W2$ time series were sampled into 1-day bins, where outliers lie more than 3$\sigma$ from the median of the bin were excluded, and the final output light curve is the unweighted mean value of each bin. We also visually inspected the {\it WISE} images to identify sources that have suspicious detections from surrounding nebulosity, saturated nearby sources, or spatially close companions. Similarly, {\it Spitzer} $I1$ to $I4$ photometry {was} obtained from the {\it Glimpse} survey \citep[][prior to  the VVV survey]{Benjamin2003}, and the 24 $\mu$m data was adopted from the {\it MIPSGAL} survey \citep{Carey2009, Gutermuth2015}. 

\begin{figure*}
\includegraphics[width=4.8in,angle=0]{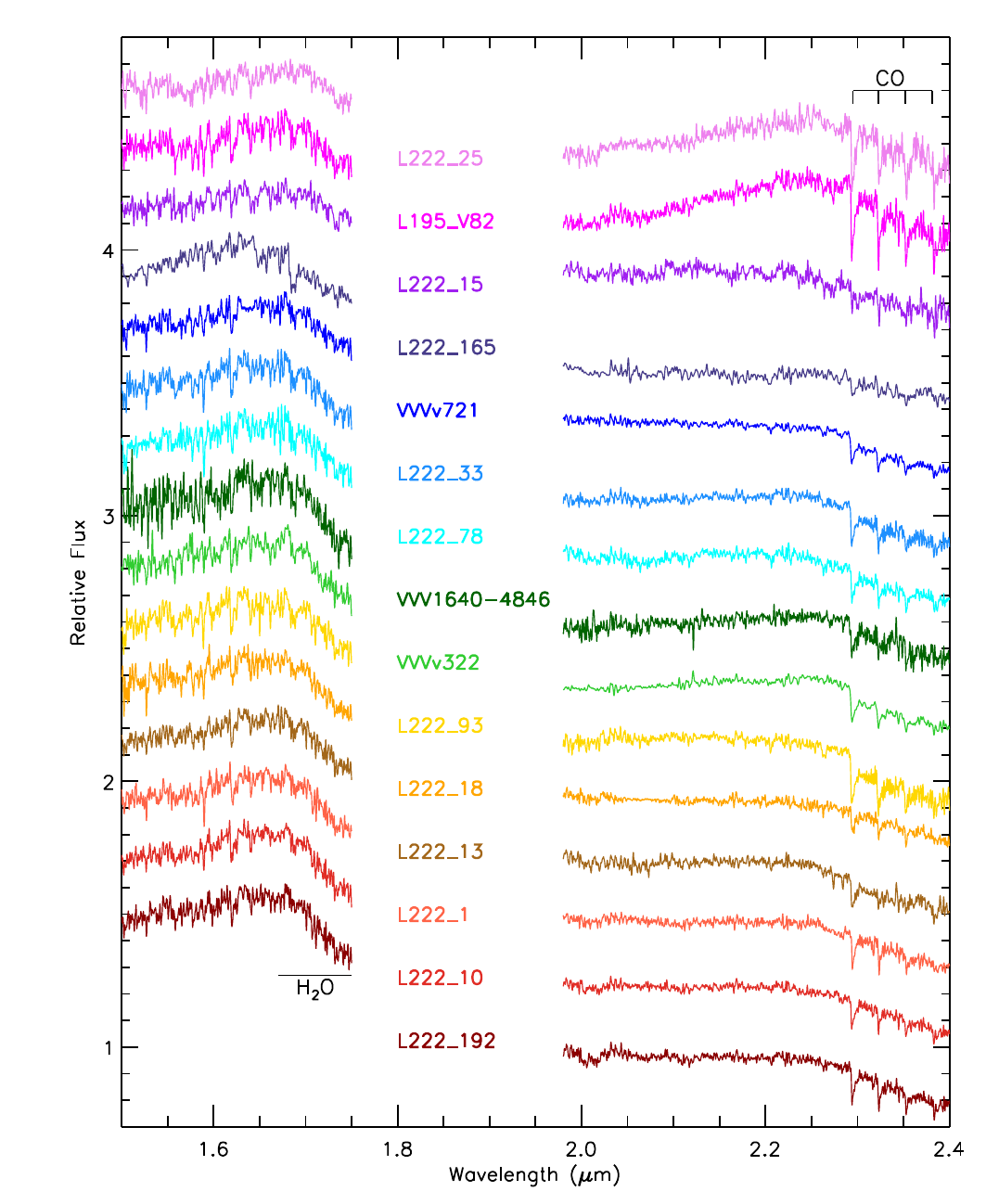}
\caption{Normalised near-infrared (1.5 to 2.4 $\mu$m) XSHOOTER/VLT spectra of thirteen new FUor-type objects originally discovered by the VVV survey. In addition, spectra of two previously identified FUors (VVVv322 and VVVv721) are also presented. All spectra were de-reddened based on the shape of their $H$ bandpass continuum, to match the spectrum of FU Orionis \citep[provided by][]{Connelley2018}. The spectra are ordered by their $H-K_s$ colour after de-reddening, with the reddest on top.}
\label{fig:FUor_spec_sum}
\end{figure*}
\begin{figure*}
\includegraphics[height=1.9in,angle=0]{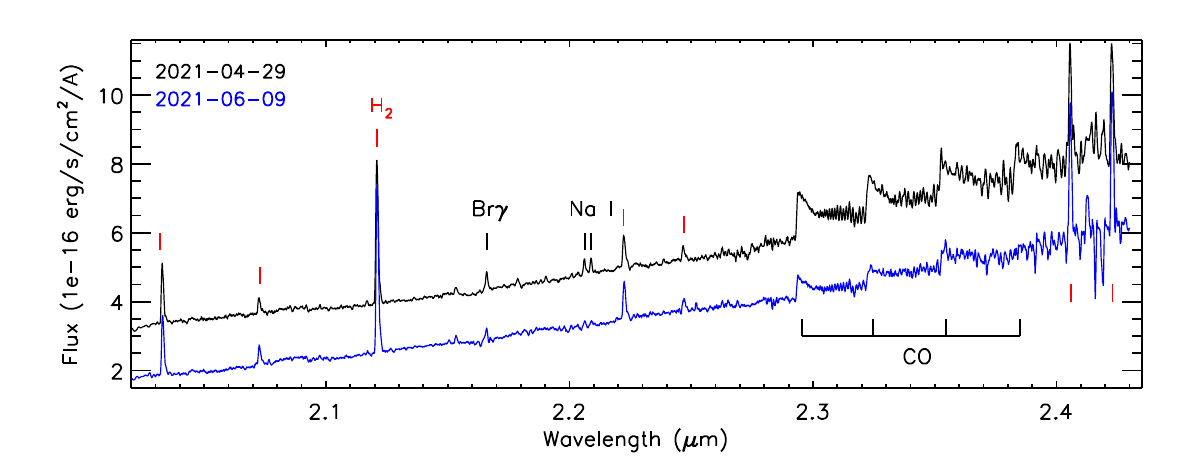}
\includegraphics[height=1.85in,angle=0]{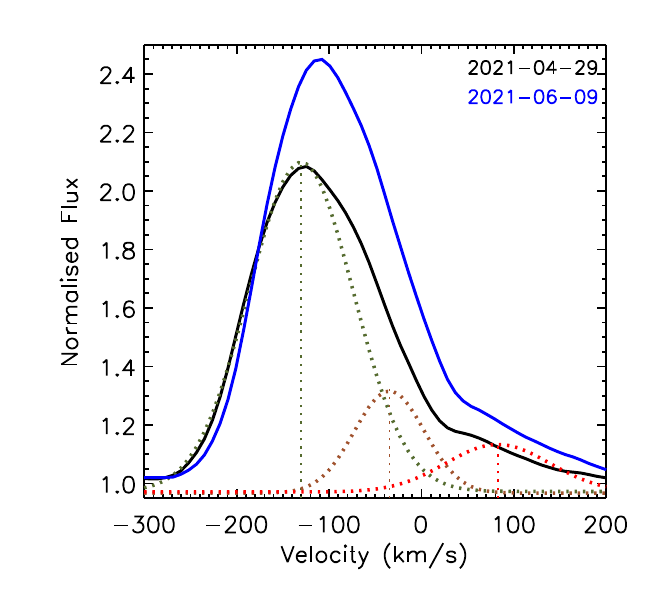}
\caption{{\it Left}: $K$ bandpass XSHOOTER spectra of L222\_167 with spectral features marked out individually. Two spectral epochs are presented in different colours. The flux of the second epoch is shifted by 1 unit smaller for better visualisation. {\it Right}: Normalised line profiles of H$_2$ S(1) lines at 2.12 $\mu m$. Two spectral epochs are shown by the black and blue solid lines. Three Gaussian components, presented as the dotted lines, are fit to the line profile of the second epoch.}
\label{fig:L195_V146}
\end{figure*}

Spatially close companions are commonly seen among our targets. In the low spatial resolution {\it WISE} images (e.g. $W1$ and $W2$ bands), sources located within 8$''$ are not well distinguished. We set a threshold of 2$''$ on the spatial distance when retrieving data from archival catalogues since the blending of two or more sources affects the location of archival detections when being extracted as a point source. Such blended sources are L222\_4, L222\_13, L222\_25, L222\_78, L222\_148 and L222\_210. In the following analysis, readers should keep in mind that the mid-infrared brightness of these targets, especially during photometric minima, is probably overestimated. The amplitudes of these sources are lower limits due to the shallower photometric minima (see the notes on Table \ref{tab:eruptivelc}). In a special case, L222\_78, we performed custom-designed PSF photometry on the binned {\it WISE} images. The details of this procedure are presented in another publication (Guo et al., submitted).

We applied the slope of the infrared spectral energy distributions (SED), namely as the spectral index $\alpha_{\rm class}$ \citep{Lada1987, Greene1994}, to identify the evolutionary stage of each source. Protostars (Class 0 \& I) have positive values as $\alpha_{\rm class} > 0.3$ while T Tauri stars (Class II \& III) have negative slopes ($\alpha_{\rm class} < -0.3$). The transitional period, also called the ``flat-spectrum'' stage named after the shape of their SEDs, is defined as $ -0.3 \leq \alpha_{\rm class}  \leq 0.3$. In this work, $\alpha_{\rm class}$ was calculated using the quiescent infrared SEDs ($2 - 24$ $\mu$m, where available). Whenever possible, we preferred the SEDs from contemporary VVV $K_s$ and {\it WISE All-Sky} survey. The line-of-sight extinction could lead to a higher $\alpha_{\rm class}$ as $K_s$ is more sensitive to extinction than the longer wavelength bands. However, after examining the SEDs (see appendix), we found that including $K_s$ has not significantly affected the result of the linear fitting, in terms of the classification of YSO evolutionary stages. The measured $\alpha_{\rm class}$ is presented in Table~\ref{tab: info}. Sources without mid-infrared detections during their quiescent stage are not measured. In total, 22 targets are classified as embedded Class I objects, and four targets are identified as ``flat-spectrum'' and Class II objects. In the next section, we will further refine the number of YSOs via spectral features. In the end, 16 targets are eventually confirmed as Class I sources, and two targets are ``flat-spectrum'' and Class II objects.


\section{Result}
\label{sec:res}
\subsection{Spectroscopic classification}

Based on the infrared spectral features, the 33 variables observed in this paper are classified into YSO and post-MS groups. Specifically, the YSO group is further divided into sub-categories based on their spectral features, which also reveal their physical conditions during the eruptive stage. 

\subsubsection{FUor-type}

During a FUor-type outburst, the mass accretion process, often enhanced by three orders of magnitudes, is no longer controlled by the stellar magnetic field \citep[see][and references therein]{Audard2014}. Instead, the circumstellar material is directly accreted onto the stellar surface, and the inner accretion disc (up to a scale of 1 AU) is heated internally by the viscous dissipation of accretion energy, with a hotter mid-plane and a cooler disc surface \citep[see][]{Zhu2007}. Numerical models show that, during a FUor-type event, the observed surface brightness of the viscous heated accretion disc could be five times greater than the stellar photosphere, dominating the infrared spectral features \citep{Liu2022}. On the observational side, FUor-type objects are diagnosed with rich absorption features in their near-infrared spectra, such as hydrogen recombination lines, the triangular-shaped $H$-band continuum due to low gravity water vapour absorption, and CO overtone bandheads beyond 2.3 $\mu$m \citep[see][]{Fischer2022}. In the optical domain, the spectra of FUors resemble supergiants with low gravity and an early spectral type (G to K) at shorter wavelengths but a later type (M) at longer wavelengths. The infrared spectra of FUors resemble those of young brown dwarfs with low gravity, although FUors have much higher bolometric luminosity and usually show strong Pa$\beta$ and He {\sc i} absorption features in the 1 to 1.3~$\mu$m region, attributed to a wind \citep[see][]{Connelley2018}.

In this work, 15 eruptive sources are identified as FUor-type objects. We de-reddened their spectra based on the triangular-shaped $H$-band continuum, to align with the spectrum of FU Orionis \citep[provided by][]{Connelley2018}, using the extinction law from \citet{WangS2019} and $A_V = 1.5$~mag estimated in \citet{Zhu2007}. Subsequently, we estimated the $K_s$-band extinction ($A_{K_s}$) for each FUor-type object by applying the minimum $\chi^2$ method to compare the de-reddened spectra with the standard $H$-band spectrum, with $A_{K_s}$ values ranging between 0 and 10 mag. However, we acknowledge that this is only a rough estimation of $A_{K_s}$ as the disc spectra of FUors are not necessarily the same from one case to another. The de-reddened $H$ and $K$-bandpass spectra of FUor-type sources are presented in Figure~\ref{fig:FUor_spec_sum}.

\begin{figure}
\includegraphics[width=3.in,angle=0]{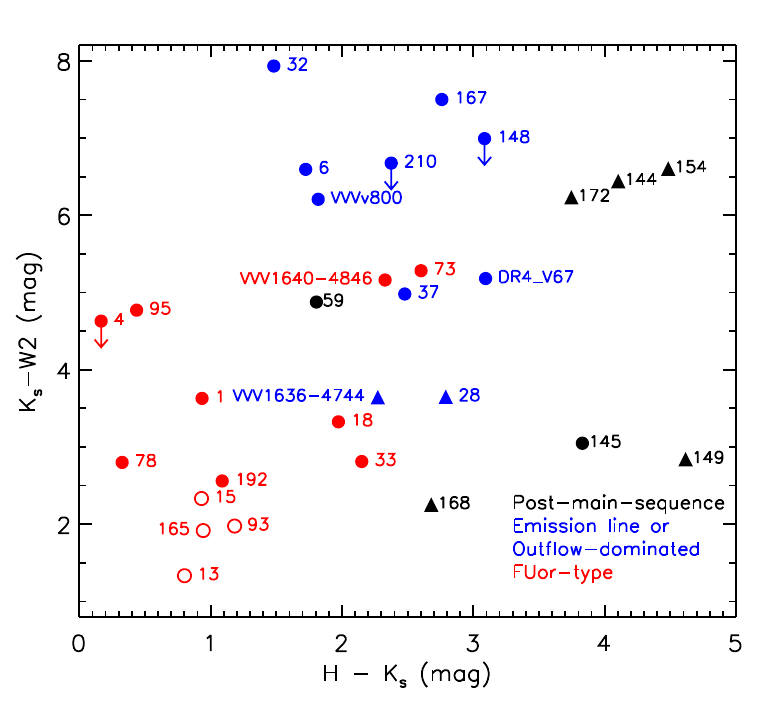}
\centering
\caption{$H-K_s$ and $K_s-W2$ colour-colour diagram of targets observed in this work (see name tags on the plot, the suffix ``L222\_'' is omitted), colour-coded by spectral features. Targets with classic eruptive $K_s$ light curves are shown by dots and others are shown by triangles. Filled dots represent YSOs with colours measured during the pre-outbursting stage. Sources without any W2-band detections are not presented. Four FUors that only have $W2$ detections during the outbursting stage are presented by open circles. Typical photometric error bars ($<$0.05 mag) are about the same size as the data points.}
\label{fig:ccd}
\end{figure}

The XSHOOTER spectra of two FUors (L222\_4 and L222\_73) were taken during their fading stages (see \ref{fig:xshooter_spec_1}), and show spectral features that slightly differ from those of classical FUors. Some weak and cool CO absorption features were detected, however, the triangular-shaped $H$-band continuum is absent or less prominent due to the low brightness in $H$, and the H$_2$ lines are seen in emission. In the case of L222\_4, the FIRE spectrum taken in 2019 matches a standard FUor-type, where deeper absorption features were observed when the $K_s$-band brightness was higher (see Figure B5, left panel). After the original 6 mag outburst in $K_s$ (the highest amplitude in this sample), L222\_4 entered a rapidly fading stage that decayed 3~mag in 5 years. The spectroscopic and photometric variability of L222\_4 resembles another rapidly faded FUor \citep[VVVv322, ][]{Contreras2017b}. It agrees with the physical model of FUors, as a warm circumstellar disc with a steeper vertical temperature gradient at the peak of the outburst.

The radial velocity (RV) of FUors is measured by the CO bandhead absorption feature. We applied the rovibrational CO model from \citet{Contreras2017b}, with free parameters including the temperature and column density of the CO gas, and the RV of the system. The best-fit models were confirmed through visual inspection. The grids of the model are $\Delta$RV = 5 km/s, smaller than the pixel size of XSHOOTER (12 km/s). The kinematic distance ($d_{\rm k}$) is then calculated using the ``Monte Carlo simulation mode'' of the online tool provided by \citet{Wenger2018}, based on the Galactic coordinates and RV of each source. In most cases, either the near or far $d_{\rm k}$ agrees with $d_{\rm SFR}$, suggesting that the variable source is associated with nearby star-forming regions. When comparing $d_{\rm k}$ with $d_{\rm SFR}$ (listed in Table \ref{tab: info}), many FUors have $d_{\rm SFR}$ beyond the 1$\sigma$ error of $d_{\rm k}$. However, $d_{\rm k}$ could be wrong if the target is in a binary system due to the orbital motions or if it has an intrinsic velocity, but our data are not enough to confirm or reject the existence of any close companions.


\subsubsection{Emission line and outflow-dominated objects}

Some YSOs exhibiting long-lasting eruptive events have spectroscopic features disagreeing with FUors. One highlighted source is V1647 Ori, whose eruptive light curve resembles FUors. However, a local dimming event with a timescale of 1000 days was seen after the original outburst. The spectra of V1647 Ori have water vapour absorption accompanied by weak CO absorption in the first outburst but with Br$\gamma$ and CO bandheads emission in the second outburst \citep[see][]{Aspin2009, Connelley2018}. Since then, eruptive sources with features between FUors and EXors were named MNors or V1647 Ori-type \citep{Contreras2017b, Hodapp2020, Fischer2022}. In the near-infrared, a group of eruptive YSOs was discovered by the VVV survey with long-lasting outbursts but EXor-like spectra (e.g. Na I doublet, Br$\gamma$ and CO bandheads emission), which are named as emission line objects in \citet{Guo2021}. Another group of young variables are classified as outflow-dominated objects since only $H_2$ and/or [Fe II] emission lines were identified from the spectra \citep[see Section 3 and Table 5 in][]{Guo2021}. In the VVV survey, these novel categories of embedded outbursting YSOs are more populated than FUors, even towards the longest duration end.  

In this paper, six sources are identified as emission line objects with signatures of magnetospheric accretion (see Figure~\ref{fig:xshooter_spec_4}). Among them, L222\_148 and L222\_167 were observed during their brightness plateau, whilst the rests were observed at least 1.5~mag fainter than their photometric maxima. CO bandheads were detected (beyond $3\sigma$) on L222\_148 and L222\_167 in emission, presenting a positive correlation between the infrared brightness and the existence of hot gas at the accretion disc, aligned with a high mass-accretion rate. One source, L222\_167 was observed on two epochs 40 days apart and the continuum on the first epoch is 13\% brighter than the second epoch (Figure~\ref{fig:L195_V146}). Stronger mass-accretion indicators are seen in the brighter epoch, whilst the fainter epoch has stronger indicators of the outflow ($H_2$ lines), suggesting non-correlation between the mass-accretion rate and the strength of outflow or there is a time delay between them. Typically, emission from a jet is rarely detected among FUor-type outbursts, as the strong accretion disrupts the magnetic field of the inner disc, preventing the launching of a jet. Complex line profiles are detected on the $H_2$ lines, including a blue-shifted high-velocity component around -130 km/s, a Gaussian component around -33 km/s, and a weak red-shifted tail. This complex line feature indicates that winds/outflows were launched towards different directions, and sometimes at perpendicular angles.

Four targets are classified as outflow-dominated objects, only having spectral indicators of stellar wind/outflows. Among them, the spectra of two YSOs (VVVv800 and L222\_120, a.k.a. DR4\_v67) were previously published in \citet{Guo2020, Guo2021}. In comparison with the latest spectra, there is no significant spectroscopic variability. In the case of L222\_32, two XSHOOTER spectra were obtained in 40 days. Despite very similar continuum levels, the $H_2$ S(1) line in the second epoch is 35\% stronger in flux and 11~km/s redder in RV than the first epoch. An interpretation is that the variability of the outflow is not correlated with or delayed response to the variability of the infrared continuum. Combined with the list published in \citet{Guo2021}, 12 VVV targets are classified as outflow-dominated sources. In most cases, the VVV light curves of outflow-dominated sources are irregular, with significant variations on (sub)year-long timescale, in contrast to classic outbursts (see definitions in LSG23). This could be a geometric effect as one is looking through the outflow, as part of the variability might be attributed to the winds/outflows lifting dust from the disc, hence affecting the extinction towards the source.

Wherever available, we measured the extinction using the flux ratio between $H_2$ 1-0 Q(3) (2.42~$\mu$m) and 1-0 S(1) (2.12~$\mu$m) lines, as it is independent of the excitation conditions \citep[Q(3)/S(1) = 0.7;][]{Turner1977}. We applied the power-law extinction function from \citet{WangS2019}. The typical precision of the $A_{K_s}$ measurement is 0.4~mag, with uncertainty rising from the flux measurement of the Q(3) line ($\sigma =$ 8\%), which is veiled by the continuum and the CO overtone emission. The colour variation before and after the outburst is discussed further in \S\ref{sec: amp}.

\subsubsection{Post-main-sequence objects}

Many post-MS variables in the sample have VVV light curve morphologies that appear either to show a dip or an eruption: this is ambiguous in some cases \citep[see examples in][]{Contreras2017b, Guo2021}. Spectra were obtained to determine which of the two is occurring or whether these sources are YSOs or giant stars. However, most such sources with ambiguous light curves were detected in a bright state by 2MASS, e.g. sources L222\_144, L222\_145, L222\_154, L222\_144, L222\_168, L222\_172, which tended to indicate that we are observing the dips due to extinction rather than photometric eruptions (see LSG23). These sources often have a red colour in the near-infrared ($H-K_s > 3.5$~mag) or lack $J$-band detections in VVV data. Mid-infrared colour-colour and colour-magnitude diagrams were applied to distinguish YSOs with post-MS sources, although no rigid boundaries could be drawn \citep[see][]{Lucas2017, Guo2022}. Due to nucleosynthesis processes and the first dredge-up, most post-MS stars display a fairly strong $^{13}$CO absorption bandhead at 2.345 $\mu$m in our intermediate resolution spectra, whereas this feature is rarely seen in the YSO spectra presented herein.

In the LSG23 catalogue, 21 sources are classified as dipping giants based on their infrared light curves, with a majority of them projected in the Nuclear Disc of the Milky Way. We obtained spectroscopic verification of eight sources (see Figure~\ref{fig:xshooter_spec_5}), including seven dipping giants and one eruptive source (L222\_59). All dipping giants have similar spectroscopic features resembling O-rich post-MS stars. The quiescent colour-colour diagrams of our targets are shown in Figure~\ref{fig:ccd}. Most post-MS sources exhibit redder $H - K_s$ colours compared to YSOs. Furthermore, it is noteworthy that FUors consistently display smaller colour indices than non-FUor eruptive YSOs, which will be explored further in \S\ref{sec: amp}. We derived the stellar parameters for post-MS sources by fitting the CO bandhead models plus equivalent width (EW$_{\rm CO}$ and EW$_{\rm Na}$) based on methodologies derived from \citet{Feldmeier-Krause2017} and \citet{Fritz2021}. The outcomes of this analysis are listed in Table \ref{tab:postms}. Most dipping giant sources have an effective temperature between 3000 to 4000 K and super solar metallicity (except L222\_168). This high metallicity is consistent with the location in the Nuclear Disc, for which the metallicity distribution is distinctly metal-rich \citep{Fritz2021}. More details are presented in the Appendix \ref{sec:postmsmeasurements}.

\begin{figure}
\centering
\includegraphics[width=3.5in,angle=0]{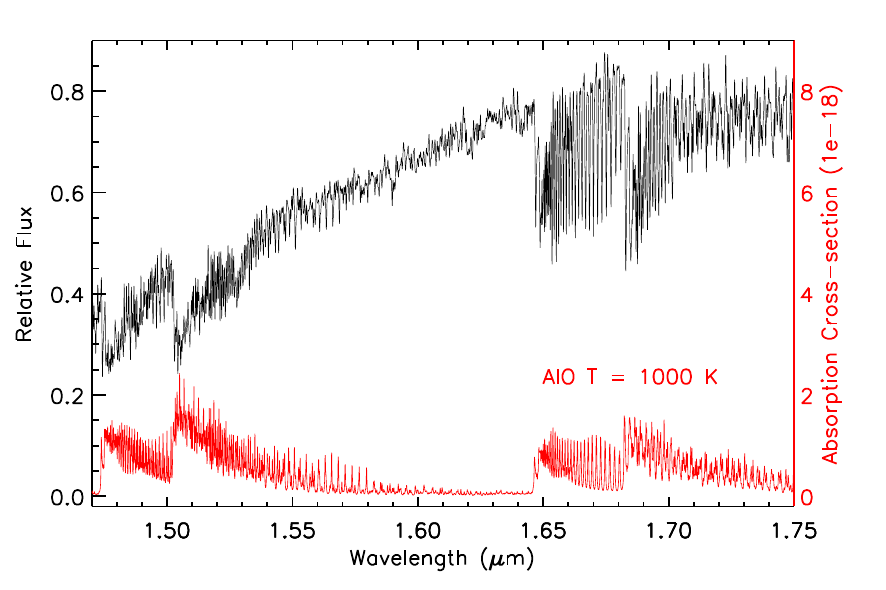}
\caption{The $H$-bandpass spectrum of L222\_59 with an analytical model of the AlO absorption bands from ExoMol \citep[red lines;][]{Bowesman2021}.}
\label{fig:L195_V50_H}
\end{figure}

We highlight an eruptive source (L222\_59, a.k.a. VVVv746) that exhibits rarely seen deep aluminium monoxide (AlO) absorption bands (see Figure~\ref{fig:L195_V50_H} and \ref{fig:xshooter_spec_5}). This source experienced a $\Delta K_s > 5$ mag outburst and reached the photometric maximum in less than three years. According to our follow-up $K_s$ photometry, after spending nearly 1000 days on its brightness plateau, L222\_59 dramatically faded nearly 3 mag in the year 2021-2022. The XSHOOTER spectrum presented in this paper was taken during the outbursting stage, with many absorption features, including cool $^{12}$CO overtone bands in the $K$ bandpass (as fitted temperature, $T = 1200$ K, assuming LTE) and AlO in the $J$ and $H$-bandpasses, confirmed by the ExoMol database \citep{Patrascu2015, TENNYSON2016, Bowesman2021} to be the A-X electronic transitions of the AlO radical. Interestingly, it also has narrow H{\sc{i}} lines in absorption, with different RV compared to the CO bandhead. These spectral features indicate the complexity of a cool envelope and a warm central star. Hitherto, such deep AlO absorption bands have only been observed in the stellar merger candidate V838 Mon \citep{Evans2003, Banerjee2005}, which exhibits a Nova-like eruption on a timescale of 50 days \citep{Munari2002}. The hypothesis of a candidate stellar merger will be further discussed in a follow-up work, including up-to-date photometric and spectroscopic data.

\begin{table}
\caption{Spectroscopic information of post-MS sources}
\renewcommand\arraystretch{1.1}
\begin{tabular}{l c c c c c c c c}
\hline
\hline
Name & t$_{\rm obs}$ & EW$_{\rm CO}$ & EW$_{\rm Na}$ & [Fe/H] & T$_{\rm eff}$ & T$_{\rm CO}$\\
\hline
 & d/m/y & \AA & \AA & & K & K \\
 \hline
L222\_59$^\star$  & 29/04/21 & 19.1  & 1.8 & -0.68 & 3648 & 1200  \\
L222\_130 & 15/05/22 & 18.2  & 5.0 & 0.62 & 3744 & 3800 \\
L222\_144 & 15/05/22 &  18.0  & 7.0 & 1.00 & 3765 & 3600 \\
L222\_145 & 16/05/22 & 22.1  & 5.2 & 0.28 & 3325 & 3800 \\
L222\_149 & 04/05/21 & 24.1   & 5.2 & 0.10 & 3112 & 2800 \\
L222\_154 & 16/05/22 &  23.2  & 6.0 & 0.30  & 3208 & 2400 \\
L222\_168 & 23/05/22 & 14.9   & 2.1 & -0.42 & 4097 & 3800 \\
L222\_172 & 16/05/22 &  17.8  & 5.9  & 0.97  & 3788 & 3600\\
\hline
\hline
\end{tabular}
\flushleft{$T_{\rm eff}$ is measured using the methods from \citet{Feldmeier-Krause2017}, with a typical uncertainty of 163 K. $T_{\rm CO}$ is measured from the analytical models of CO bandheads \citep{Contreras2017b}, with an error bar of 200 K.
$^\star$: the measurements of $T_{\rm eff}$ and [Fe/H] are not reliable due to the unique spectral features of L222\_59, e.g. circumstellar CO.}
\label{tab:postms}
\end{table}

\begin{figure*}
\includegraphics[height=3.in,angle=0]{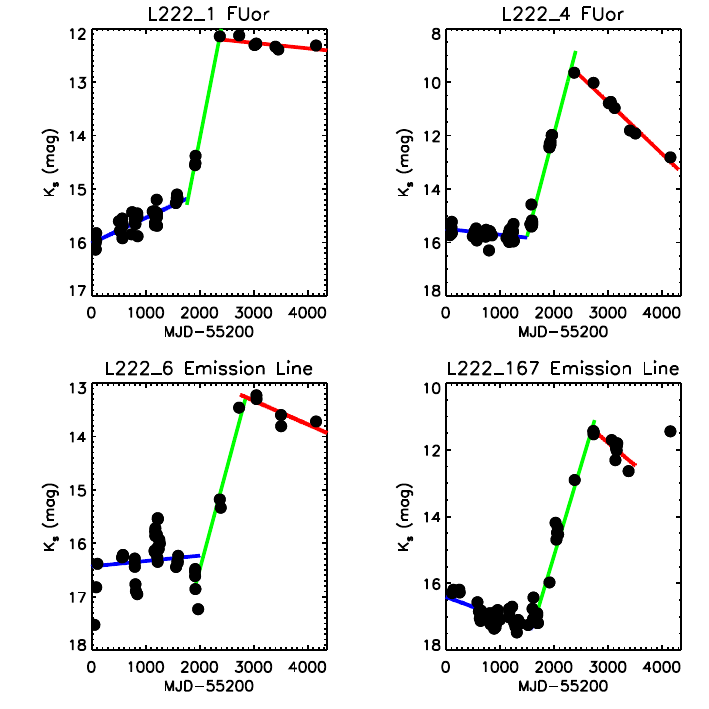}
\includegraphics[height=3.05in,angle=0]{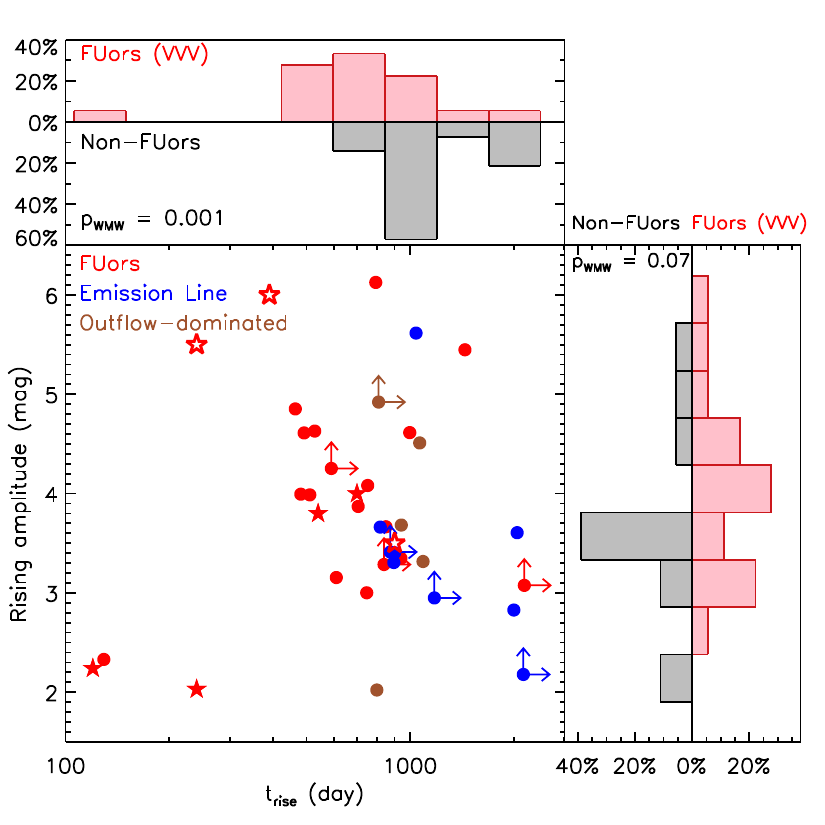}
\caption{{\it Left}: $K_s$ light curves of four eruptive YSOs observed in this work. Linear functions are applied to fit the quiescent/pre-eruptive (blue), rising/eruptive (green), and post-outbursting (red) stages.  {\it Right}: The rising timescales and amplitudes of large amplitude and long-lasting eruptive YSOs. Archival FUors are shown as comparisons, among which the amplitudes of three FUors were measured in optical (open stars) and the rest were measured in $K_s$ (filled stars). Outbursts started in the pre-VVV era were labelled with lower limits. The distributions of rising timescales and amplitudes of VVV sources (Archival sources are not included) are presented in the top and right panels. The possibility from the Wilcoxon-Mann-Whitney test ($p_{WMW}$) for each distribution is labelled on the histogram.}
\label{fig:eruptive_time}
\end{figure*}

\section{Discussion}
\label{sec:dis}

In the following sections, we present analyses based on the photometric and spectroscopic behaviours of eruptive YSOs discovered in the VVV survey. Individual characters, such as the rising timescales of the outburst, and the near to mid-infrared amplitudes, are measured to distinguish FUors from other eruptive events. 

\subsection{Eruptive samples}

We selected a sample of 32 long-lasting eruptive events discovered by the VVV survey, with near-infrared spectroscopic confirmation by our group. To minimise the selection bias, we included all sources with $\Delta K_s \geq 2$~mag and have eruptive events lasting longer than 1000 days. Among them, 18 sources are spectroscopically confirmed as FUors and the rest are either emission line objects or outflow-dominated sources (non-FUors). This selection of confirmed long-lasting eruptive YSOs is roughly consistent with the sample discussed in LSG23, although the samples in this work include three emission line objects with light curve morphologies different from the ``classic outbursts''. This is to maintain enough non-FUors in our sample. However, the results of the measured distribution of several characters (between FUors and non-FUors) are not severely impacted if these three sources were not included. According to their quiescent SEDs and estimated distance, the vast majority of these sources are embedded low-mass Class I protostars. Our sample forms a valuable laboratory to investigate outbursting events on embedded sources, especially to explore the relationship between photometric and spectroscopic features. The $K_s$ light curves of these sources, including data from both the VVV survey and SOFI observations, are shown in Figure~\ref{fig:eruptive_lc}. We note that three FUors previously identified in the VVV survey are not included in this sample, due to the absence of $K_s$ detections during the rising stage, hence their eruptive stages are poorly described. 

The complexity of our eruptive sample plays a key role in our following statistical analysis. A few factors could affect the detection of long-duration outbursts, such as saturation in the VIRAC catalogue during the outbursting stage, eruption prior to the beginning of the VVV survey (the year 2010), and lower amplitude events that are yet to be confirmed by spectroscopic follow-ups. In another paper of this series (Contreras Pe\~na et al., submitted), we will further discuss the occurrence rate of eruptive behaviours on Class I protostars, by applying mid-infrared SED-based catalogues \citep[e.g. SPICY,][]{Kuhn2021} and including archival $K_s$-band detections (e.g. 2MASS) to form a two-decade-long timeline. 

\subsection{Near-infrared light curves of eruptive samples}
\label{sec: EruptiveTimescales}

Eruptive events on YSOs are known to have various photometric characteristics, due to different triggering mechanisms and disc configurations \citep{Vorobyov2021}. For instance, FU Ori reached the peak brightness in eight months while the rising stage of V1057 Cyg lasted 15 years \citep{Herbig1977eruptive}. Here, we apply linear functions to describe the quiescent, rising, decaying, and post-outbursting stages of the aforementioned 32 eruptive events (see examples Figure~\ref{fig:eruptive_time}). Two parameters were defined to quantify the rising stage:  the rising amplitude ($\Delta K_{\rm s, rise}$) and the rising timescale ($t_{\rm rise}$). Specifically, $\Delta K_{\rm s, rise}$ is the $K_s$ amplitude of the rising stage, measured from the linear fitting result. The $t_{\rm rise}$ is calculated as $\Delta K_{\rm s, rise}$ divided by the measured linear slope (mag/d) of the rising light curve. The measured $t_{\rm rise}$ and $\Delta K_{\rm s, rise}$ are presented in the right panel of Figure~\ref{fig:eruptive_time}, with numbers listed in Table~\ref{tab:eruptivelc}. 


\begin{figure}
\centering
\includegraphics[height=2.8in,angle=0]{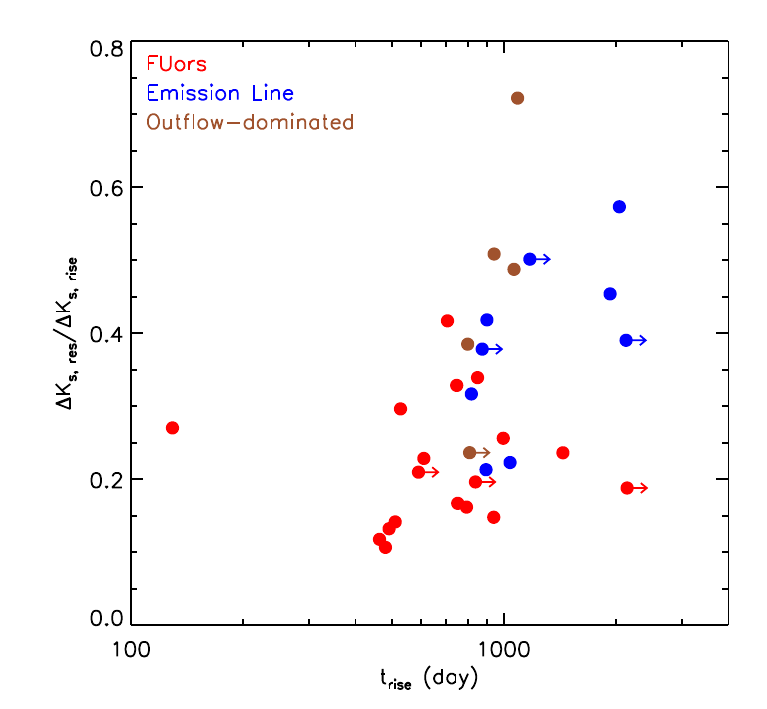}
\caption{The ratio between the short-term residual amplitude and the rising amplitude versus the rising timescale of eruptive sources. Colour codes and lower limits are as same as in Figure~\ref{fig:eruptive_time}.}
\label{fig:residual}
\end{figure}

\begin{figure*}
\includegraphics[height=2.8in,angle=0]{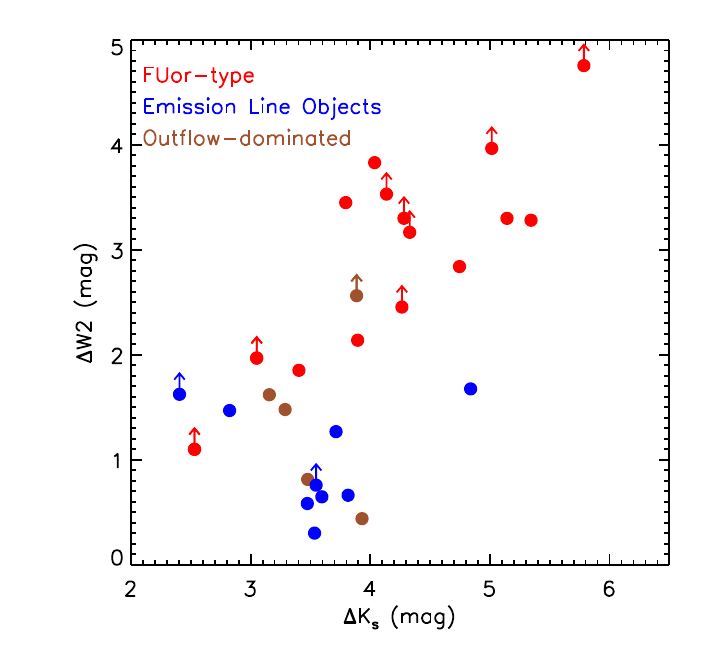}
\includegraphics[height=2.8in,angle=0]{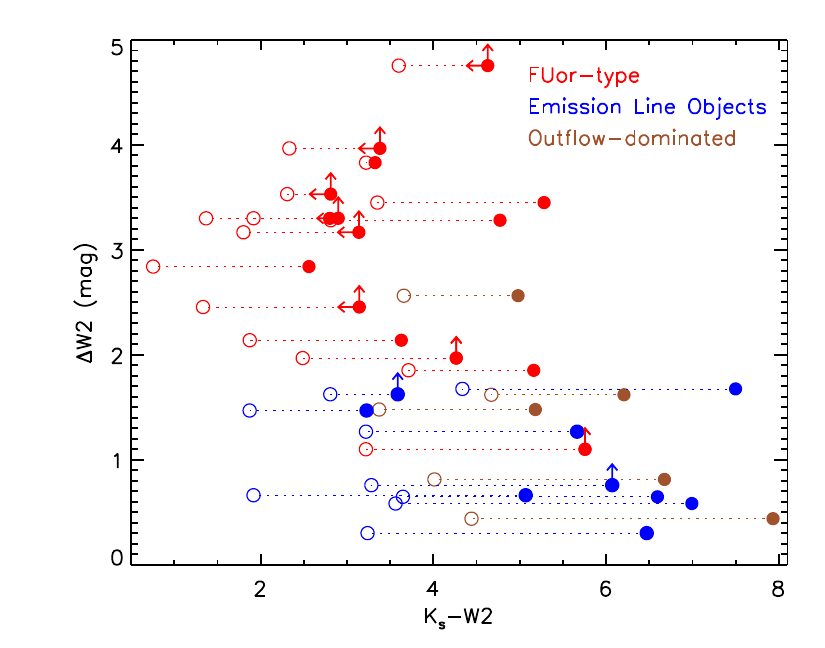}

\caption{{\it Left}: Contemporary near-infrared ($K_s$) and mid-infrared ($W2$) amplitudes of long-term eruptive YSOs. Sources without quiescent detections are marked as lower limits. Photometric uncertainties are smaller than the symbols. {\it Right}: $K_s - W2$ colour versus the mid-infrared amplitudes.The quiescent and in-outburst colours are presented by filled and open dots, respectively.}
\label{fig:wiseamp}
\end{figure*}

We present the statistical differences of $t_{\rm rise}$ and $\Delta K_{\rm s, rise}$ between FUors and non-FUors (emission line/outflow-dominated objects). In general, we find that FUors experienced shorter $t_{\rm rise}$ compared to non-FUors. Among the examined samples, 14/18 FUors have $t_{\rm rise}$ < 850 days (mean at 747~d), whilst all non-FUors have $t_{\rm rise}$ > 800 days (mean at 1038~d). To support this finding, we employed the Wilcoxon-Mann-Whitney test \citep{mann1947test}, a non-parametric rank-based test. The possibility of FUor and non-FUor sharing the same distribution in $t_{\rm rise}$ is only $p_{\rm WMW} = 0.001$. There should not be any selection bias on the $t_{\rm rise}$ among our sample, as we examined almost all high-amplitude eruptive candidates, however, the sparse cadence of the VVV survey (after 2015) limits our capability to measure any timescale less than 1 year. Among the longest-duration objects, we detect eruptive sources with $t_{\rm rise} > 2000$ d, including two sources that entered their outbursting stage before 2010. We find that FUors and non-FUors share a comparable range of $\Delta K_{\rm s, rise}$, although FUors are more abundant towards the highest amplitude end, as 61\% of FUors and 21\% of non-FUors have $\Delta K_{\rm s, rise} > 3.8$ mag. However, FUors and non-FUors are not distinguishable at the intermediate amplitude range ($2.0 < \Delta K_{\rm s, rise} \leq 3.8$ mag). The $p_{\rm WMW}$ on the $\Delta K_{\rm s, rise}$ distribution is 0.07, larger than the one measured on $t_{\rm rise}$. Since there is no strongly justified null hypothesis regarding the relative amplitudes of the FUors and non-FUors, we regard this $p_{\rm WMW}$-value as fairly good evidence that the amplitude distributions are different.

We further compare the $t_{\rm rise}$ between VVV and FUors discovered from optical time series, including FU Ori, V1057 Cyg, V960 Mon, V2775 Ori, Gaia17bpi, Gaia18dvy, and Gaia21bty \citep{Kenyon2000, Connelley2018, Hillenbrand2018, Szegedi-Elek2020, Siwak2023}. Despite three fast-eruptive sources with $t_{\rm rise}$ < 300 days, other archived FUors share similar $t_{\rm rise}$ with the FUors discovered in the infrared (see more discussion on this topic in LSG23). However, some theoretical works predicted that outbursts triggered by outside-in migrated instabilities should have shorter $t_{\rm rise}$ in the optical than in the infrared, in the order of years \citep{Cleaver2023}. Intriguingly, our VVV FUors have two outliers, one has a rapid rise followed by a rapid decay (VVVv322), and the other (VVVv721) experienced a long rising stage as $t_{\rm rise}$ > 2140 days.  The long rise time of VVVv721 indicates that the instability originated further out from the accretion disc or else the viscosity parameter was smaller \citep[see discussions in][]{Liu2022}.  



Some eruptive sources exhibit long-term variability during their quiescent stage, such as L222\_1, L222\_10, L222\_25 and VVV1640-4846. All of these sources have a gradual increase in $K_s$ before the main outburst, which can be interpreted as the piling up of warm disc material. This interpretation aligns with the first stage of the MRI-GI model, in which disc material is first gathered outside the magnetic dead zone through GI \citep{Bourdarot2023}.  We have seen different cooling slopes among VVV FUors, including a few rapidly decaying sources (e.g. VVVv322, L222\_4, VVV1640-4846) that differ from FU Ori, and even faster than the fading stage of V1057 Cyg. Apart from VVVv322, we have not observed other FUors returning completely to the quiescent brightness or the re-establishment of magnetospheric accretion. 

Short-term variations are detected in eruptive YSOs with timescales from hours to hundreds of days, in addition to their overall eruptive behaviour, on both FUors \citep[e.g. V1037 Cyg][]{Szabo2021} and non-FUors \citep[e.g. V899 Mon][]{Ninan2015}. We subtracted the long-term linear trends in the lightcurves to measure the short-timescale residual variability ($\Delta K_{\rm s, res}$), as the residual amplitude within a time window of two years in each measurement (see Figure \ref{fig:residual}). In general, FUors have smaller $\Delta K_{\rm s, res}$ than non-FUors, consistent with their physical nature, as an abrupt and persistent enhancement of the mass accretion rate which efficiently heated the inner accretion disc. Short-timescale extinction and accretion variabilities are not commonly seen among FUors in our sample. On the contrary, some local dipping events are observed on non-FUors, such as the quiescent stage of L222\_6, and the risings stage of Stim 5, DR4\_v10, DR4\_v34 and DR4\_v67. These features could be attributed to extinction events \citep[like V582 Aur][]{Abraham2018} or the discontinued accretion stream. We infer that the reduction of line-of-sight extinction may in fact contribute to some high-amplitude ($\Delta K_s$) variation seen among non-FUors.

\subsection{Mid-infrared amplitude and colour of eruptive samples}
\label{sec: amp}
\begin{figure}
\centering
\includegraphics[height=3in,angle=0]{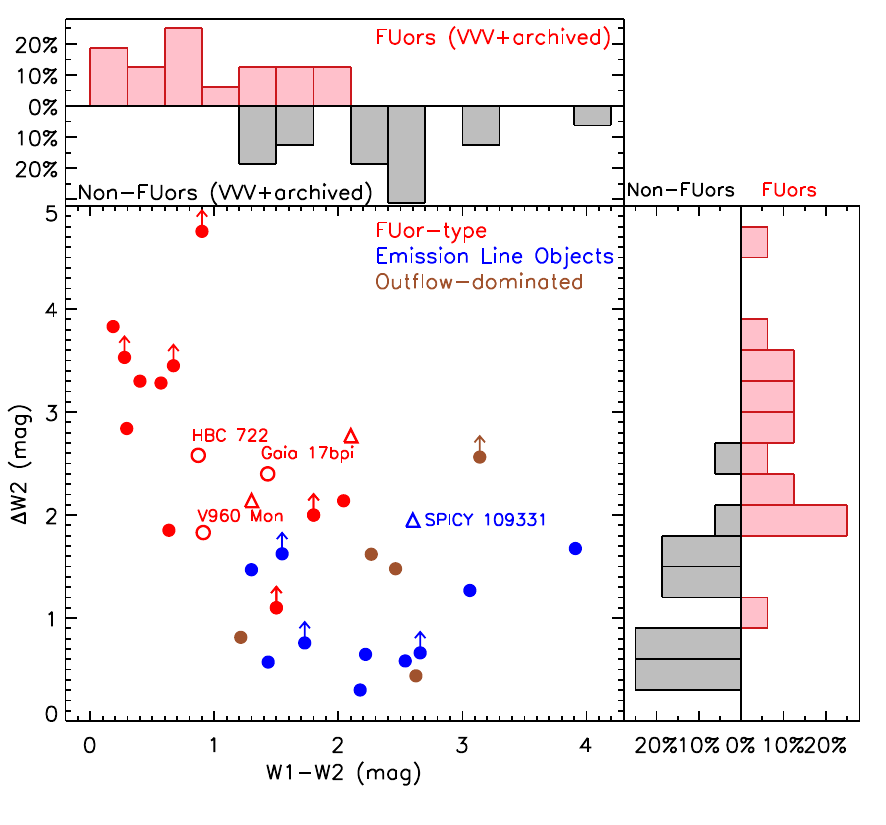}
\caption{Quiescent $W1$-$W2$ colour and $\Delta W2$ of eruptive YSOs. Sources without {\it WISE} detections during their quiescent stage are shown with lower limits. Archived eruptive YSOs are presented as open dots and triangles. The typical uncertainties on the colour and amplitude are 0.15 mag. Distributions of $W1$-$W2$ and $\Delta W2$ are presented as histograms.}
\label{fig:wisecolour}
\end{figure}

Many eruptive YSOs show large amplitudes across a wide range of wavelengths, from optical to mid-infrared \citep[e.g.][]{Kospal2016, Lucas2020, Contreras2023}. \citet{Guo2021} summarised that eruptive YSOs have comparable amplitudes in the near- and mid-infrared, in contrast to the expectations from variable extinction. In this section, we will compare the contemporary amplitudes and colours of these eruptive events, from $K_s$ to $W2$, to further investigate the physical nature. Since the VVV and {\it NEOWISE} time series have different cadences, we measured the contemporary amplitudes via linear interpolation, based on the photometric maximum identified from $W2$-band light curves. For sources without {\it WISE} detections during the quiescent stage, we converted the fluxes measured in {\it Spitzer} filters to {\it WISE} filters using the empirical photometric transformations from \citet{Antoniucci2014}. In two cases (L222\_10 and L222\_25), there are no archival quiescent mid-infrared detections, hence the $\Delta W2$ and $K_s - W2$ can not be measured. The results are listed in Table~\ref{tab:eruptivelc}. 

The correlation between $\Delta W2$ and $\Delta K_s$ is shown in Figure~\ref{fig:wiseamp}. FUors exhibit systematically higher $\Delta W2$ compared to non-FUors, indicating efficient heating of the circumstellar disc during the outburst. This aligns with the spectral feature of FUors, where their optical to mid-infrared emission is primarily from a self-luminous inner accretion disc. By contrast, there are no emission line objects with $\Delta W2 > 2$ mag, even among those with $\Delta K_s > 3.5$ mag. The infrared colour indices are presented in Figure~\ref{fig:wiseamp} and Figure~\ref{fig:wisecolour}. Lower limits are put on objects that erupted before the VVV survey. All FUors have $K_s - W2 < 6$~mag at the quiescent stage, whilst the majority of long-term non-FUors have a redder $K_s - W2$ colour. Similar trends are seen in the $W1-W2$ colour. During the outbursting stage,  most non-FUors are still redder than most FUors although the gap is smaller.
This colour difference is consistent with the $H - K_s$ colour distribution shown in Figure~\ref{fig:ccd}, where most FUors are bluer than non-FUors during the quiescent phase. As comparisons, we added some archived eruptive YSOs in Figure~\ref{fig:wisecolour}, including three FUors and three recently confirmed eruptive objects from the {\it NEOWISE} data (with prefix SPICY, Contreras Pe\~na et al., submitted).


 Due to the logarithmic nature of the magnitude measurement, the amplitude is smaller when the source is already bright in a particular band. This aligns with the low $\Delta W2$ seen among red non-FUors ($K_s - W2 > 6$ mag). However, the origin of their substantial excess in $W2$ is still uncertain. These objects could be heavily obscured during the quiescent stage. The high line-of-sight extinction may arise from a dusty envelope, an edge-on circumstellar disc, or even the foreground cloud. However, the disc inclination and random foreground extinction cannot account for the fact that most non-FUors exhibit a redder colour than most FUors (see distribution in Figure \ref{fig:wisecolour}). Nevertheless, the extinction measured from in-outburst spectra between FUors and non-FUors is comparable (see Table~\ref{tab:obs}).  The reduction of extinction during the outburst might play a role here, where the stellar outflow clears the immediate vicinity around the star, and results in a greater $\Delta K_s$ than $\Delta W2$. We also admit that there is a selection bias in our spectroscopic sample (i.e. only sources with high $\Delta K_s$ were selected from LSG23). Alternatively, a red colour index typically indicates an earlier stellar evolution stage, suggesting that most long-lasting emission line objects could be Class I systems characterized by massive accretion discs and envelopes. 

In summary, we {find} FUors have bluer $K_s - W2$ and $W1 - W2$ colours than most non-FUors during the quiescent phase. However, we can not simply draw a solid conclusion that there are no heavily embedded FUors or that all FUors are slightly more evolved than other eruptive sources, due to the limited size of samples and contribution from variable extinction. In a forthcoming work, Morris et al, (in prep) will obtain spectroscopic confirmation of a group of extremely red candidate FUors discovered from the mid-infrared {\it NEOWISE} time series.

\begin{table}
\caption{Variation amplitudes and pre-outbursting colour of eruptive YSOs}
\renewcommand\arraystretch{1.4}
\begin{tabular}{p{1.5cm} p{1.05cm} p{0.45cm} p{0.65cm} p{0.65cm} p{0.35cm} c}
\hline
\hline
Name & Type & $t_{\rm rise}$ & $\Delta K_{\rm s, rise}$ &  $\Delta K_{\rm s, res}$ &  $\Delta W2$ & $K_s - W2$\\
\hline
& & day &~mag &~mag &~mag &~mag \\
\hline
L222\_1 & FUors &  609 & 3.2 & 0.7 & 2.1 & 3.6 \\
L222\_4$^{\star}$ & FUors &  793 & 6.1 & 1.0 & >4.8 & <4.6\\
L222\_10& FUors &  939 & 3.3 & 0.5 & - & -\\
L222\_13$^{\star}$ & FUors &  481 & 4.0 & 0.4 & >2.5 & <3.1\\
L222\_15 & FUors & 995 & 4.6 & 1.2 & >4.0 & <3.4\\
L222\_18$^{\dagger}$ & FUors &  510 & 4.0 & 0.6  & 3.8 & 3.3\\
L222\_25 & FUors &  849 & 3.7 & 1.2 & - & -\\
L222\_33$^{\dagger}$ & FUors &  751 & 4.1 & 0.7 & >3.5 & 2.8\\
VVV1640-4846$^{\dagger}$ & FUors &  747 & 3.0 & 1.0 & 1.9 & 5.2\\
L222\_73$^{\ddagger}$ & FUors &  >589 & >4.3 & 0.9 & >3.4 & 5.3\\
L222\_78 & FUors &  500 & 4.6 & 0.6 & 3.3 & 2.8\\
L222\_93 & FUors &  527 & 4.6 & 1.4 & >3.2 & <3.1\\
L222\_95$^{\dagger}$ & FUors & 1439 & 5.4 & 1.3 & 3.3 & 4.8 \\
L222\_165 & FUors &  705 & 3.9 & 1.6 & >3.3 & <2.9\\
L222\_192$^{\dagger}$ & FUors &  463 & 4.9 & 0.6 & 2.8 & 2.6 \\
VVVv322 & FUors &  129 & 2.3 & 0.6 & - & 3.3 \\
VVVv721$^{\ddagger}$ & FUors & >2140 & >3.1 & 0.6 & >2.0 & 4.3 \\
DR4\_v20$^{\ddagger}$ & FUors &  >838 & >3.3 & 0.6 & >1.1 & 5.8\\
VVVv270 & Em. Line &  895 & 3.3 & 0.7 & 1.3 & 5.7 \\
VVVv631 & Em. Line & 1926 & 2.6 & 1.2 & 1.5 & 3.2\\
L222\_6  & Em. Line &  900 & 3.4 & 1.4 & 0.6 & 6.6 \\
L222\_148$^{\star}$ & Em. Line & 2040 & 3.6 & 2.1  & >0.7 & <6.9 \\
L222\_167 & Em. Line & 1038 & 5.6 & 1.3 & 1.7 & 7.5\\
DR4\_v10$^{\ddagger}$ & Em. Line & >1173 & >2.9 & 1.5 & >0.8 & 6.1\\
DR4\_v34 & Em. Line &  817 & 3.7 & 1.2 & 0.3 & 6.5\\
Stim1$^{\ddagger}$ & Em. Line & >2126 & >2.2 & 0.8 & >1.6 & 3.6\\
Stim5$^{\ddagger}$ & Em. Line &  >874 & >3.4 & 1.3 & >0.7 & 5.1 \\
L222\_32 & Outflow & 1088 & 3.3 & 2.4  & 0.4 & 7.9\\
L222\_37$^{\ddagger}$ & Outflow &  >808 & >4.9 & 1.2 & >2.6 & 5.0\\
L222\_120 & Outflow &  808 & 4.5 & 2.2 & 1.8 & 4.8\\
L222\_210$^{\star}$ & Outflow &  941 & 3.7 & 1.9 & >1.0 & <6.3\\
VVVv800 & Outflow &  800 & 2.0 & 0.8 & 1.6 & 6.2\\
\hline
\hline
\end{tabular}
\flushleft{This table contains 32 large-amplitude and long-duration outbursting YSOs discovered in the VVV survey \citep[][{LSG23}, and this paper]{Contreras2017b, Guo2021}. \\
$\star:$ Eruptive objects without clear detections from the {\it WISE} images due to blending with a spatially nearby source. \\
$\dagger:$ Sources without $W2$ detections during the photometric minima, but have {\it I2}-band detections from {\it Spitzer}.\\
$\ddagger:$ Eruptive events started their rising stage before the VVV survey.
}
\label{tab:eruptivelc}
\end{table} 

\section{The periodic variation on L222\_148}
\label{sec: L222_148}
\begin{figure*}
\includegraphics[height=3.3in,angle=0]{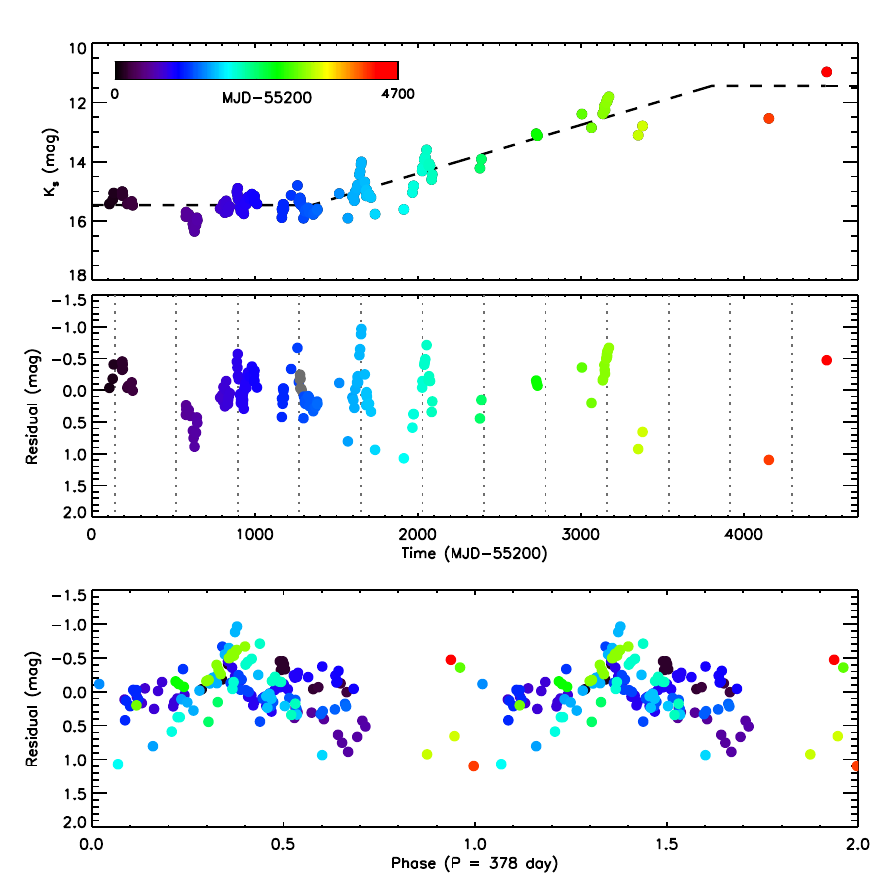}
\includegraphics[height=3.3in,angle=0]{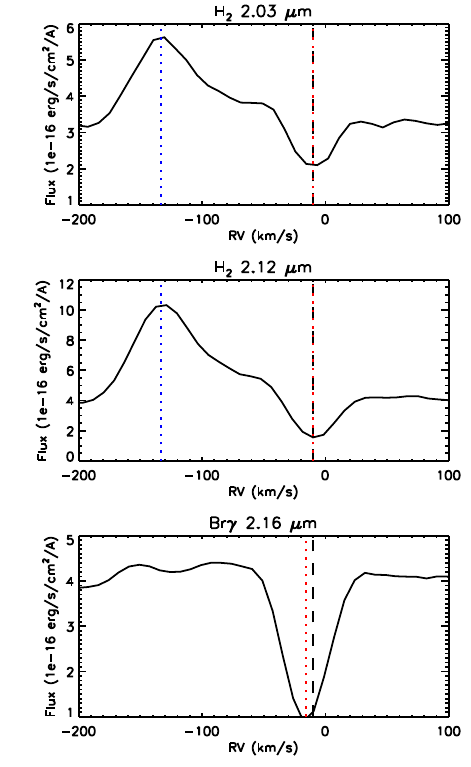}
\caption{{\it Upper Left}: $K_s$ light curve of L222\_148, colour coded by the observation time. Overall linear trends are shown by the dashed lines. {\it Middle Left}: residual light curves after removing the linear trends. The locations of periodic photometric maxima are marked by dotted vertical lines. {\it Lower Left}: phase-folded residual light curves. {\it Right}: profiles of $H_2$ emission lines and the Br$\gamma$ line. The radial velocity of CO bandhead emission is marked by the black dashed line. The central wavelengths of the emission/absorption features are shown by the dotted lines.}
\label{fig:L195_V129}
\end{figure*}

One emission line object, L222\_148, exhibits repeated lower amplitude variation in addition to the overall long-term eruptive behaviours. We applied an analytical fit to the $K_s$ light curves of L222\_148, composed of a sinusoidal function and three linear slopes as the quiescent stage, rising stage and the brightness plateau (see Figure~\ref{fig:L195_V129}). A 378-day quasi-periodic signal is detected with $\Delta K_s \sim$ 2.0 mag after applying the generalised Lomb-Scargle periodogram \citep[][]{Zechmeister2009} to the residual light curve. Since the photometric minima are poorly covered, the M-index is not accurately calculated, hence we did not recognise this target as a periodic outbursting candidate \citep[][see the definition of M-index herein]{Guo2022}. The hybrid photometric behaviour of L222\_148 resembles the eruptive source LkH$\alpha$ 225 South \citep{Hillenbrand2022a}, with a period of 43 days and an overall 7 mag eruption in optical bands. The photometric period of L222\_148 is likely attributed to the Keplerian rotation of an asymmetric structure in the circumstellar disc, which may be triggered by a secondary body in the system, similar to the 130 candidate periodically outbursting YSOs discovered in \citet{Guo2022}.

The near-infrared spectrum of L222\_148, taken during the outbursting stage, shows emission features associated with the magnetospheric accretion process (e.g. Na{\sc ii} and CO bandheads) and coupled with strong stellar outflow (H$_2$ line series). In contrast, the Br$\gamma$ line is in absorption. Inverse P Cygni profiles are seen on the H$_2$ lines, suggesting a combination of blue-shifted stellar outflow coupled with a stream of cold disc materials (see the right panels of Figure~\ref{fig:L195_V129}). The RV of the absorption components are consistent with the CO bandhead emission ($\sim -10$ km/s). In a recent spectrum of this target, we do not detect any absorption features around H$_2$ lines, and the Br$\gamma$ line (previously in absorption) is in emission. We infer that the changing of the spectral features results from the evolution of the inner disc structure, which might be triggered by gravitational instabilities or perturbation from a secondary object. The new spectrum will be published in a follow-up work. 

This unique target, L222\_148, provides us with a complex eruptive phenomenon, as a slow-rising accretion burst happens on a possible stellar binary or a star-brown dwarf system. We have detected the hot inner accretion disc, a cold material stream, and stellar wind/outflow from the near-infrared spectrum. It is an intriguing case to explore the physics corresponding to variable mass accretion processes on timescales from years to a decade. In addition, we can not rule out the possibility that this object is in a stellar binary (explains the two spectral components), in which one source exhibits periodic variation and the other source undergoes a slow-rising outburst, like the case of Z CMa \citep{Bonnefoy2017}. 

\section{Summary}
\label{sec:con}

In our series of works, over two hundred high-amplitude variable stars were identified from the decade-long VVV/VVVX $K_s$ time series (see LSG23). Among them, FUor-type outbursts are seen among the highest amplitude and longest duration events. In this work, we have presented near-infrared spectroscopic follow-up observations of 33 sources using the XSHOOTER spectrograph on VLT and FIRE on the Magellan telescope. These sources are categorised into a few groups based on their spectral features and the nature of their variability. We summarise our findings as follows. 

\begin{itemize}

\item Fifteen new FUor-type objects are confirmed by unique spectral signatures, such as CO absorption bands beyond 2.3 $\mu$m and the triangular-shaped $H$-band continuum. 


\item Four targets are classified as outflow-dominated with $H_2$ emission line series, and seven targets are identified as emission line objects with signatures of magnetospheric accretion. Most of these sources have a longer duration than the optically seen EXors.

\item Eight sources, including seven dipping giants, are identified as post-main-sequence sources with the existence of $^{13}$CO absorption features and extremely red near-infrared colours (five of them have $H - K_s > 3.5$). The rarely seen deep near-infrared AlO absorption bands are detected on an eruptive source, L222\_59.

\item We form a sample of 32 spectroscopically confirmed long-duration eruptive YSOs originally identified from the VVV survey. We find that in the near-infrared, FUors have faster eruptive timescales and predominate the highest amplitude end ($\Delta K_s > 3.8 mag$). In the mid-infrared, FUors have systematically higher $W2$ amplitude than other eruptive objects. It suggests that FUors can heat their inner accretion disc more efficiently.

\item FUors have comparable near-infrared and mid-infrared amplitudes. By contrast, most non-FUors have much lower amplitudes in the mid-infrared than in the near-infrared, with redder $K_s - W2$ and $W1 - W2$ colours, indicating early evolutionary stages or high line-of-sight extinction.

\item In \citet{Guo2021}, we found that most multi-year duration eruptive events on YSOs are emission line objects. However, in this work, we notice that the distribution is reversed towards the highest amplitude end, as FUors are more abundant among the eruptive samples with $\Delta K_s > 4$ mag.

\end{itemize}

In summary, this work significantly increased the total number of spectroscopically known FUor-type events. Our photometric and spectroscopic surveys provide several key characteristics of the eruptive timescales and amplitudes of eruptive events in two sub-categories. To further reveal the physical nature behind these eruptive events, high-cadence multi-wavelength observations and numerical simulations are desired.

\section{Data Availability}
The {\it WISE} and {\it Spitzer} data underlying this article are publicly available at the IRSA server \url{https://irsa.ipac.caltech.edu/Missions/wise.html}, \url{https://irsa.ipac.caltech.edu/Missions/spitzer.html}, 
and \url{https://irsa.ipac.caltech.edu/Missions/2mass.html}. The VVV and VVVX data are publicly available at the ESO archive \url{http://archive.eso.org/cms.html}. The VIRAC2$\beta$ version of the VVV/VVVX light curves has not yet been publicly released but is available on request to the first author. Raw spectra are available on the ESO archive service when released to the public. Reduced spectra are provided at  \url{http://star.herts.ac.uk/~pwl/Lucas/GuoZ/VVVspec/}. 

\section*{Acknowledgements}

We thank the anonymous referee for providing valuable comments and suggestions to improve this work. ZG is supported by the ANID FONDECYT Postdoctoral program No. 3220029. ZG and KM acknowledge support by ANID, -- Millennium Science Initiative Program -- NCN19\_171.

ZG, PWL, and CJM acknowledge support by STFC Consolidated Grants ST/R00905/1, ST/M001008/1 and ST/J001333/1 and the STFC PATT-linked grant ST/L001403/1. This work has made use of the University of Hertfordshire's high-performance computing facility (\url{http://uhhpc.herts.ac.uk}).

We gratefully acknowledge data from the ESO Public Survey program ID 179.B-2002 taken with the VISTA telescope, and products from the Cambridge Astronomical Survey Unit (CASU). This work contains data from ESO programs 105.20CJ and 109.233U. 

 J.B. and R.K. thank the support from the Ministry for the Economy, Development and Tourism, Programa Iniciativa Cientifica Milenio grant IC120009, awarded to the Millennium Institute of Astrophysics (MAS). 
D.M. gratefully acknowledges support from the ANID BASAL projects ACE210002 and FB210003, from Fondecyt Project No. 1220724, and from CNPq Brasil Project 350104/2022-0.
  J.A.-G. acknowledges support from Fondecyt Regular 1201490 and from ANID's Millennium Science Initiative ICN12\_009, awarded to the Millennium Institute of Astrophysics (MAS).
ACG acknowledges support from INAF-GOG "NAOMY: NIR-dark Accretion Outbursts in Massive Young stellar objects" and PRIN 2022 20228JPA3A - PATH.
SY and JT thank the support of the European Research Council (ERC) under the European Union’s Horizon 2020 research and innovation programme through Advance Grant number 883830 and of STFC under the project ST/R000476/1.
AA acknowledge the support through a Fellowship for National Ph.D. students from ANID, grant number 21212094.
CCP was supported by the National Research Foundation of Korea (NRF) grant funded by the Korean government (MEST) (No. 2019R1A6A1A10073437)
Support for M.C. is provided by ANID’s Millennium Science Initiative through grant ICN12\_009, awarded to the Millennium Institute of Astrophysics (MAS); by ANID/FONDECYT Regular grant 1231637; and by ANID's Basal grant FB210003. K. M. is funded by the European Union (ERC, WANDA, 101039452). Views and opinions expressed are however those of the author(s) only and do not necessarily reflect those of the European Union or the European Research Council Executive Agency. Neither the European Union nor the granting authority can be held responsible for them.

This research has made use of the NASA/IPAC Infrared Science Archive, which is funded by the National Aeronautics and Space Administration and operated by the California Institute of Technology.

\bibliographystyle{mnras}
\bibliography{reference}

\appendix
\section{Near to mid-infrared SEDs}

Here we present the spectral energy distribution of our targets between 1 to 24 $\mu$m. Wherever possible, we applied contemporaneous $J$, $H$, $K_s$ photometry from VVV or NTT/SOFI observations, and contemporaneous mid-infrared data from {\it ALLWISE} (preferred) or {\it Spitzer} surveys. The spectral index ($\alpha$) is measured as the slope of the linear fit to the quiescent SED between 2-24 $\mu$m. There are eight sources without the measurement of $\alpha$, including three sources that were only detected by WISE or {\it Spitzer} during their photometric maxima, three sources that were blended during their photometric minima and two sources without any mid-infrared detections. The SEDs of these eight sources are not shown. The spectral index is not measured on sources without detections beyond 5 $\mu$m.

\begin{figure*}
\includegraphics[width=1.55in,angle=0]{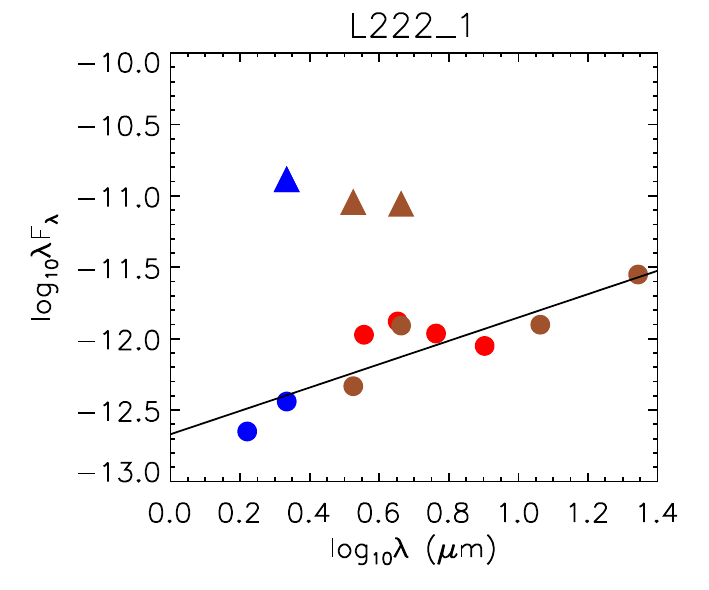}
\includegraphics[width=1.55in,angle=0]{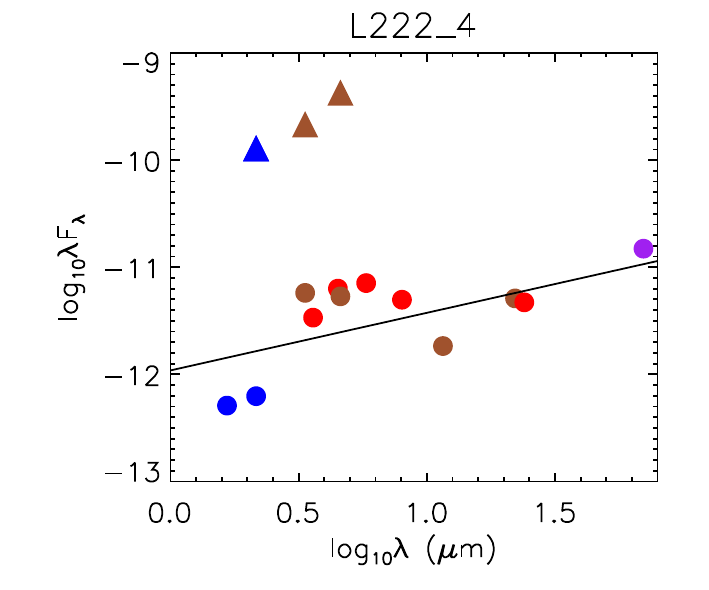}
\includegraphics[width=1.55in,angle=0]{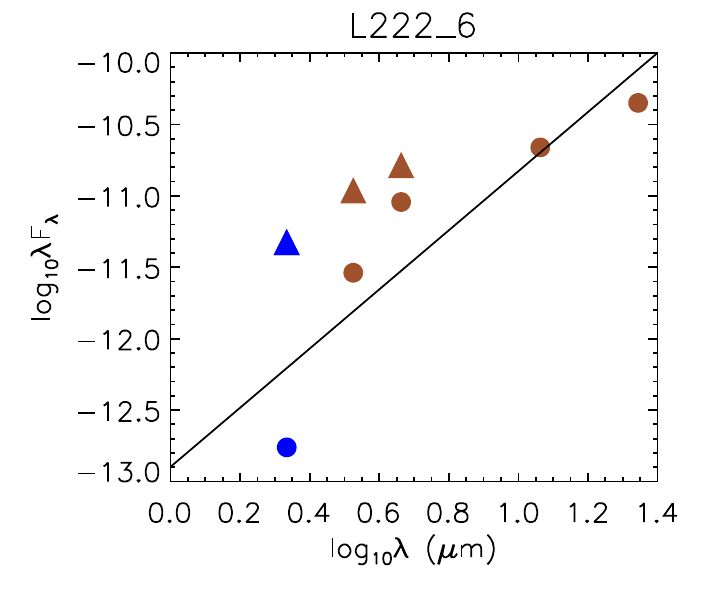}
\includegraphics[width=1.55in,angle=0]{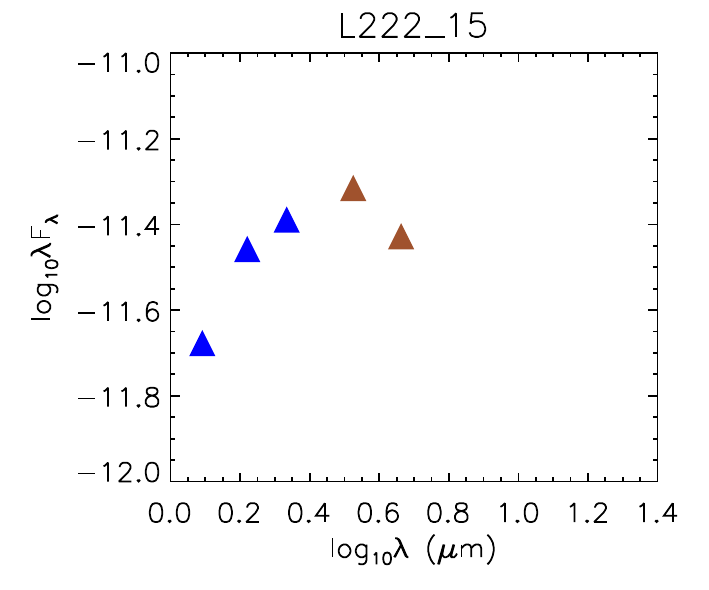}
\includegraphics[width=1.55in,angle=0]{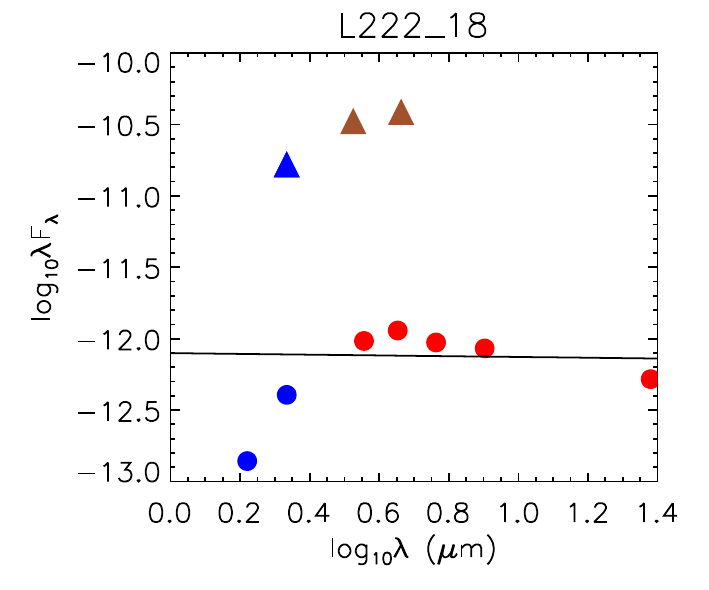}
\includegraphics[width=1.55in,angle=0]{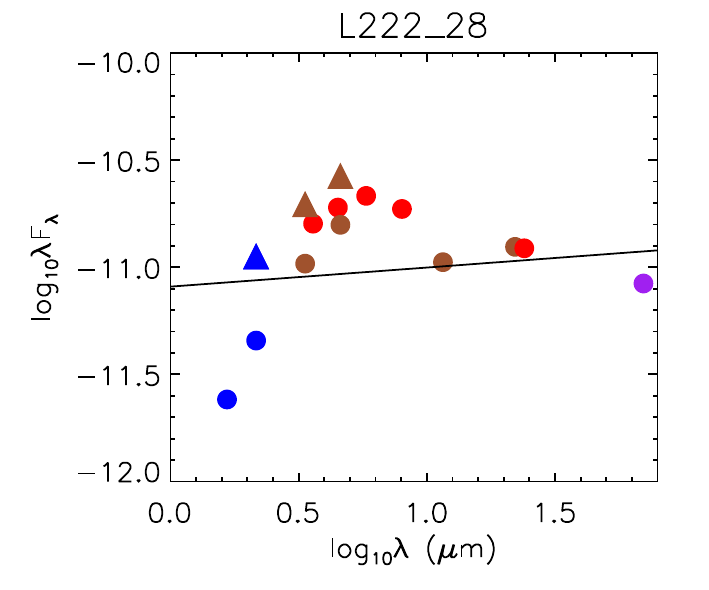}
\includegraphics[width=1.55in,angle=0]{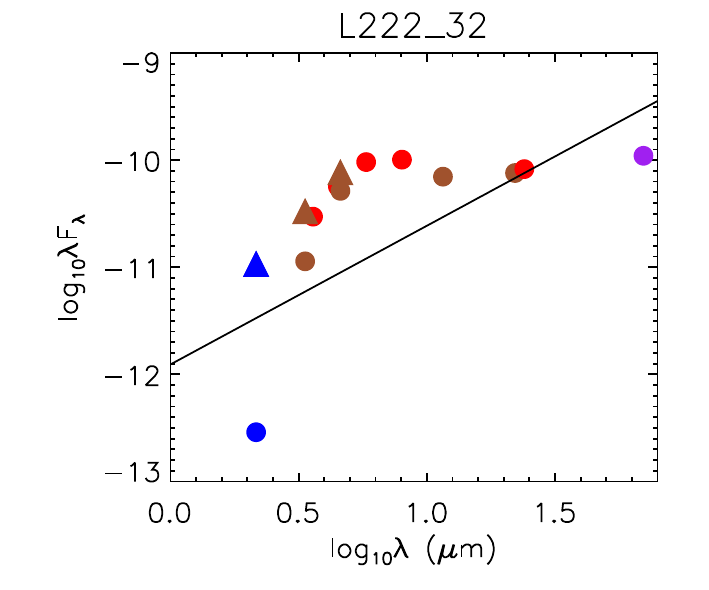}
\includegraphics[width=1.55in,angle=0]{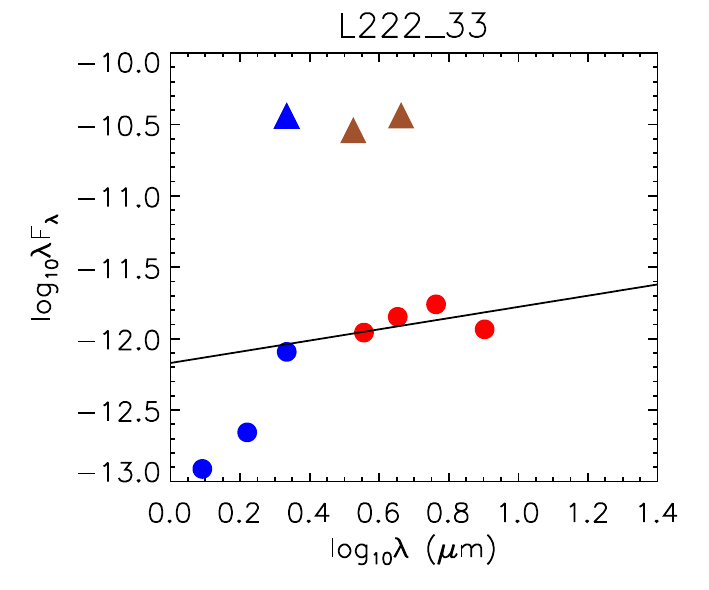}
\includegraphics[width=1.55in,angle=0]{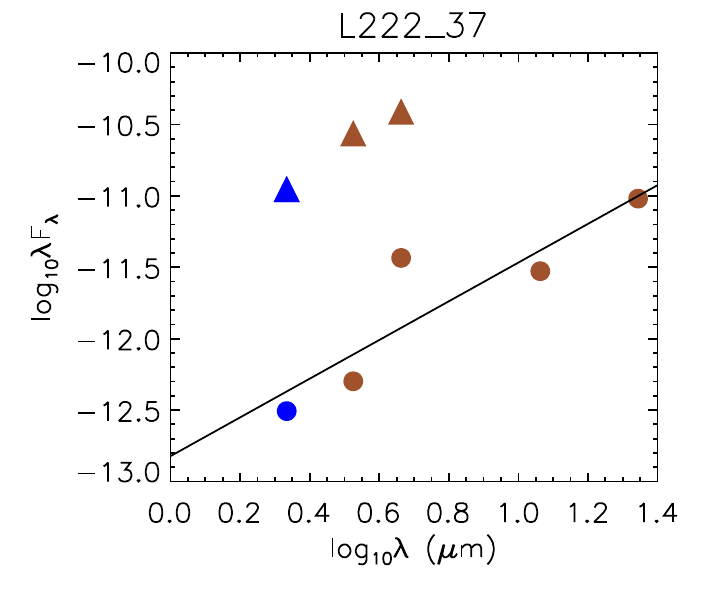}
\includegraphics[width=1.55in,angle=0]{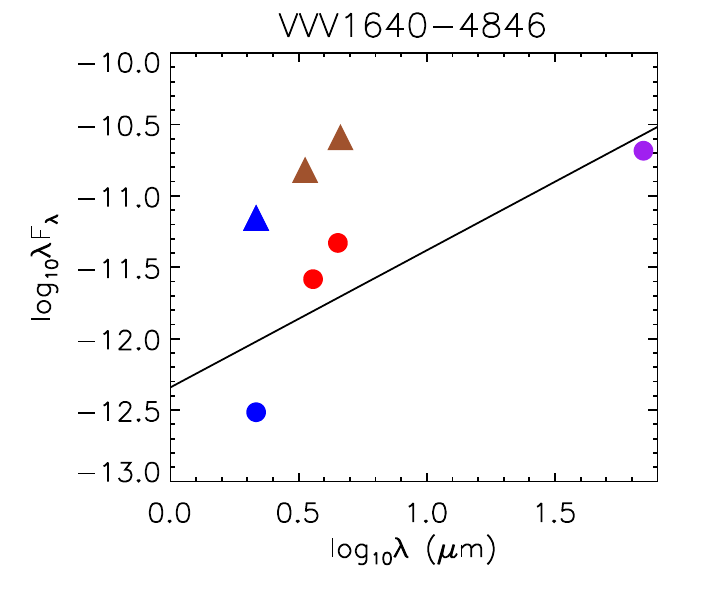}
\includegraphics[width=1.55in,angle=0]{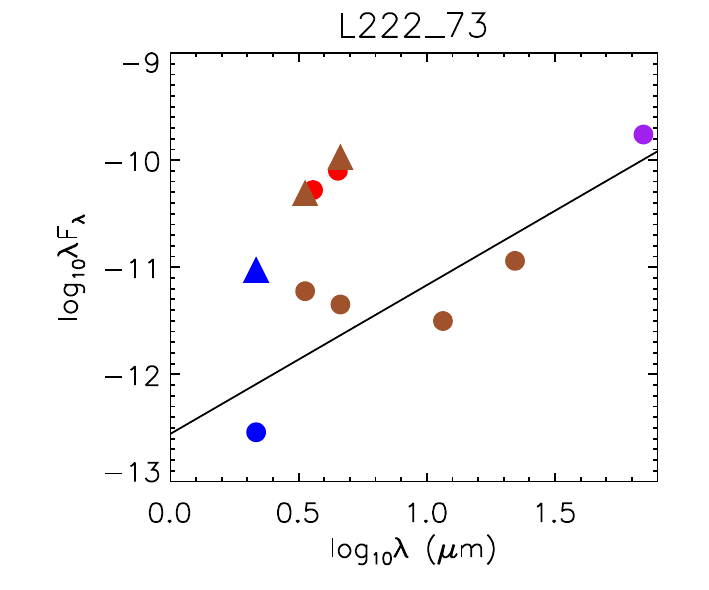}
\includegraphics[width=1.55in,angle=0]{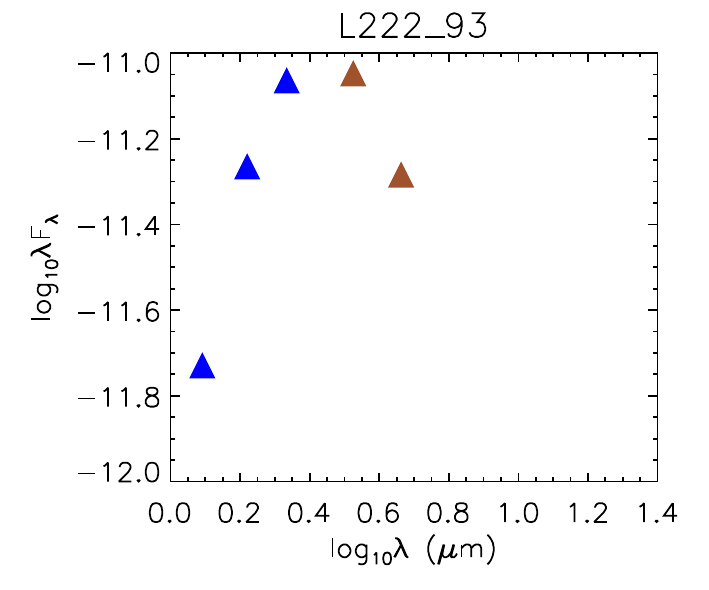}
\includegraphics[width=1.55in,angle=0]{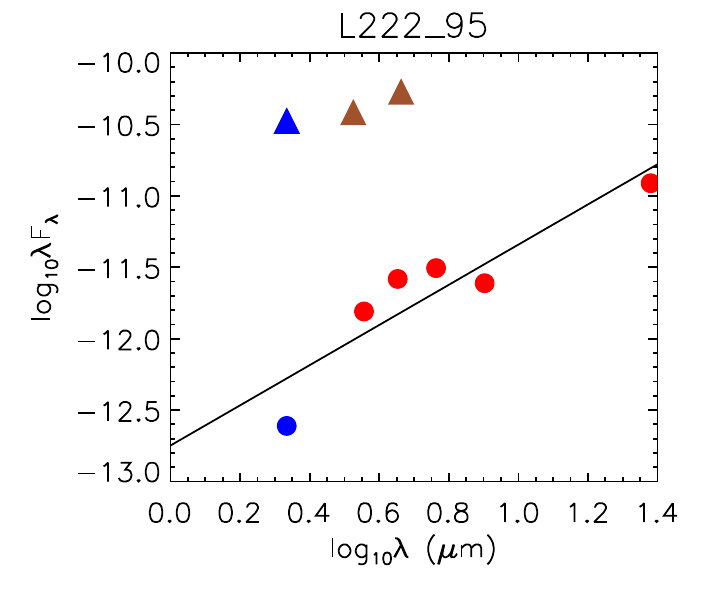}
\includegraphics[width=1.55in,angle=0]{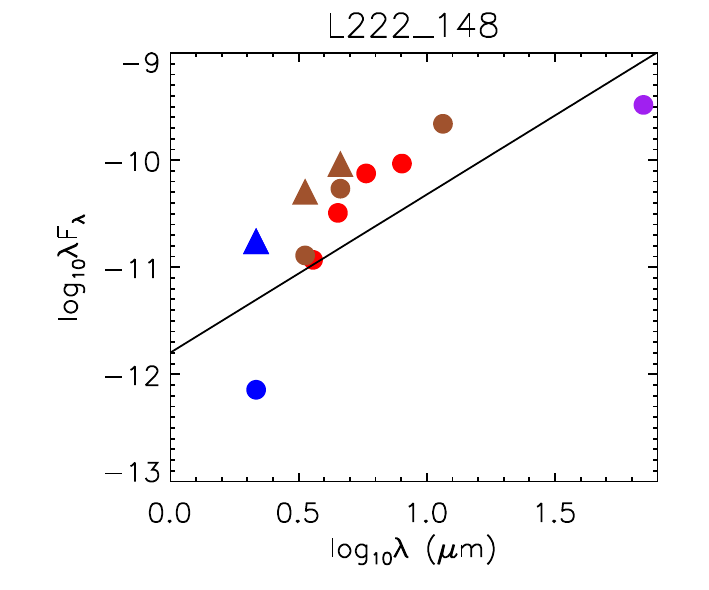}
\includegraphics[width=1.55in,angle=0]{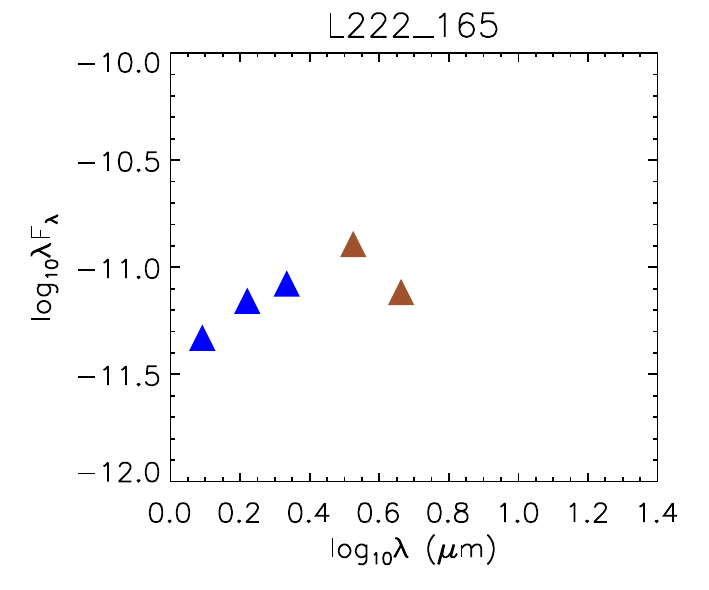}
\includegraphics[width=1.55in,angle=0]{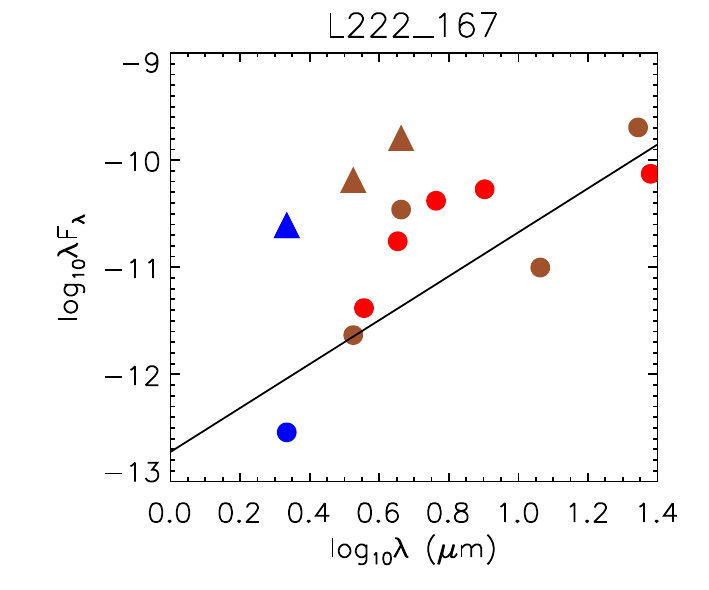}
\includegraphics[width=1.55in,angle=0]{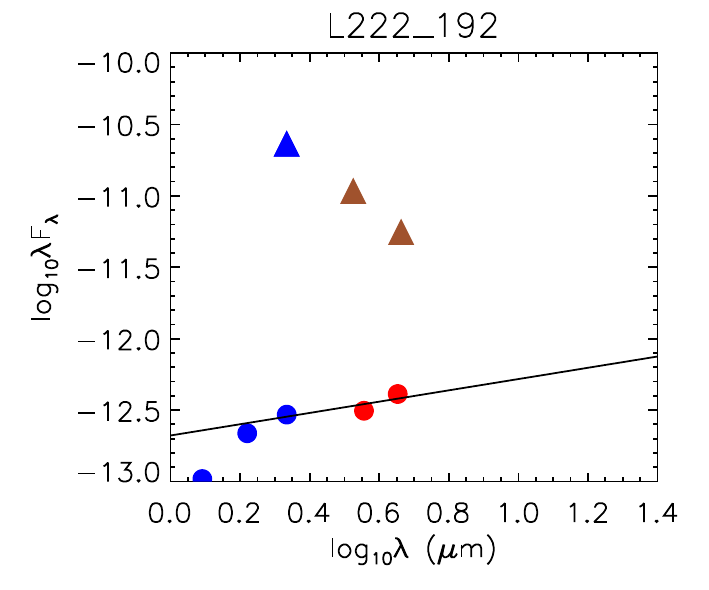}
\includegraphics[width=1.55in,angle=0]{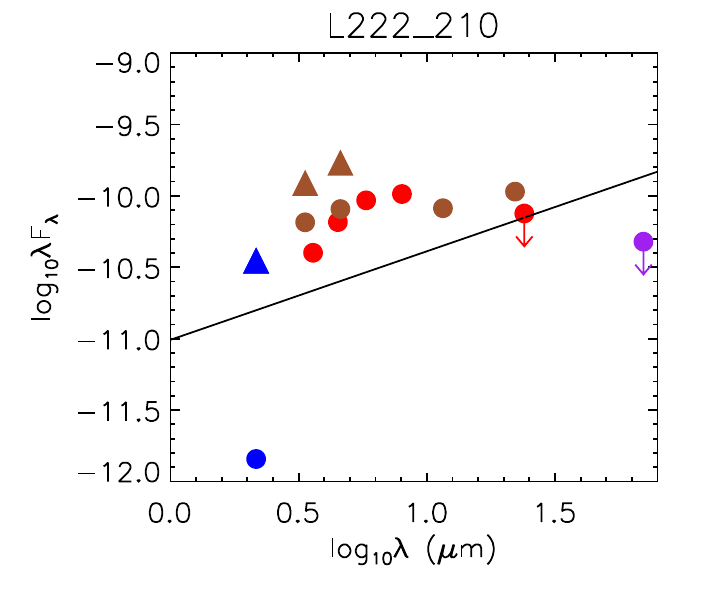}
\includegraphics[width=1.55in,angle=0]{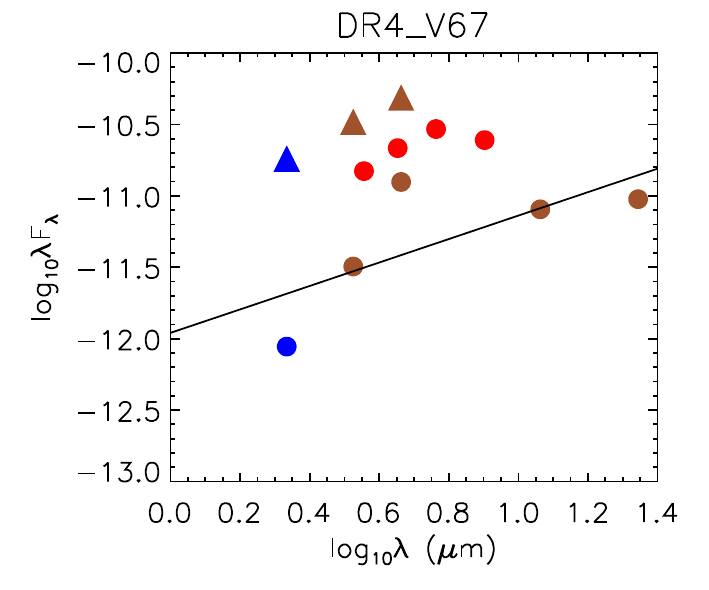}
\includegraphics[width=1.55in,angle=0]{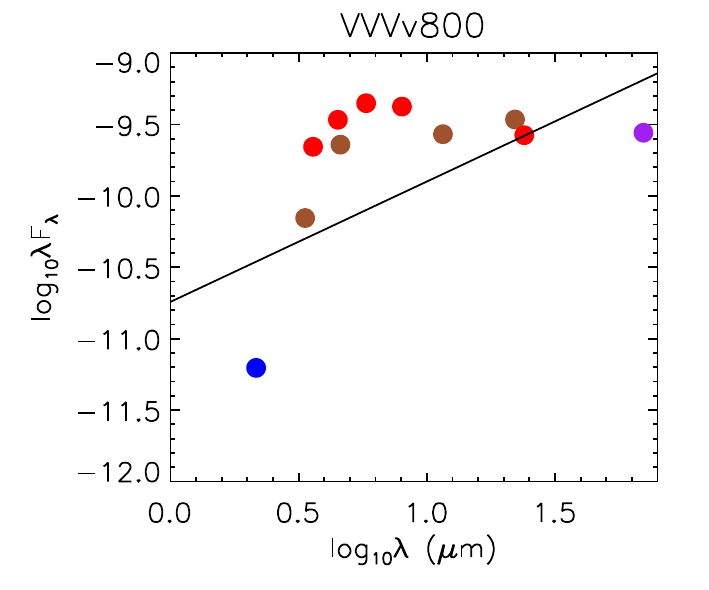}
\includegraphics[width=1.55in,angle=0]{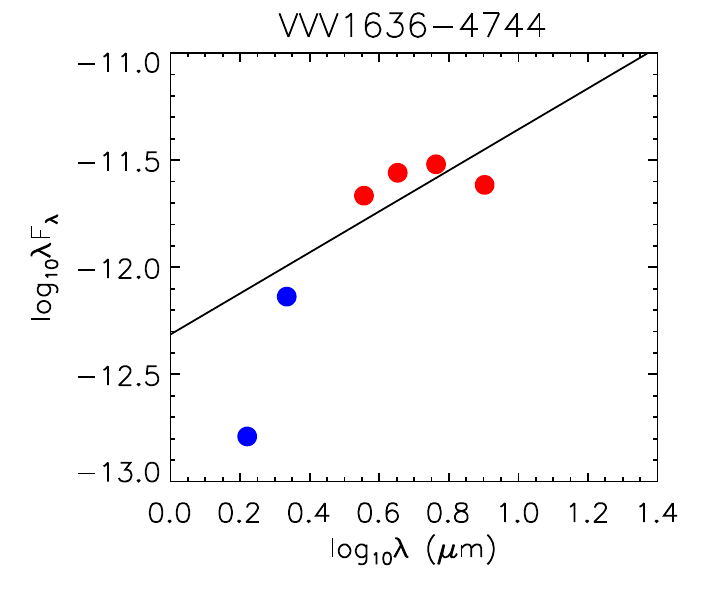}
\includegraphics[width=1.55in,angle=0]{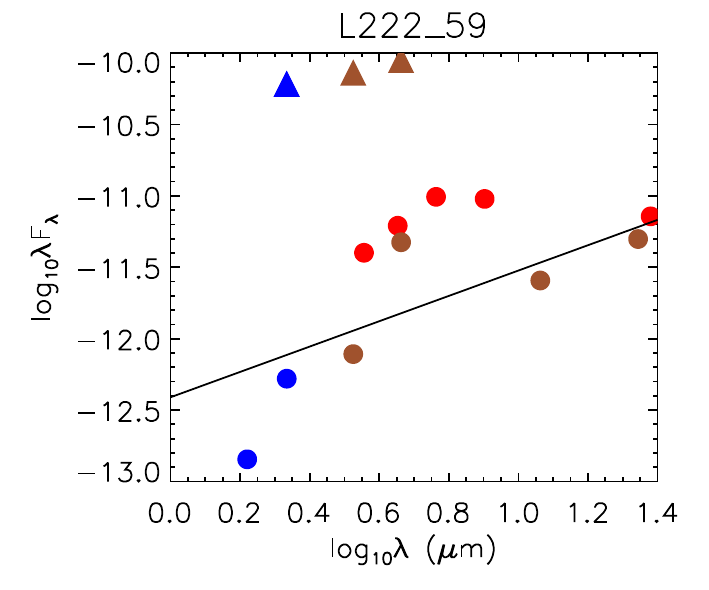}
\includegraphics[width=1.55in,angle=0]{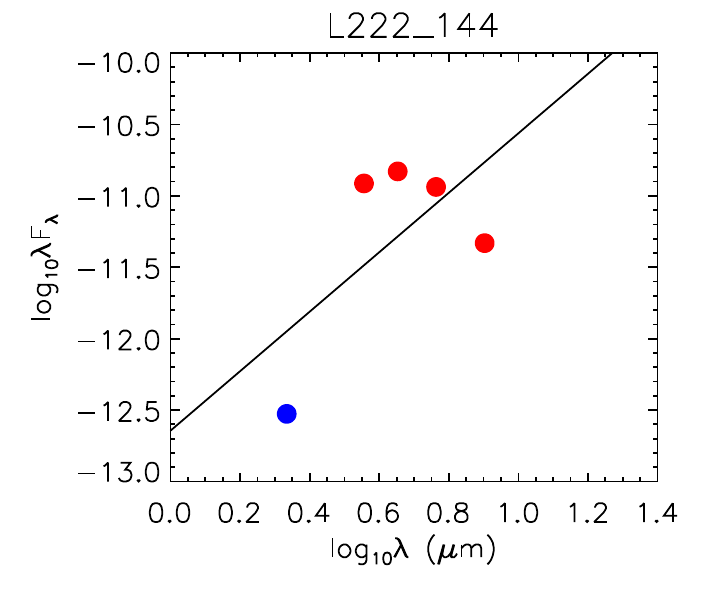}
\includegraphics[width=1.55in,angle=0]{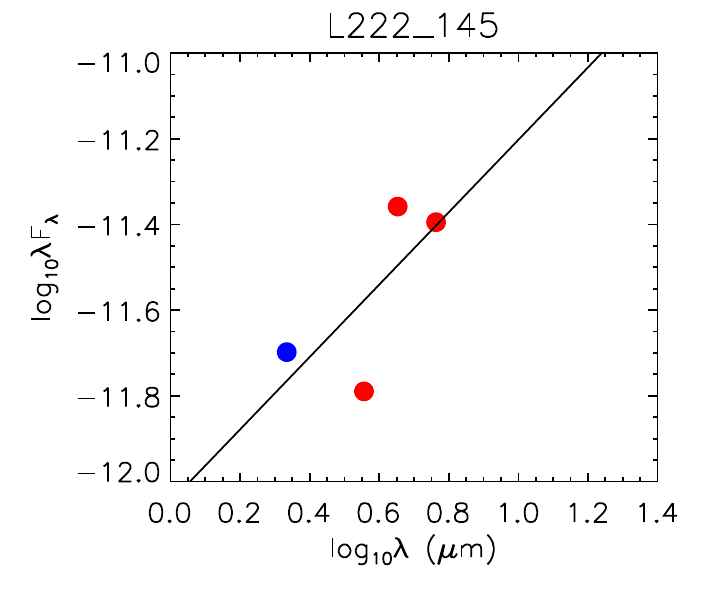}
\includegraphics[width=1.55in,angle=0]{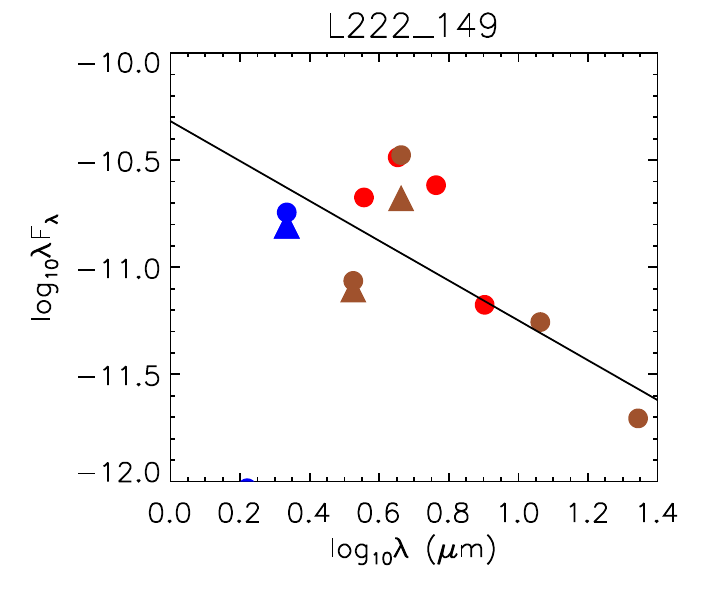}
\includegraphics[width=1.55in,angle=0]{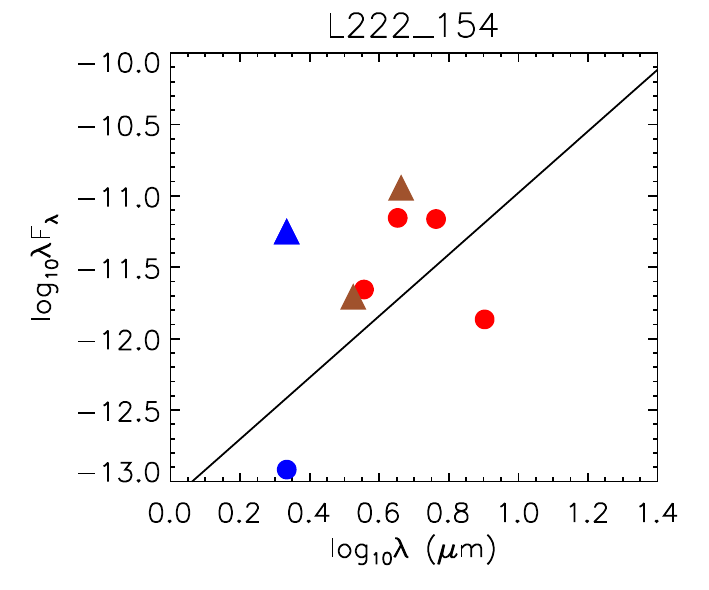}
\includegraphics[width=1.55in,angle=0]{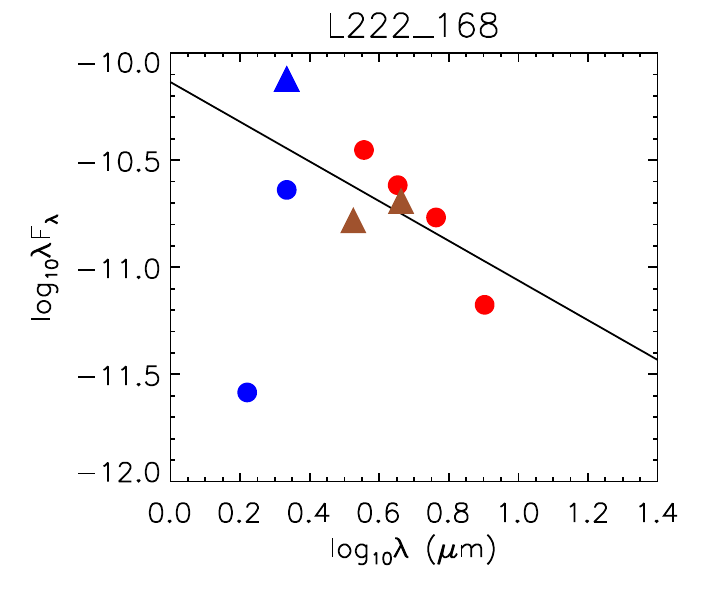}
\includegraphics[width=1.55in,angle=0]{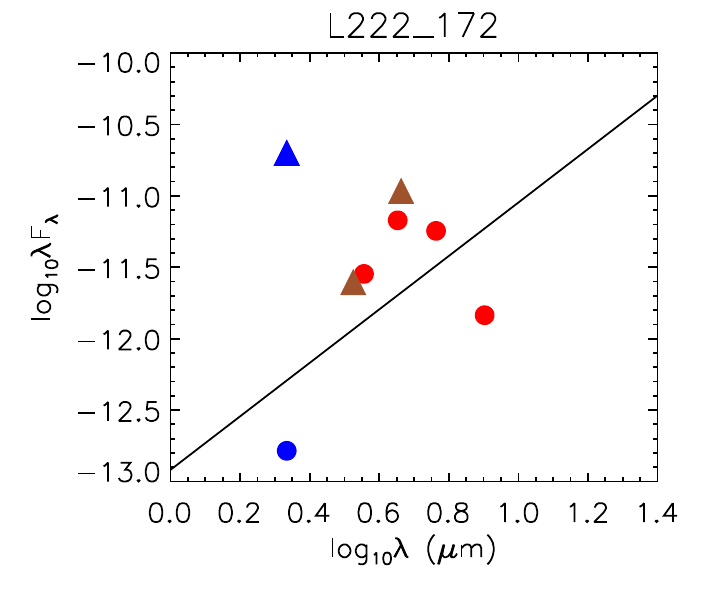}
\caption{Near to mid-infrared SEDs of 28 out of 33 sources observed in this work. Data obtained during the photometric minima are shown by dots while data obtained during photometric maxima are presented by triangles. The data point is colour-coded by the origin, as VVV or NTT/SOFI: blue; {\it WISE}: brown; {\it Spitzer}: red. Linear slopes are fit to the quiescent SEDs at $\lambda > 2 \mu$m, among sources having detections beyond 5~$\mu$m.}
\label{fig:sed}
\end{figure*}

\section{Near-infrared spectra}

We present the near-infrared spectra obtained for this work. Sources are sorted into four groups, FUors, emission line objects, outflow-dominated sources, and post-main-sequence objects. For each spectrum, we also present the $K_s$ light curves obtained from the VVV survey and our follow-up observations on NTT/SOFI. The date of the spectroscopic observation is specified in each subplot by the vertical lines. For the best visualization, all spectra are smoothed with a boxcar average of 5 pixels. 

\begin{figure*}
\includegraphics[width=1.7in,angle=0]{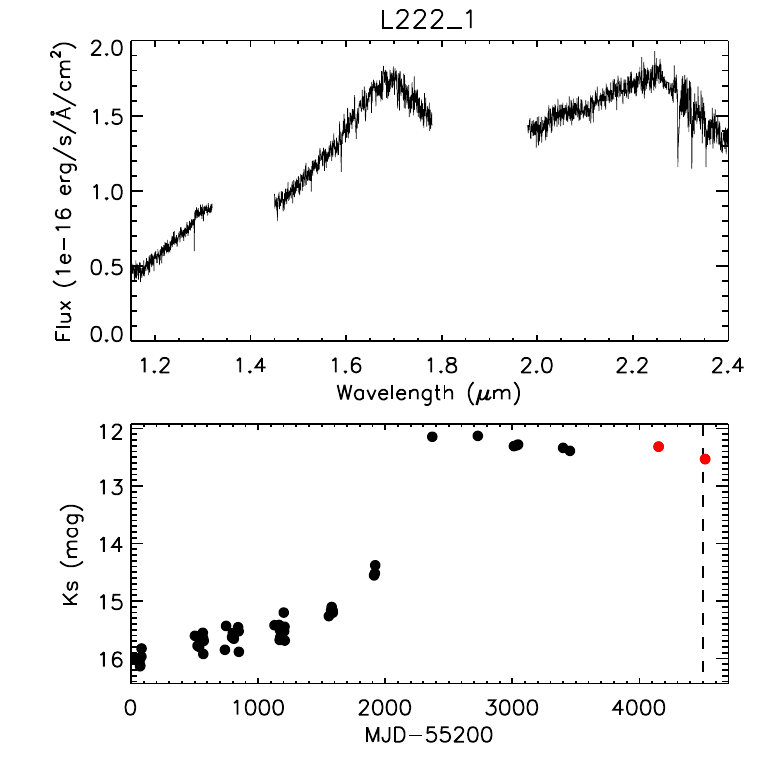}
\includegraphics[width=1.7in,angle=0]{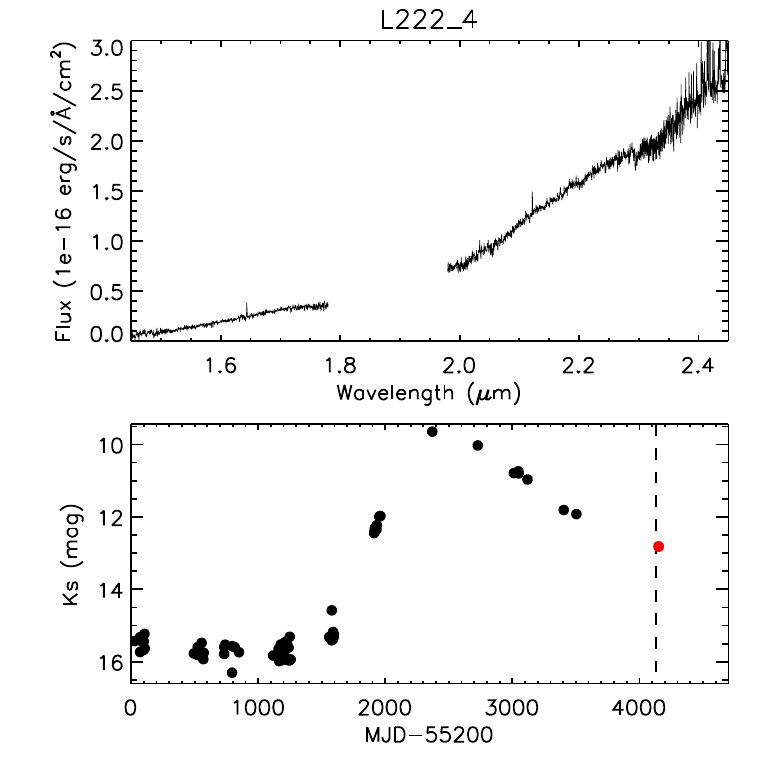}
\includegraphics[width=1.7in,angle=0]{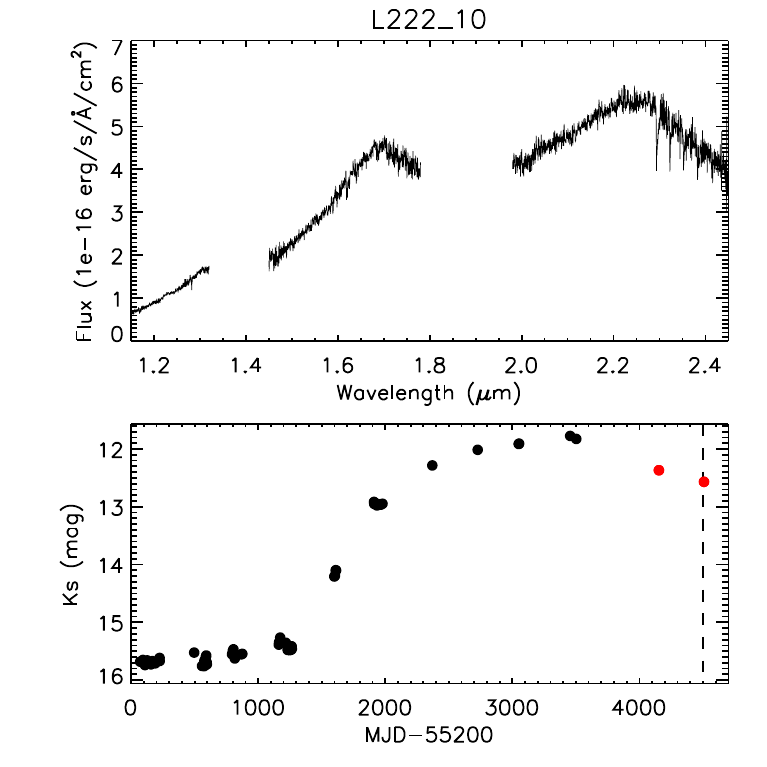}
\includegraphics[width=1.7in,angle=0]{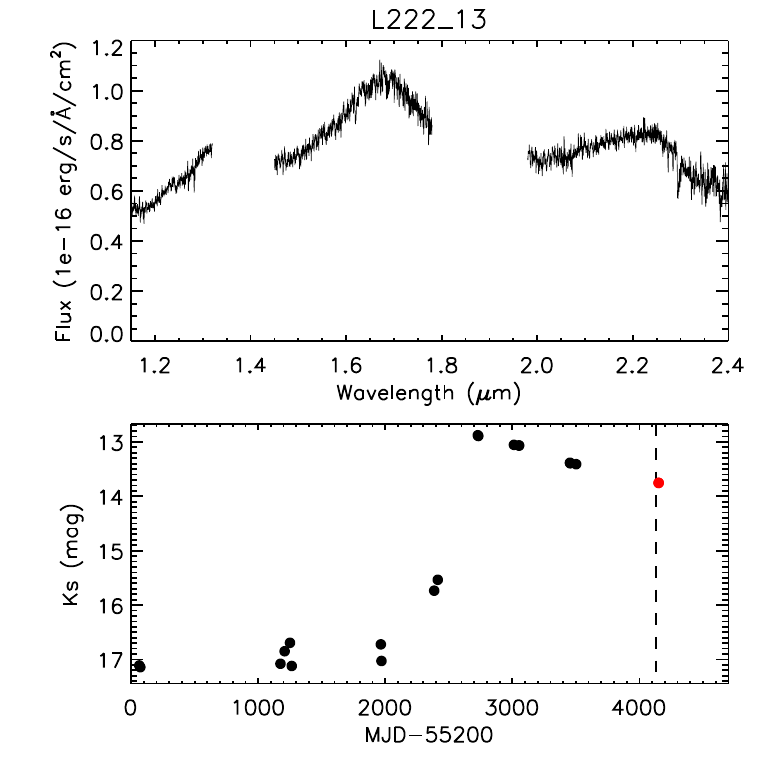}
\includegraphics[width=1.7in,angle=0]{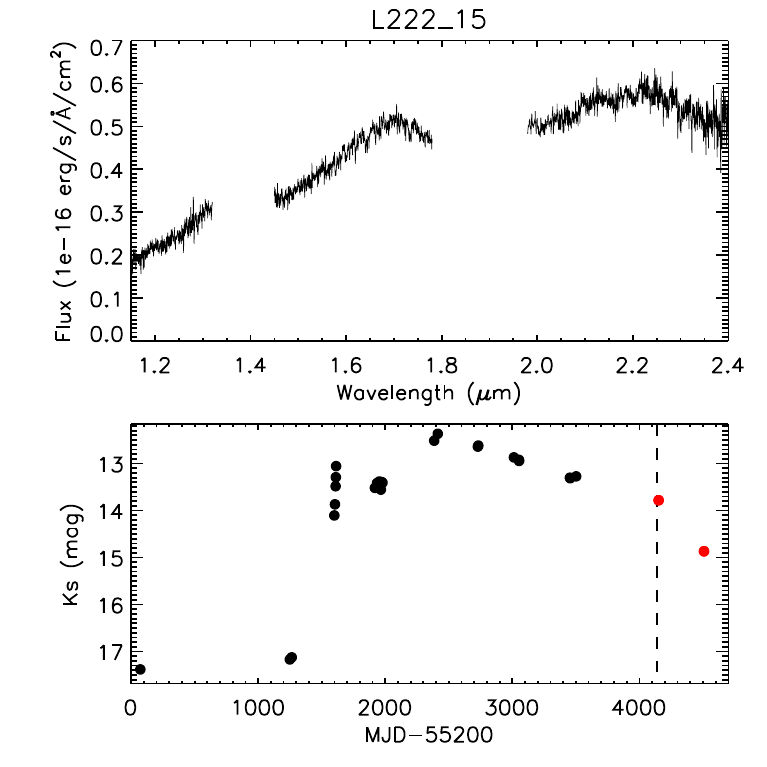}
\includegraphics[width=1.7in,angle=0]{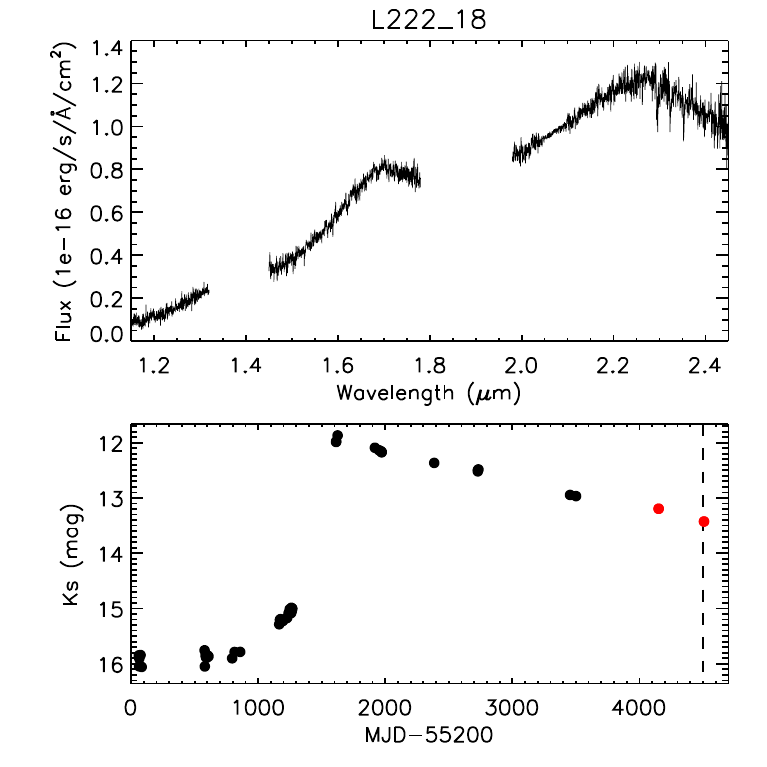}
\includegraphics[width=1.7in,angle=0]{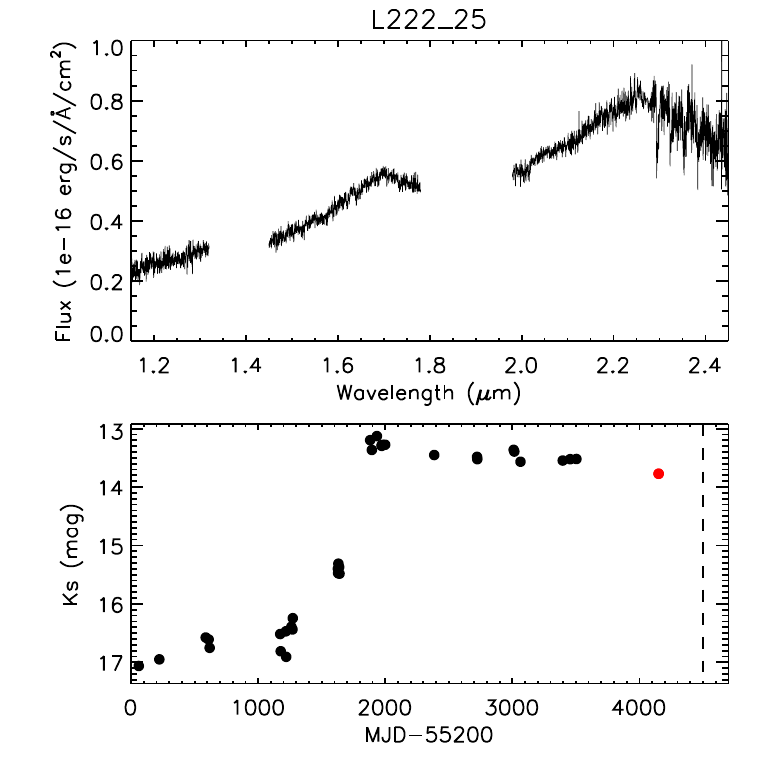}
\includegraphics[width=1.7in,angle=0]{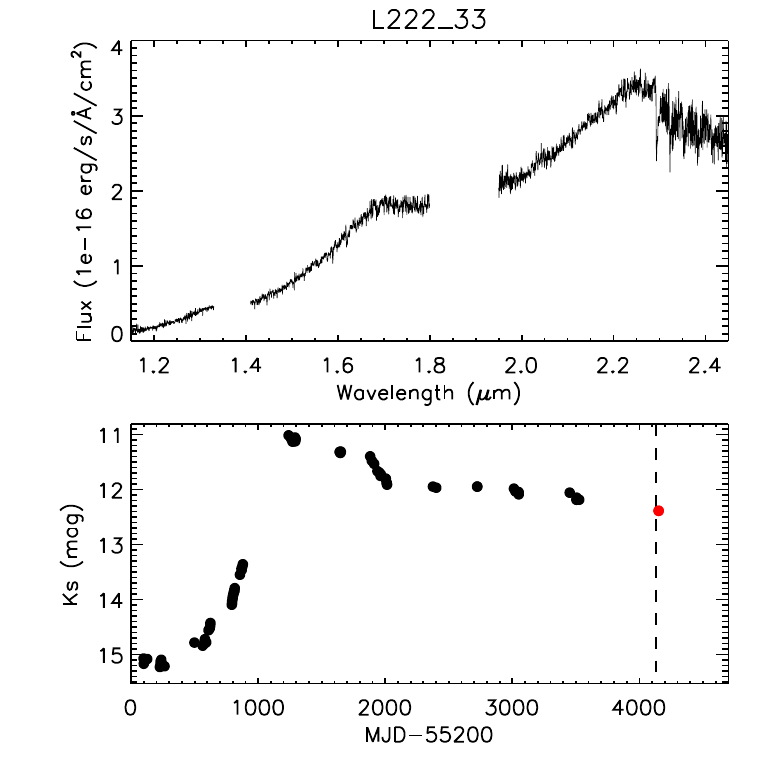}
\includegraphics[width=1.7in,angle=0]{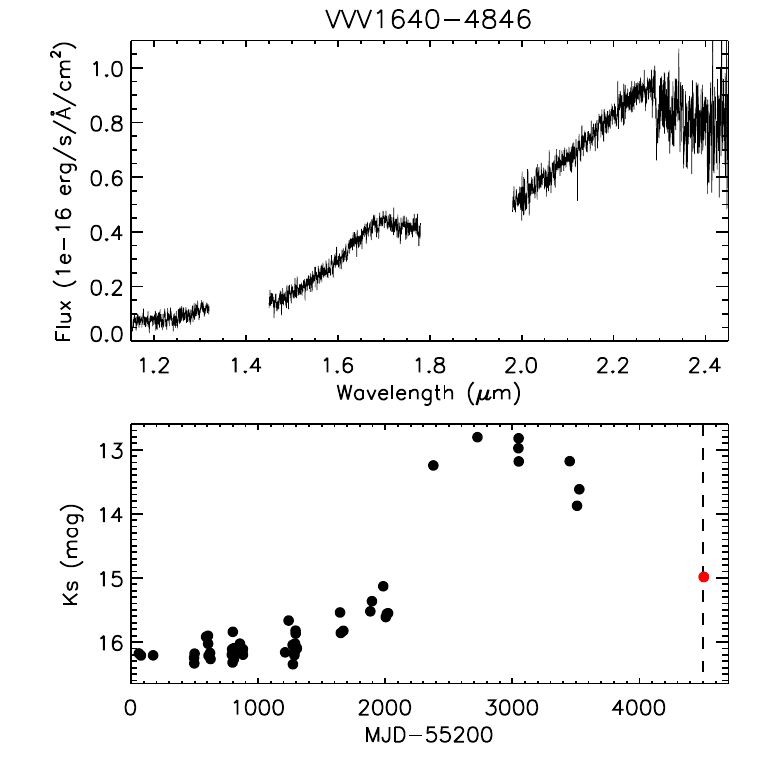}
\includegraphics[width=1.7in,angle=0]{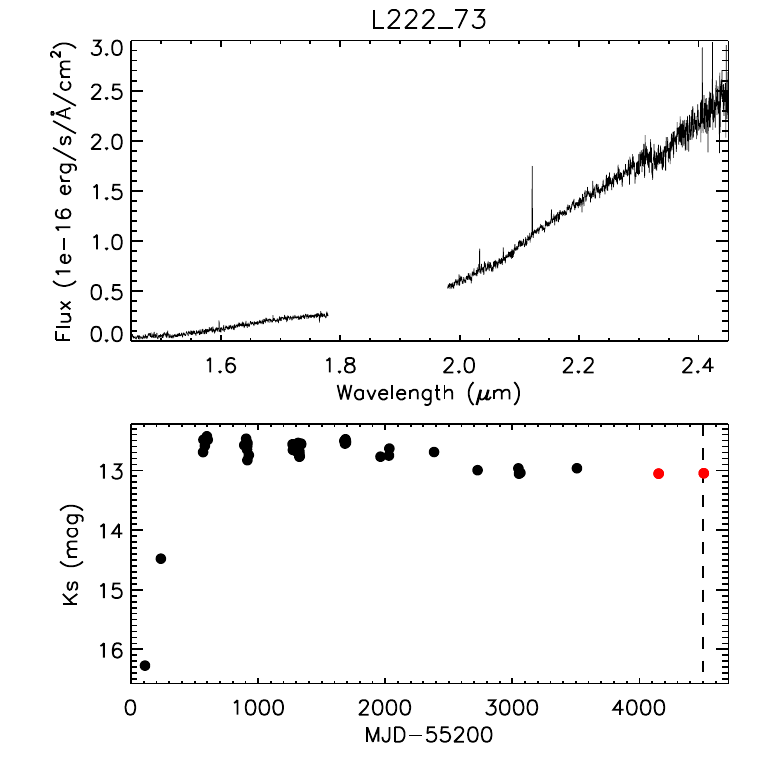}
\includegraphics[width=1.7in,angle=0]{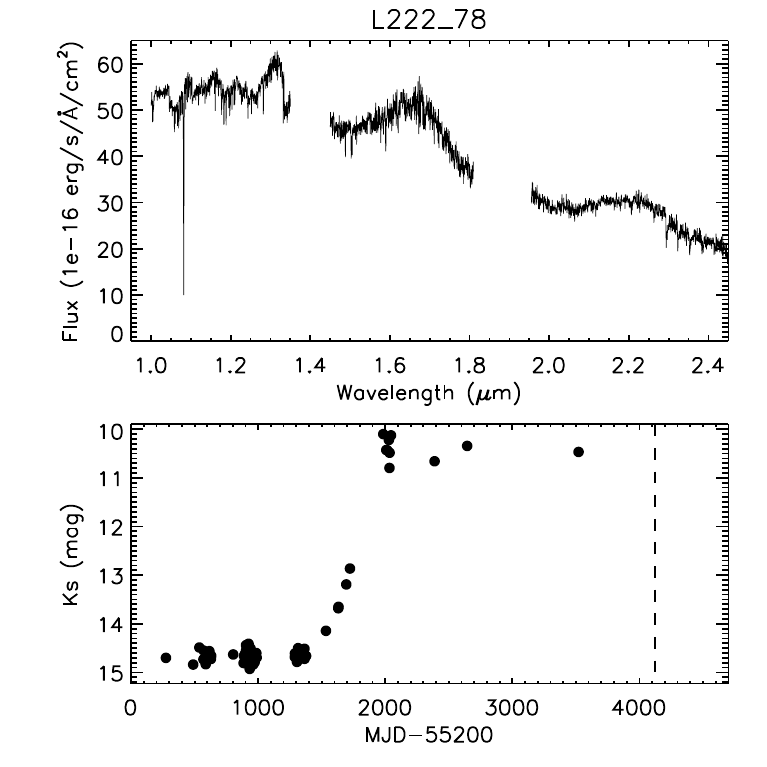}
\includegraphics[width=1.7in,angle=0]{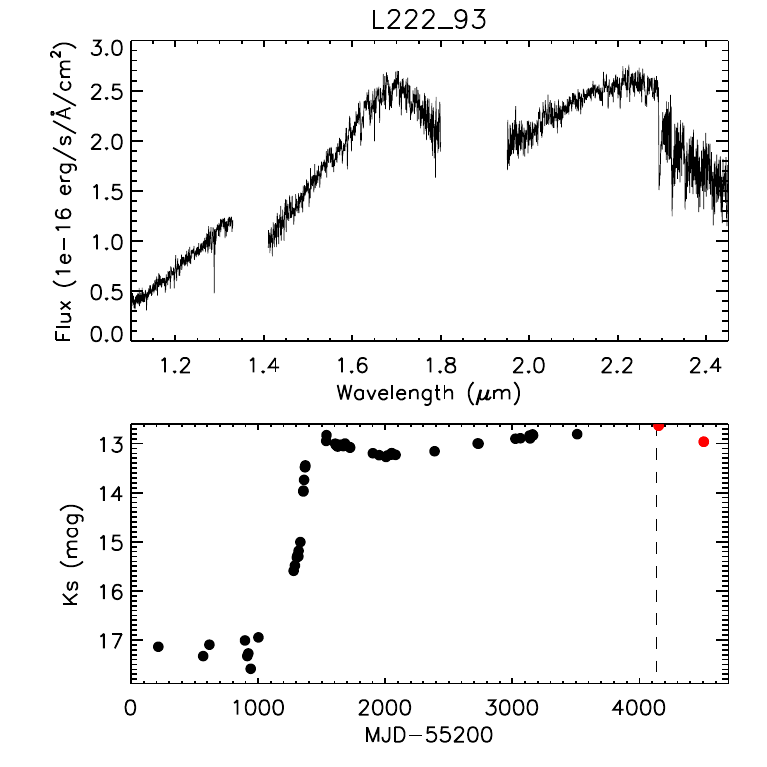}
\includegraphics[width=1.7in,angle=0]{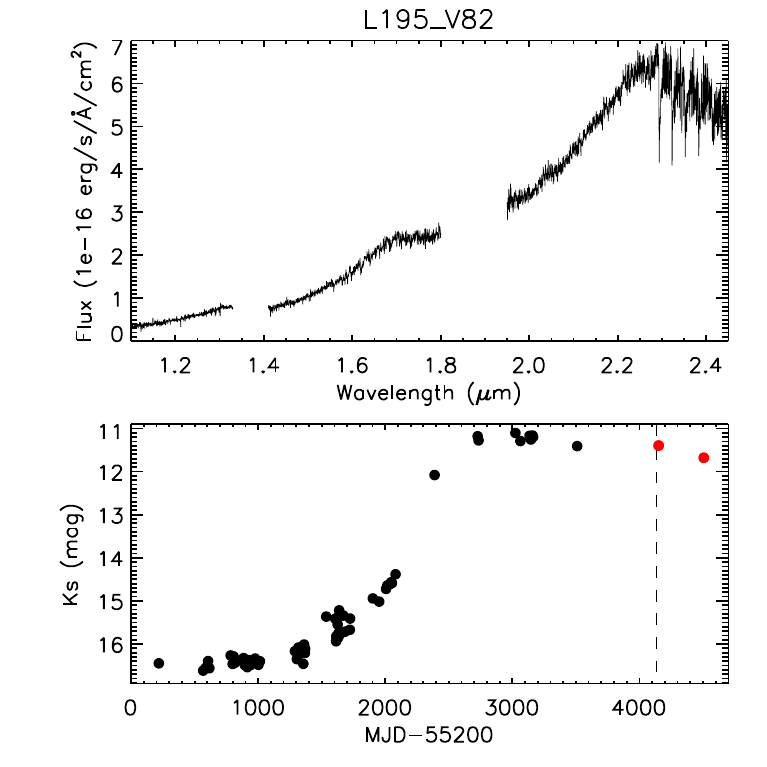}
\includegraphics[width=1.7in,angle=0]{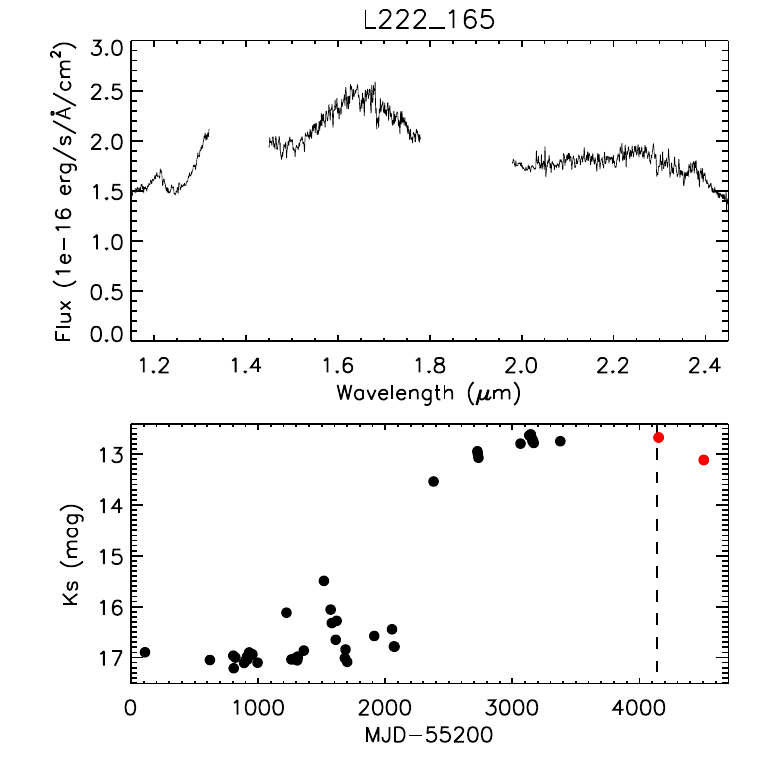}
\includegraphics[width=1.7in,angle=0]{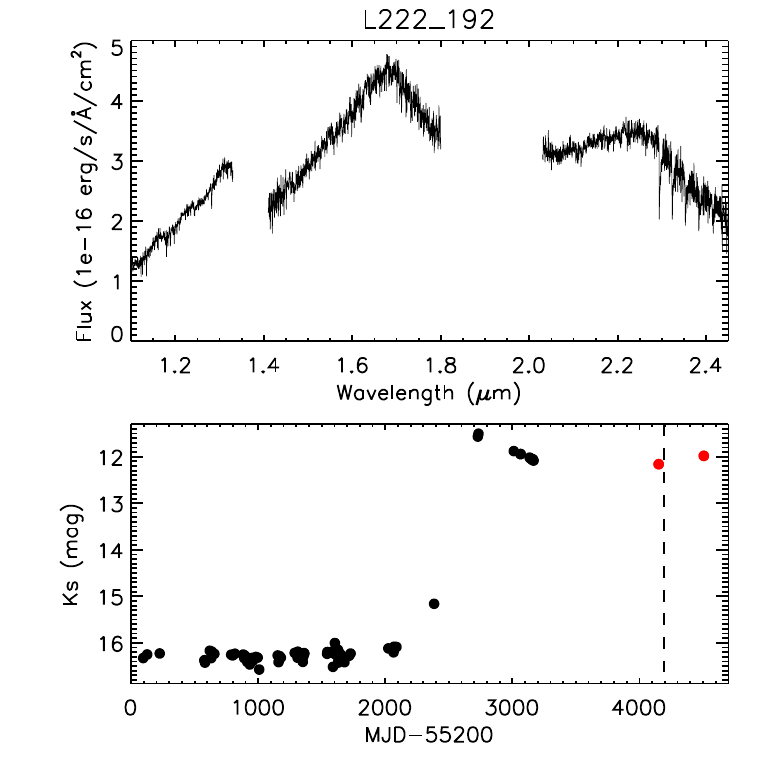}

\caption{XSHOOTER spectra and $K_s$ light curves of FUor-type objects confirmed in this paper. The detections from NTT/SOFI are marked in red. The observation time of each spectrum is presented as the dashed line in the lower panels.}
\label{fig:xshooter_spec_1}
\end{figure*}

\begin{figure*}
\includegraphics[width=1.9in,angle=0]{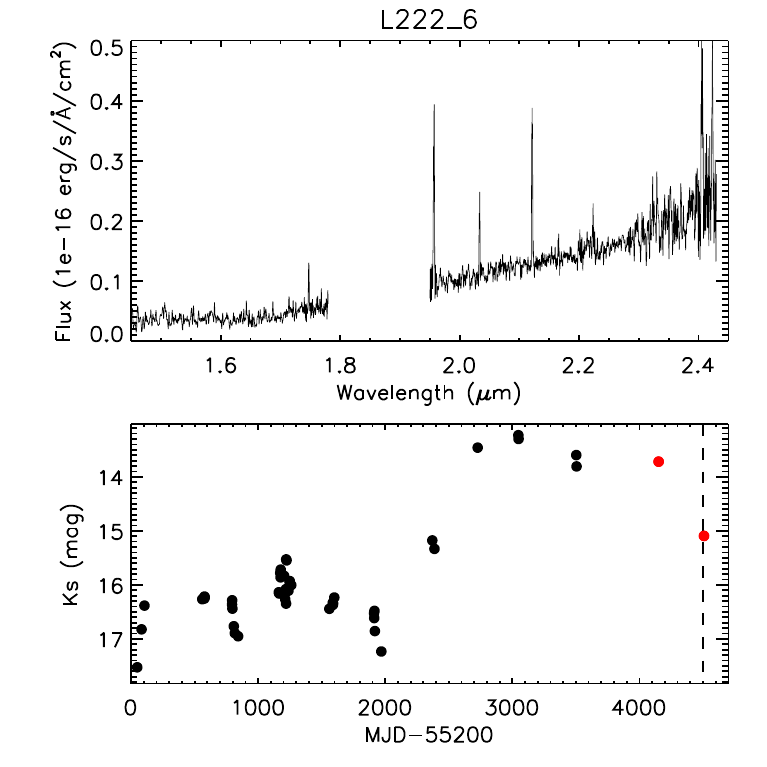}
\includegraphics[width=1.9in,angle=0]{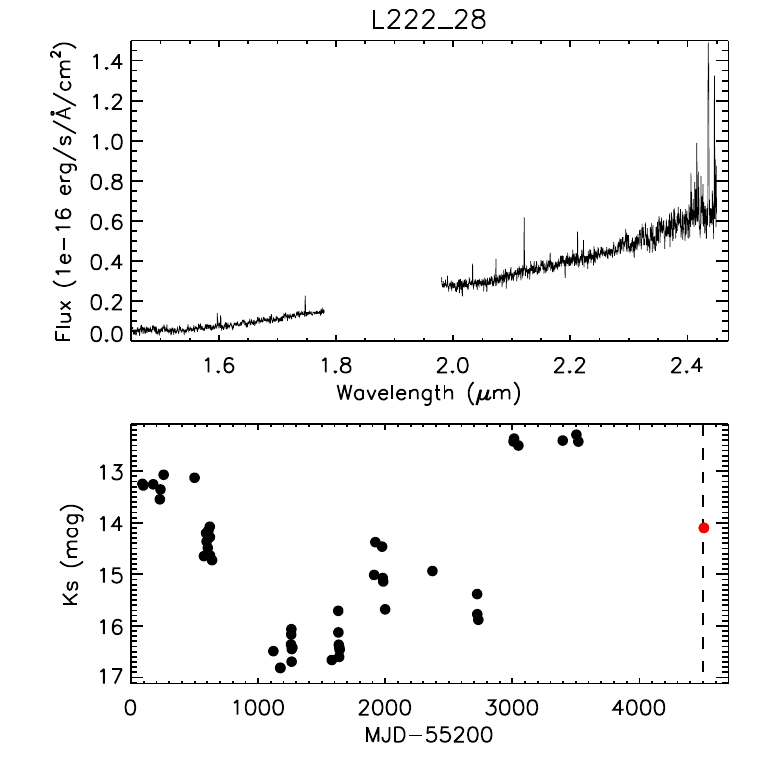}
\includegraphics[width=1.9in,angle=0]{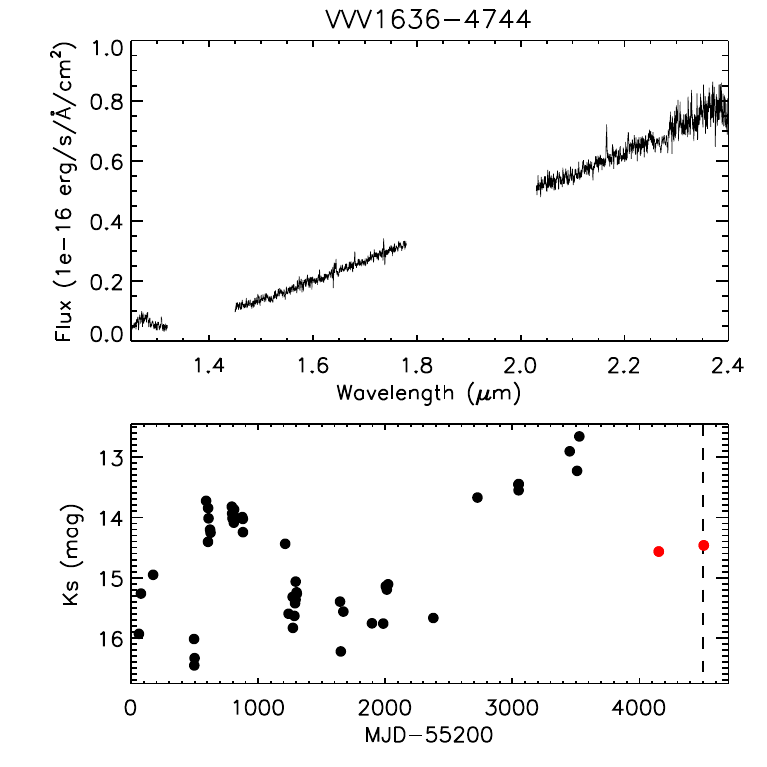}
\includegraphics[width=1.9in,angle=0]{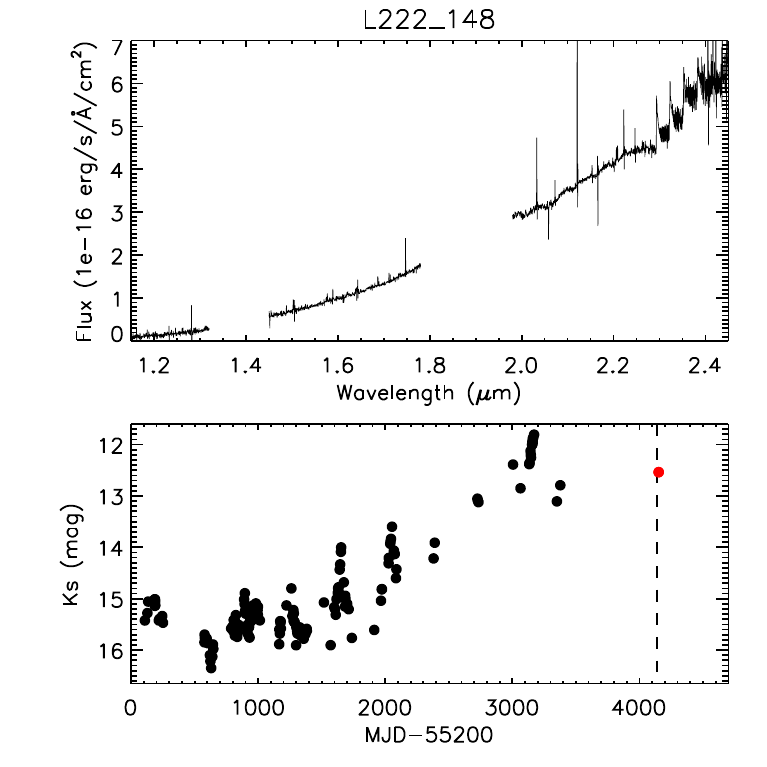}
\includegraphics[width=1.9in,angle=0]{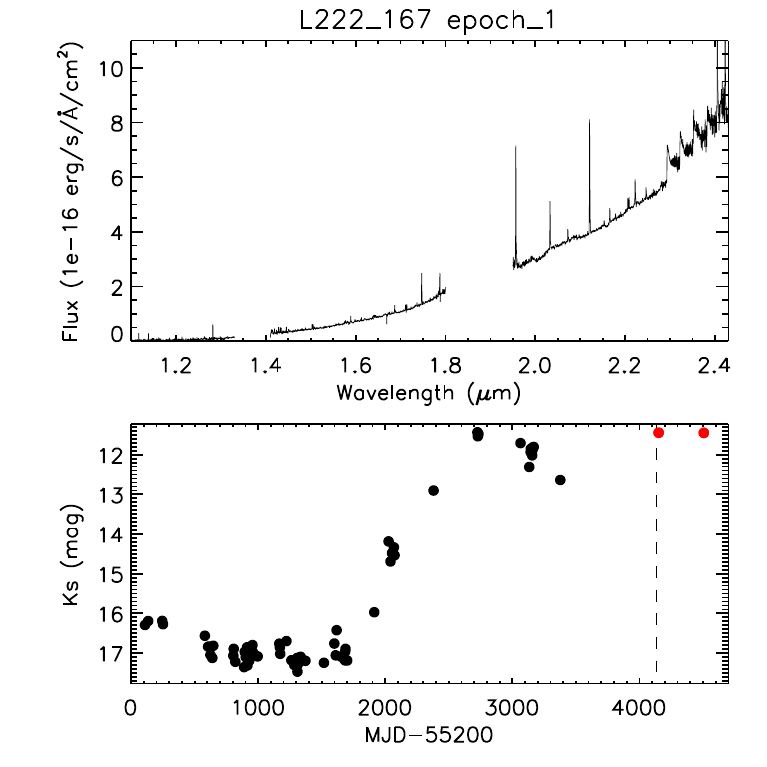}
\includegraphics[width=1.9in,angle=0]{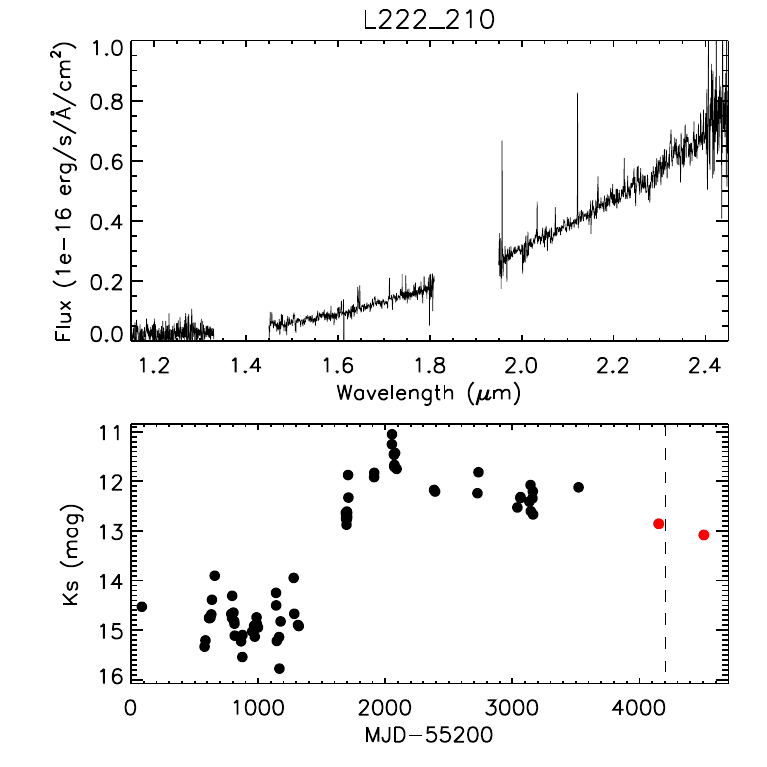}
\caption{XSHOOTER spectra and $K_s$ light curves of emission line objects. Symbols are the same as in Figure~\ref{fig:xshooter_spec_1}.}
\label{fig:xshooter_spec_3}
\end{figure*}
\begin{figure*}
\includegraphics[width=1.9in,angle=0]{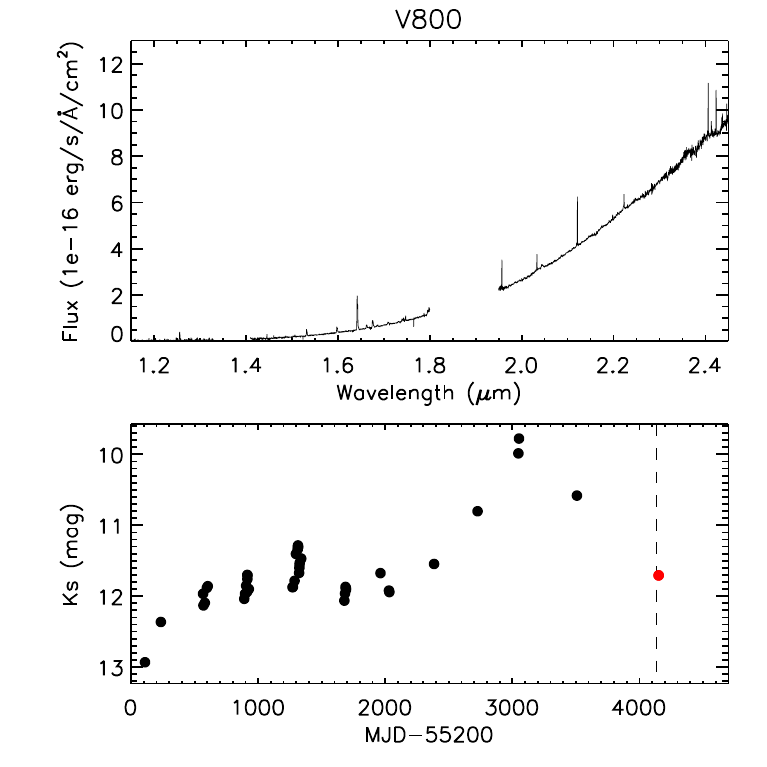}
\includegraphics[width=1.9in,angle=0]{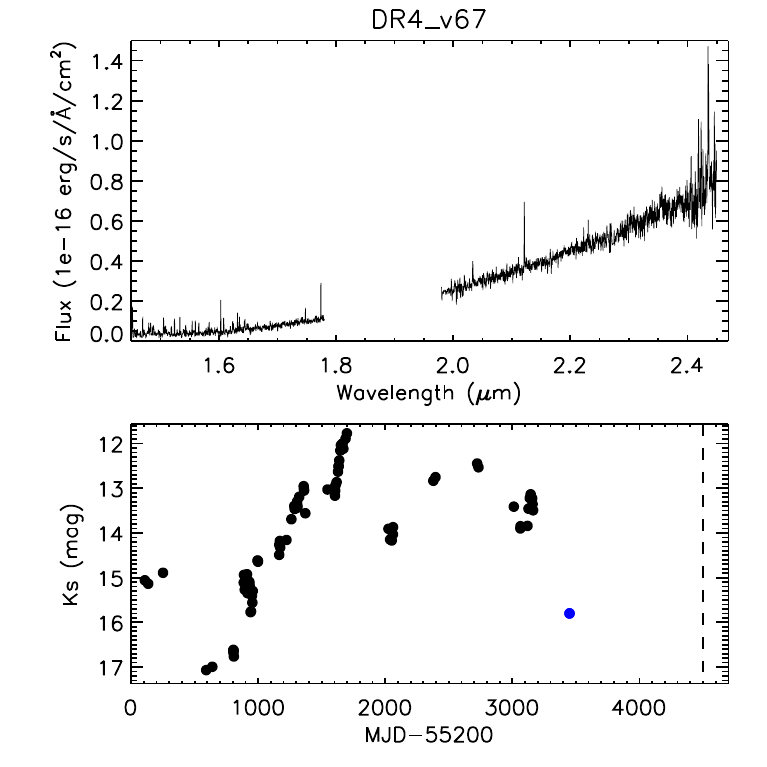}
\includegraphics[width=1.9in,angle=0]{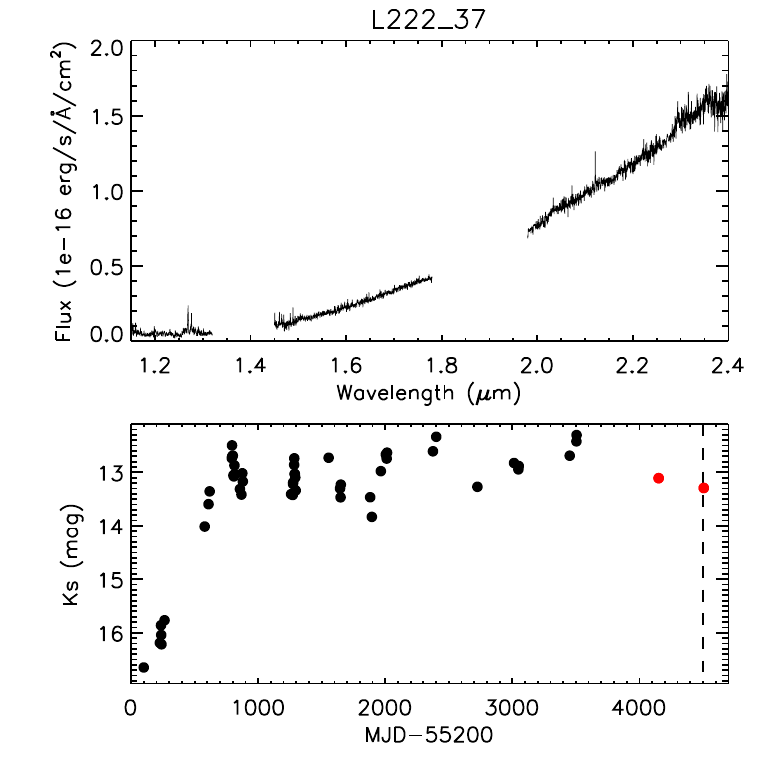}
\includegraphics[width=1.9in,angle=0]{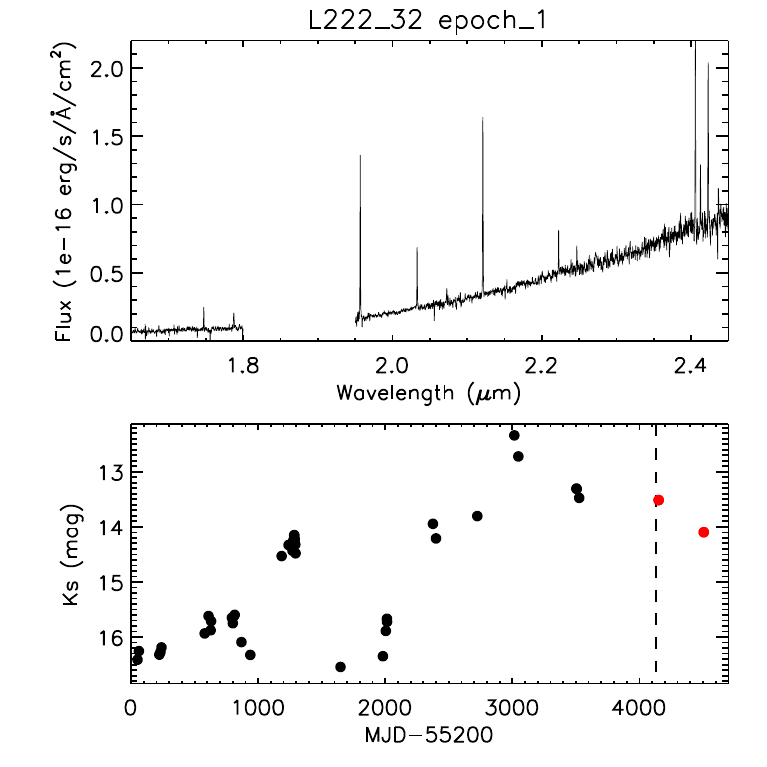}
\includegraphics[width=1.9in,angle=0]{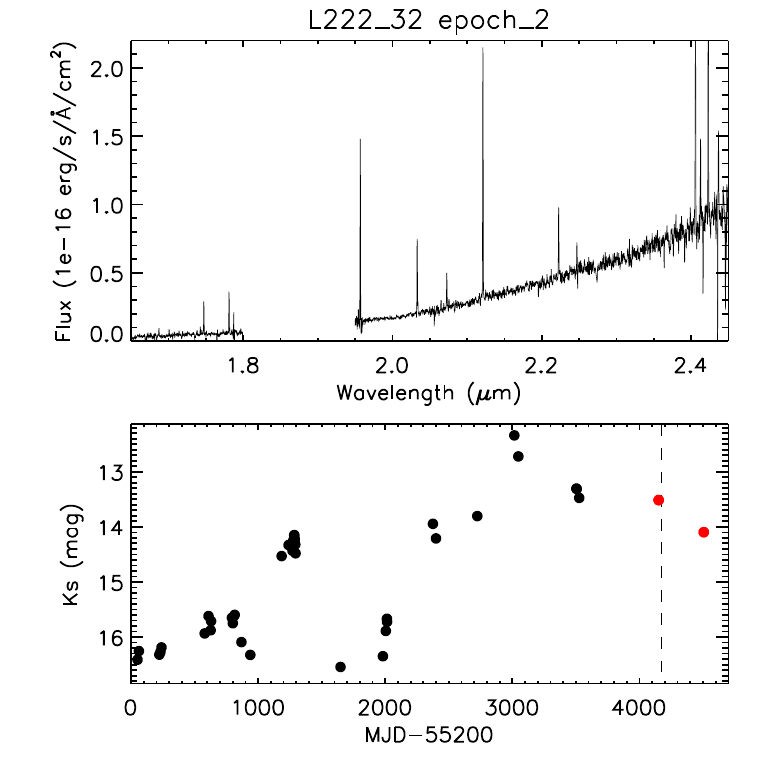}
\includegraphics[width=1.9in,angle=0]{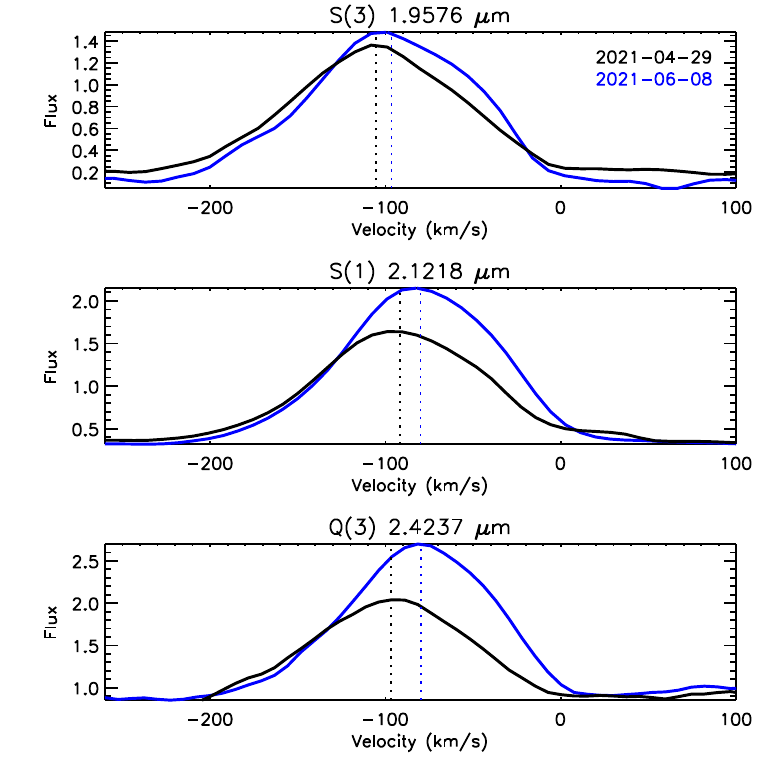}
\caption{XSHOOTER spectra and $K_s$ light curves of outflow dominated objects. Symbols are the same as in Figure~\ref{fig:xshooter_spec_1}. {\it Bottom Right:} Line profiles of three $H_2$ lines from two spectral epochs of L222\_32. The vertical dashed lines show the radial velocities of Gaussian fittings.}
\label{fig:xshooter_spec_4}
\end{figure*}
\begin{figure*}
\includegraphics[width=1.7in,angle=0]{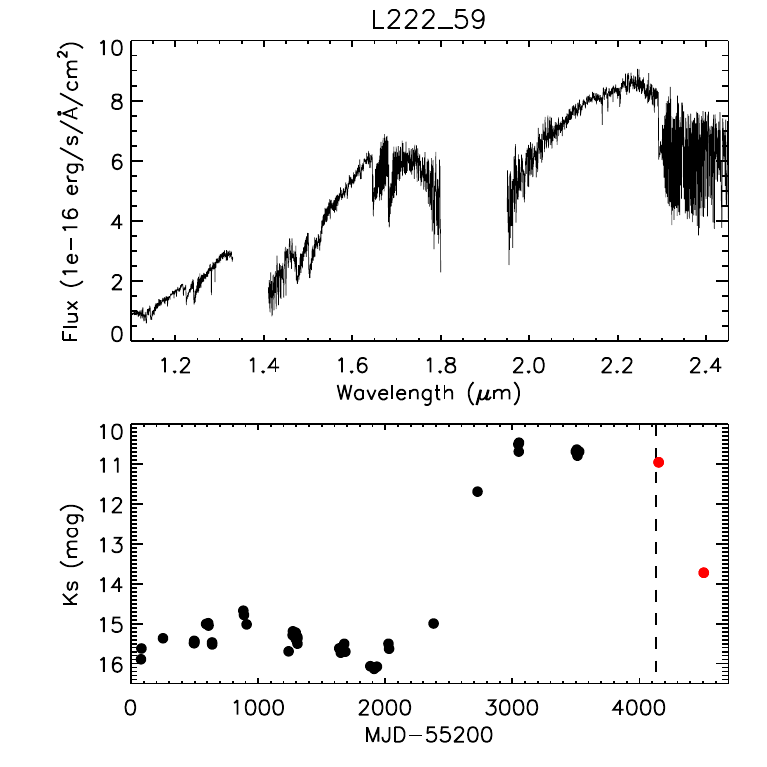}
\includegraphics[width=1.7in,angle=0]{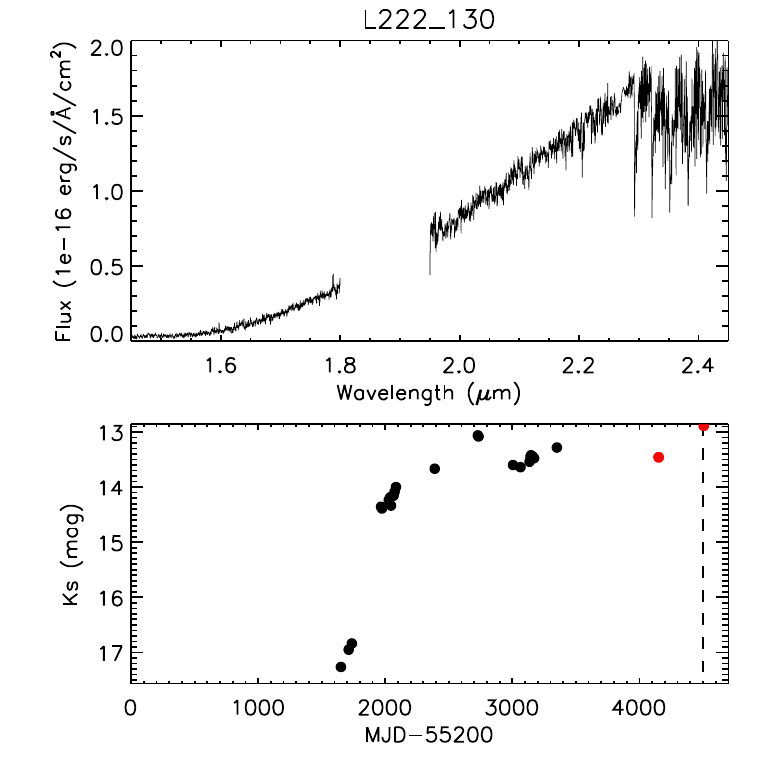}
\includegraphics[width=1.7in,angle=0]{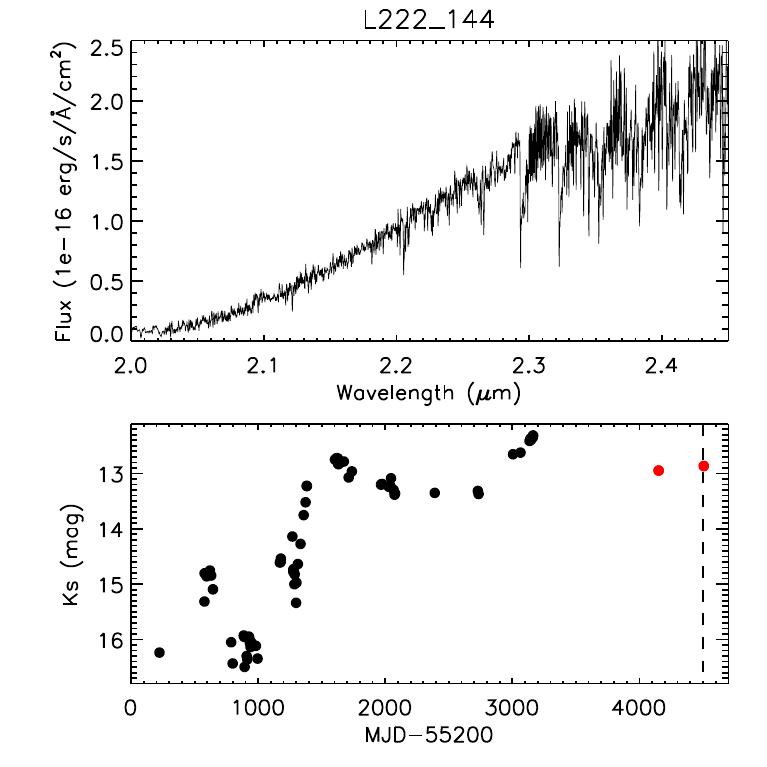}
\includegraphics[width=1.7in,angle=0]{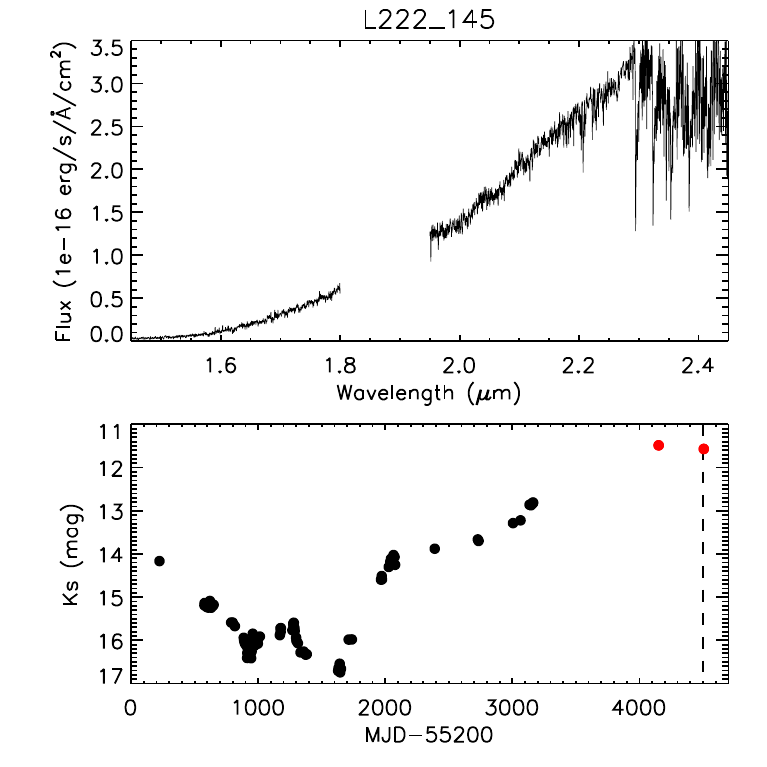}
\includegraphics[width=1.7in,angle=0]{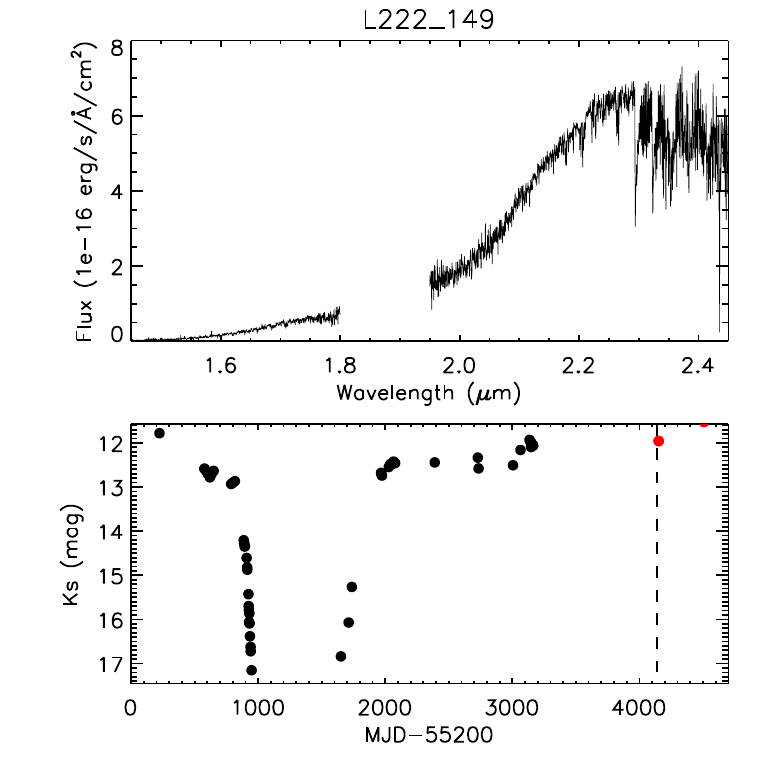}
\includegraphics[width=1.7in,angle=0]{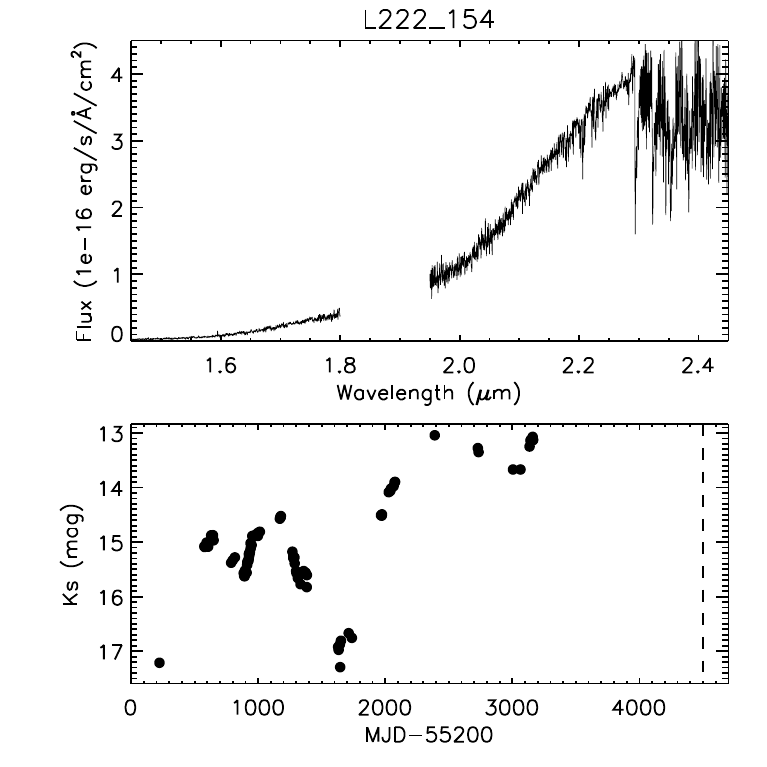}
\includegraphics[width=1.7in,angle=0]{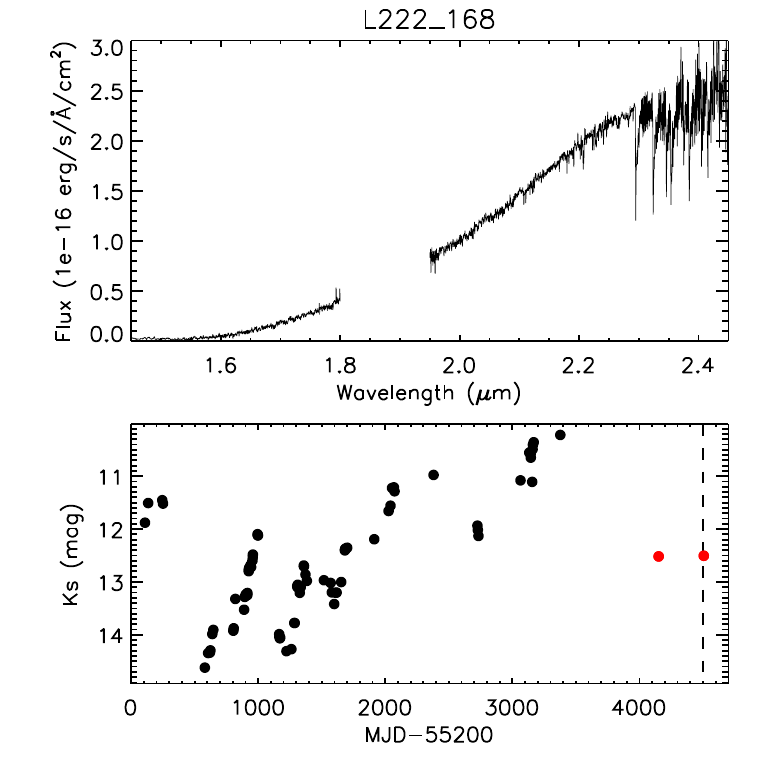}
\includegraphics[width=1.7in,angle=0]{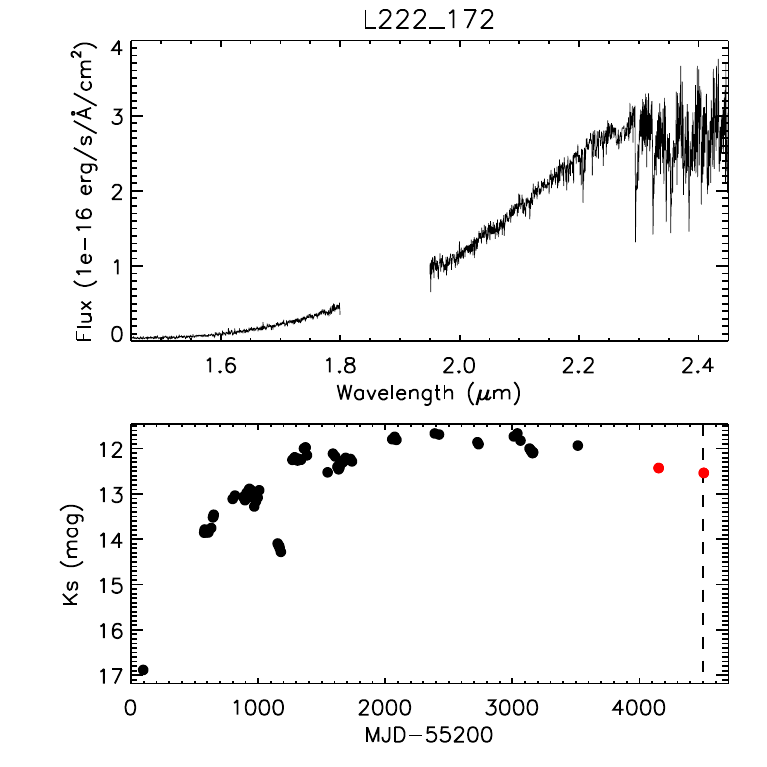}
\includegraphics[width=3.3in,angle=0]{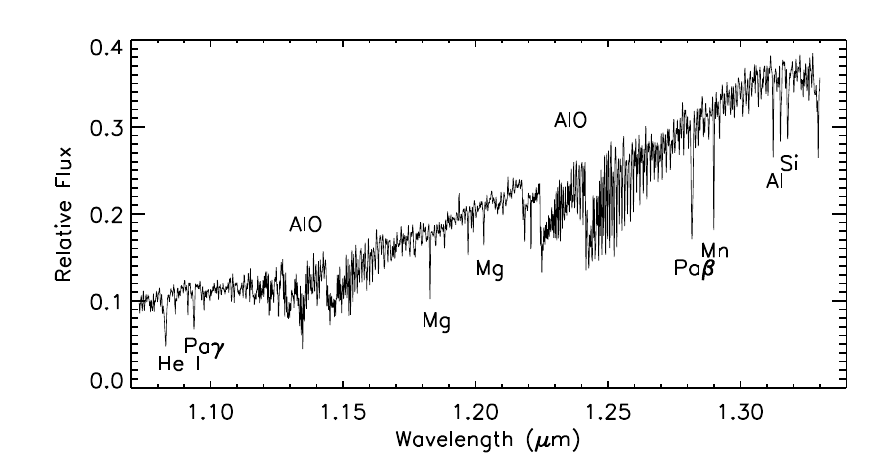}
\includegraphics[width=3.3in,angle=0]{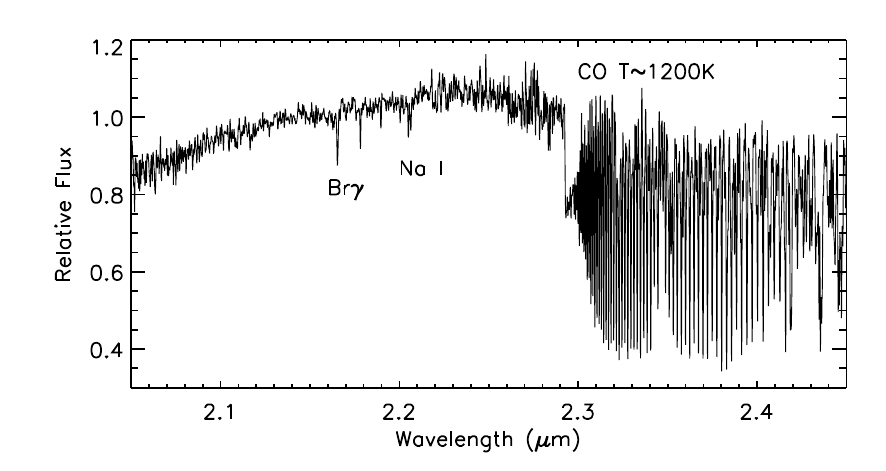}

\caption{XSHOOTER spectra and $K_s$ light curves of post-main-sequence giants. Symbols are the same as in Figure~\ref{fig:xshooter_spec_1}. The $J$ and $K$-bandpass spectra of L222\_59 are presented in the bottom two plots.}
\label{fig:xshooter_spec_5}
\end{figure*}
\begin{figure*}
\includegraphics[width=2.5in,angle=0]{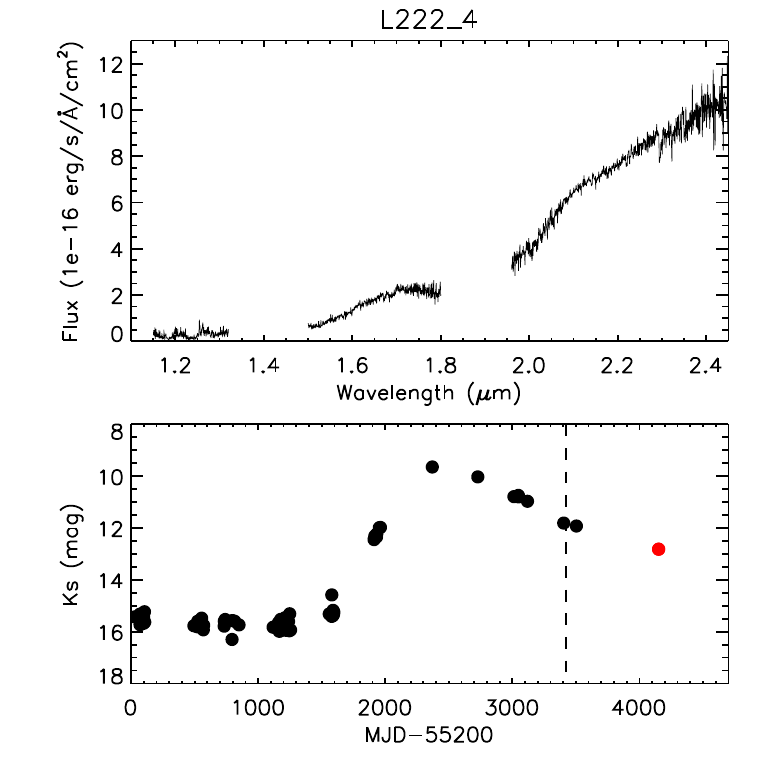}
\includegraphics[width=2.5in,angle=0]{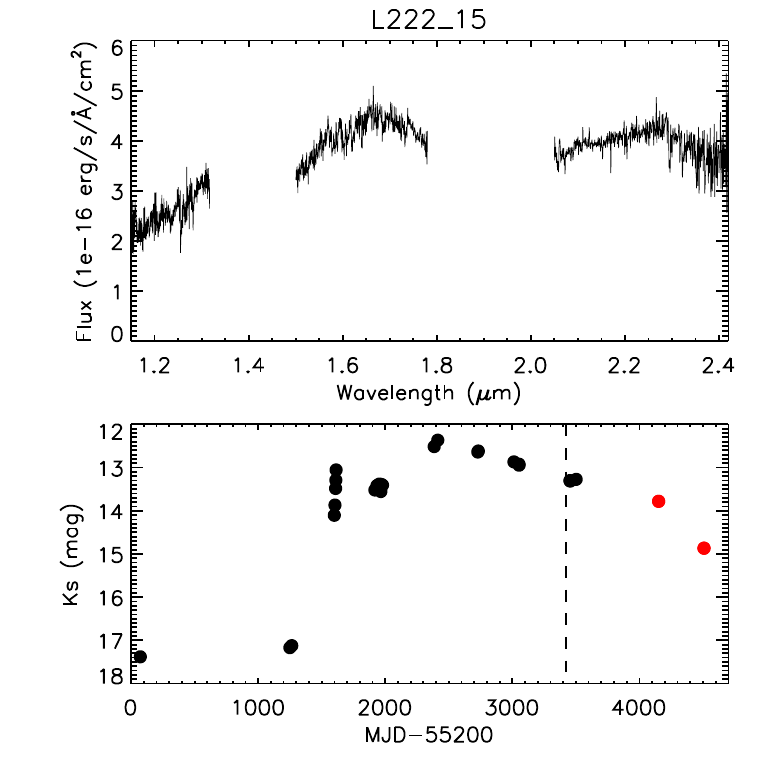}
\caption{Magellan/FIRE spectra of two FUors taken in 2019. Symbols are the same as in Figure~\ref{fig:xshooter_spec_1}. These two spectra were originally published by R. Kurtev as a poster at the Crete III conference in 2019 without permanent records.}
\label{fig:fire_spec}
\end{figure*}

\section{The near-infrared light curves of eruptive objects}
\begin{figure*}
\includegraphics[width=7in,angle=0]{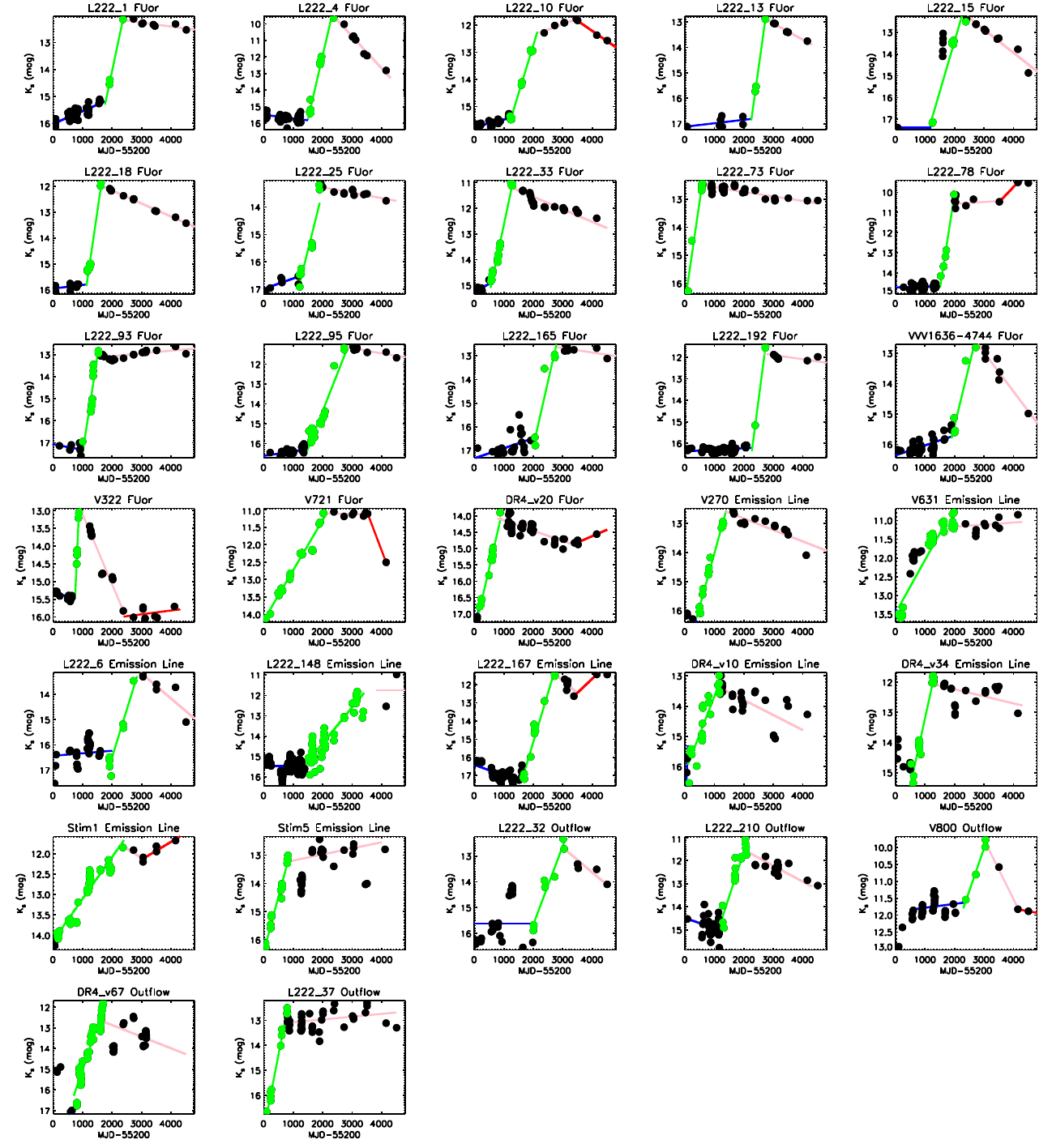}
\caption{$K_s$ light curves of long-duration eruptive YSOs discovered from the VVV survey. Linear functions are applied to fit different stages on the light curve. The eruptive stage, both the light curve and the best fit, is highlighted in green colour.}
\label{fig:eruptive_lc}
\end{figure*}

We present the $K_s$-band light curves, from both the VVV survey and our NTT/SOFI follow-up observations, of 32 eruptive objects that exhibit long-duration outbursts (see Figure~\ref{fig:eruptive_lc}). In most cases, we artificially divided the light curves into four stages: pre-outbursting/quiescent, rising, decaying and post-outbursting stages. The post-outbursting stage is only applied to sources that exhibit further eruptive or rapid decaying stage after the original outburst/photometric plateau. As mentioned in the main text, we used linear fits to measure the amplitude and duration of each stage. We noticed that not all light curves are well-presented by linear trends, especially the ones with outstanding short-timescale variability as opposed to the overall eruptive behaviour, such as L222\_6, L222\_32 and L222\_148. 

\section{Measurements of stellar parameters of post-MS stars}
\label{sec:postmsmeasurements}
In this section, we briefly describe our methods to fit stellar parameters on the post-MS stars. First, we applied the theoretical models of the CO bandhead emission beyond 2.29 $\mu$m. This model was initially designed in \citet{Contreras2017b}, using the rovibrational states and coefficients from \citet{Farrenq1991}. Three free parameters were initially designed in the model, including the radial velocity (RV), the temperature of CO gas ($T_{\rm CO}$), and the column density of CO molecules ($N_{\rm CO}$). To measure these parameters from the observed spectra, we generated synthetic spectra from the models, with $T_{\rm CO}$ ranging between 800 to 6000 K (100 K grids) and $N_{\rm CO}$ between 1e20 and 5e21 cm$^{-2}$ (1e20 cm$^{-2}$ grids). We first find the RV of the CO absorption following the methods described in the main text. Then, we removed the spectral continuum by applying polynomial fits and used the median flux between 2.285 and 2.290 $\mu$m to normalise the spectrum. Finally, we applied the minimum-$\chi^{2}$ method to locate the best-fitting models among our synthetic spectra. In the last step, to reduce the noise level, we smoothed both the observed and synthetic spectra by five pixels. The measured CO temperatures are presented in Table 3 of the main text. The typical error of this fitting is $\pm$200 K, mainly constrained by the signal-to-noise ratio of the spectra and the accuracy of the continuum fitting. Among these post-MS sources, L222\_59 has a unique CO morphology that indicates a very cool effective temperature (1200 K) and high column density (8e20 cm$^{-2}$). This cool $T_{\rm CO}$ is consistent with the AlO model presented in Figure 4 of the main article. Notably, such a low temperature implies the feature arises in circumstellar matter rather than the stellar photosphere. 

We applied a method based on the equivalent widths (EW) of the CO first overtone bands and Na {\sc i} absorption features to measure the metallicity,
as described in \citet{Fritz2021}. The measured EWs and [Fe/H] are listed in Table 3, with a typical error bar of 10\% introduced by the standard deviation of the continuum flux.  Notably, in the table, we changed the EW of absorption features to positive, to be consistent with standard practice. In addition, we used the method provided by \citet{Feldmeier-Krause2017}, to calculate the effective temperature of the stellar atmosphere ($T_{\rm eff}$) from the EW of CO bandhead, as
\begin{equation}
    T_{\rm eff}=5677.0 \,K - 106.3*EW_{\rm CO}.
\end{equation}
The intrinsic scatter of this method is 163~K according to \citet{Feldmeier-Krause2017}, which is similar to the uncertainty introduced by the measurements of EW. The measured $T_{\rm eff}$ are also presented in Table 3 of the main article. Except for one source with very cool CO, L222\_59, the $T_{\rm eff}$ measured from $EW_{\rm CO}$ are reasonably similar to $T_{\rm CO}$ measured from the shape of the CO absorption features, considering the errors. $T_{CO}$ is slightly lower than $T_{eff}$ in 5/7 cases, as one would expect for an absorption feature. According to the PARSEC stellar evolutionary models \citep[see][and references therein]{Pastorelli2019}, sources with $T_{\rm eff}$ ranging between 3000 to 4000 K have typical intrinsic $H-K_s$ colours between 0.15 to 0.35 mag. In comparison with the observed $H-K_s$ colours from the VVV survey (listed in Table 1), we found that these post-main-sequence sources are indeed severely embedded. Applying the extinction law from \citet{WangS2019}, 
$A_V = 18.8 E(H - K_s)$, the average extinction of these sources are $A_V = 57$ mag or $A_{K_s} = 4.5$ mag.
\end{document}


\appendix
\section{Star forming regions associated with each YSO}
Here we present a table containing the name and location of variable YSOs in this work that have star-forming regions (SFR) located within 10 arcmin distance, according to our search on Simbad. The kinematic distance of the closest SFR to individual variable YSO is accepted as the d$_{\rm SFR}$.

\begin{table}
    \centering
    \begin{tabular}{cccccccc}
    \hline
    \hline
            Name & R.A. (deg) & Dec (deg) & SFR & d (arcmin) & RV (km/s) & $d_{\rm near}$ (kpc) & $d_{\rm far}$ (kpc)\\
    \hline
        L222\_1 & 175.78945 & -62.35367 & [RC2004] G295.1-0.6-5.2  & 3.0 & 28.4 & - & 10.1 \\
        L222\_4 & 185.22517 & -62.63942 & [MML2017] 2257 & 5.1 & -9 & 0.59 & 7.79 \\
        L222\_6   & 193.73921 & -61.04417 & PGCC G303.38+01.79 & 2.3 & - & 4.76 & - \\
        L222\_13 & 203.04039 & -62.73005 & AGAL 307.616-00.259\_S  & 3.0 & -36.7  & 3.33 & 7.05 \\
        L222\_15 & 204.54729 & -62.48267 & AGAL 308.399-00.176\_S & 7.4 & -24.5 & 2.02 & 8.54 \\
        L222\_25 & 230.39117 & -57.88889 & [MML2017] 3366 & 6.4 & -51.6 & 3.44 & 9.93 \\
        L222\_32 & 239.45987  & -53.95964 & AGAL 328.254-00.532\_S & 1.4 & -54.0 & 3.19 & 11.26 \\
        L222\_33 & 239.85950 & -51.95328 & AGAL 329.717+00.806\_S & 3.2 & -49.9 & 3.43 & 11.25 \\
        VVV1640-4846 & 250.0490 & -48.7815 & GAL 336.51-01.48 & 4.0 & -25 & 2.1 & 13.5 \\
        L222\_73 & 258.65950 & -38.49144 & [MML2017] 3296 & 4.1 & -89 & 6.2 & 10.5 \\
        L222\_93 & 261.73025 & -34.14661 & [WHR97] 17234-3405 & 0.8 & -50.0 & 7.1 & 12.9 \\
        L222\_95 & 262.28616 & -33.52969 & [MML2017] 4244  & 4.9 & -62.5 & 6.5 & 10.4 \\
        L222\_148 & 266.64096 & -29.37922 & [KC97c] G359.7-0.4  & 0.1 & -12.0  & 7.8 & 9.10 \\
        L222\_167 & 267.60937 & -28.87519 & [AAJ2015] G000.583-00.870 & 0.8 & 14.2 & 7.8 & 8.4 \\
        L222\_168 & 267.66667 & -28.06083 & [MML2017] 1208 & 6.5 & 167 & 8.39 & 8.60  \\
        L222\_192 & 269.43525 & -24.34244 & [MML2017] 533 & 3.2 & 46 & 6.1 & 10.8 \\
    \hline
    \hline
    \end{tabular}
    \caption{Spectroscopically confirmed YSOs and their closest (projected distance) star-forming regions.}
    \label{tab:my_label}
\end{table}